\documentclass[12pt]{article}
\usepackage{blindtext}
\usepackage[utf8]{inputenc}
\usepackage{natbib}
\usepackage{amsmath}
\usepackage[letterpaper,top=2cm,bottom=2cm,left=3cm,right=3cm,marginparwidth=1.75cm,margin=1in ]{geometry}
\usepackage{hyperref}
\usepackage{graphicx}
\usepackage{graphics}
\usepackage{tabularx}
\usepackage{mwe}
\usepackage{authblk}
\usepackage{arydshln}
\usepackage{float}
\usepackage[toc,page]{appendix}
\usepackage{graphbox}
\usepackage{calrsfs}
\DeclareMathAlphabet{\mathcal}{OMS}{zplm}{m}{n}
	
\usepackage{xcolor}
\usepackage{multirow}
\usepackage{amsthm}

\usepackage{adjustbox}
\newcolumntype{H}{>{\setbox0=\hbox\bgroup}c<{\egroup}@{}}
\allowdisplaybreaks
\DeclareMathOperator*{\T}{\top}

\DeclareMathOperator*{\argmin}{arg\,min}

\newtheorem{prop}{Proposition}

\newtheorem{lem}{Lemma}
\newtheorem{remark}{Remark}

\usepackage{amsfonts}
\title{ \bf Fusion regression methods with repeated functional data }
\author[1,2*]{Issam-Ali Moindji\'e}
\author[1,2,3]{Cristian Preda}
\author[1,2]{Sophie Dabo-Niang}

\affil[1]{
MODAL team, Centre Inria
de l'Université de Lille, Villeneuve d'Ascq-59650,  France}

\affil[2]{CNRS UMR 8524-Laboratoire Paul Painlevé, Université de Lille,  Villeneuve d'Ascq-59650, France}

\affil[3]{
Institute of Statistics and Applied Mathematics of the Romanian Academy, Bucharest-050711, Romania}

\affil[*]{\small Corresponding author, email: \href{mailto:issam-ali.moindjie@inria.fr}{issam-ali.moindjie@inria.fr}}
\date{}
\begin{document}

\maketitle

\begin{abstract}

{
Linear regression and classification methods with repeated functional data are considered. For each statistical unit in the sample, a real-valued parameter is observed over time under different conditions related by some neighborhood structure (spatial, group, etc.). 
Two regression methods based on fusion penalties are proposed to consider the dependence induced by this structure. 
These methods aim to obtain parsimonious coefficient regression functions, by determining if close conditions are associated with common regression coefficient
functions. 
The first method is a generalization to functional data of the variable fusion methodology based on the 1-nearest neighbor. 
The second one relies on the group fusion lasso penalty which assumes some grouping structure of conditions and allows for homogeneity among the regression coefficient functions within groups. Numerical simulations and an application of
electroencephalography data are presented.} 
\end{abstract}

\section{Introduction}

\begin{figure}[H]
    \centering
    \scriptsize
    \begin{tabular}{c c }
         \includegraphics[scale=.3, align=c]{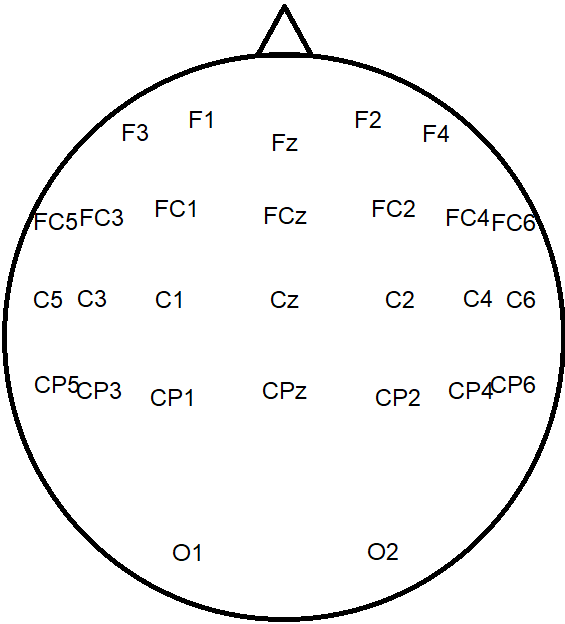} & 
    \begin{tabular}{c c c c } 
         F1: & \includegraphics[align=c, scale=.1]{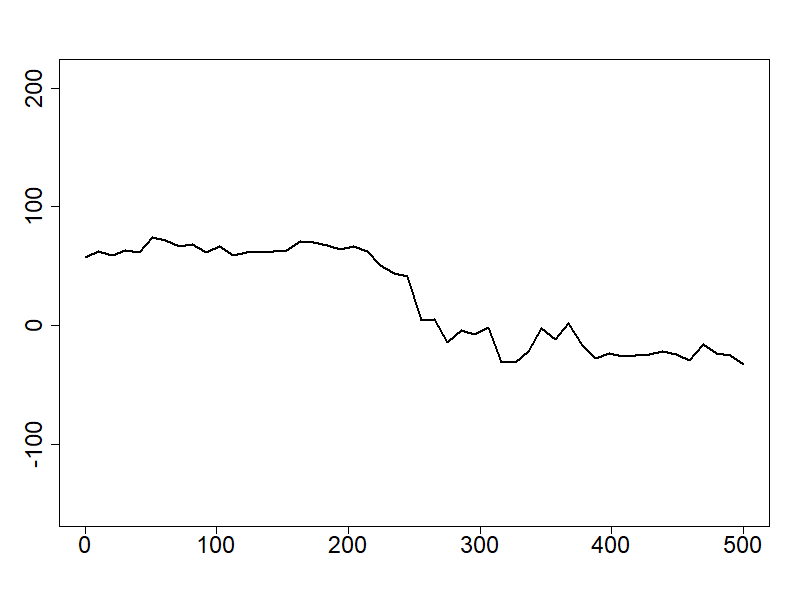} & F2: & \includegraphics[align=c, scale=.1]{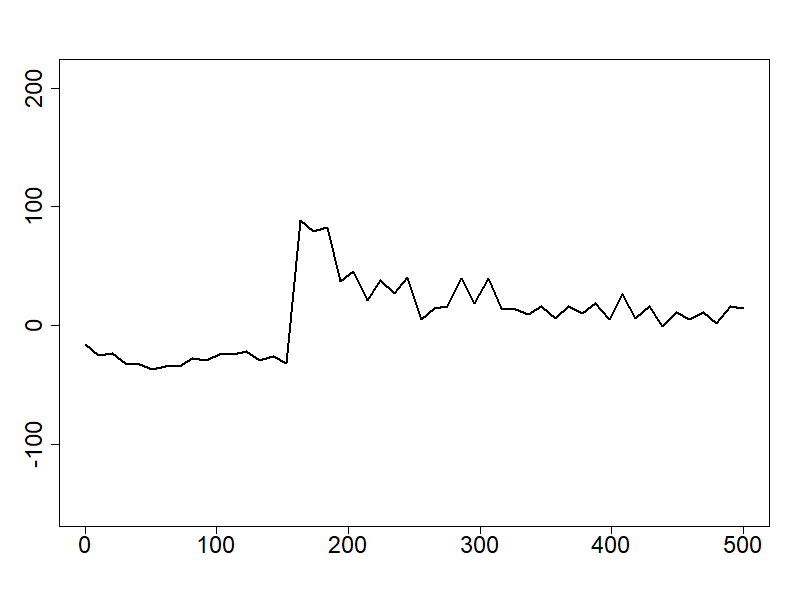}\\
         F3: & \includegraphics[align=c, scale=.1]{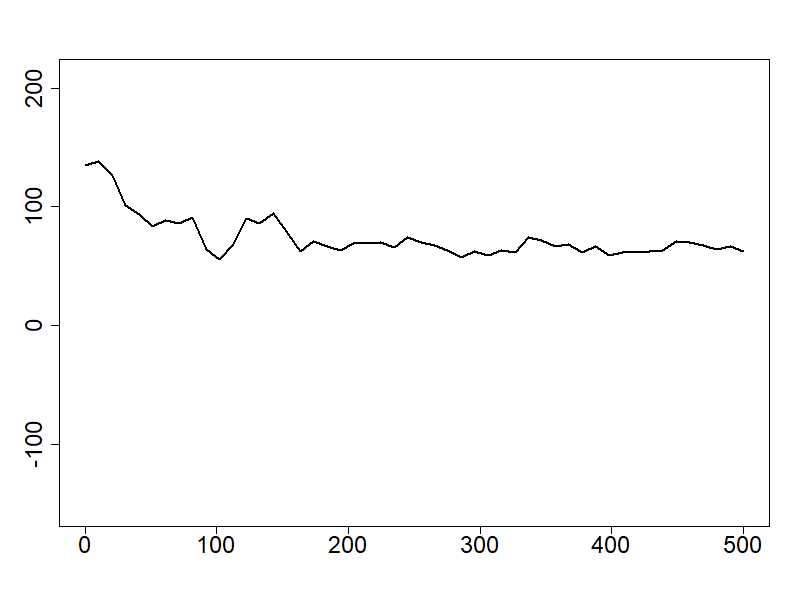} &F4: & \includegraphics[align=c, scale=.1]{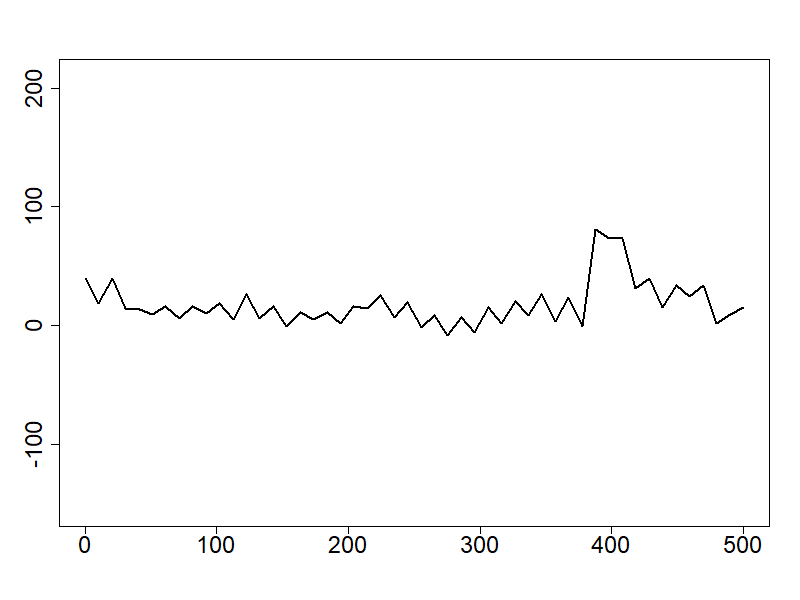}  \\ 
          & $\vdots $ & & $\vdots$ \\ 
           O1: & \includegraphics[align=c, scale=.1]{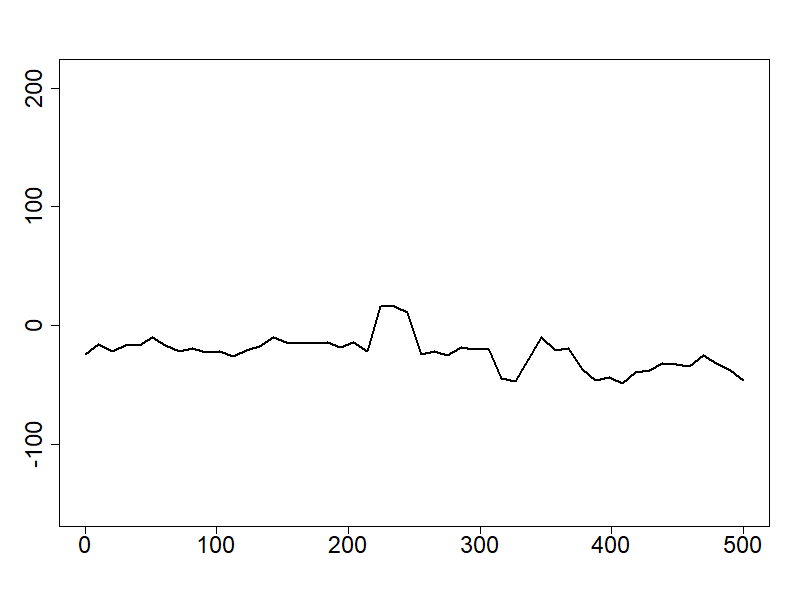} & O2: &      \includegraphics[align=c, scale=.1]{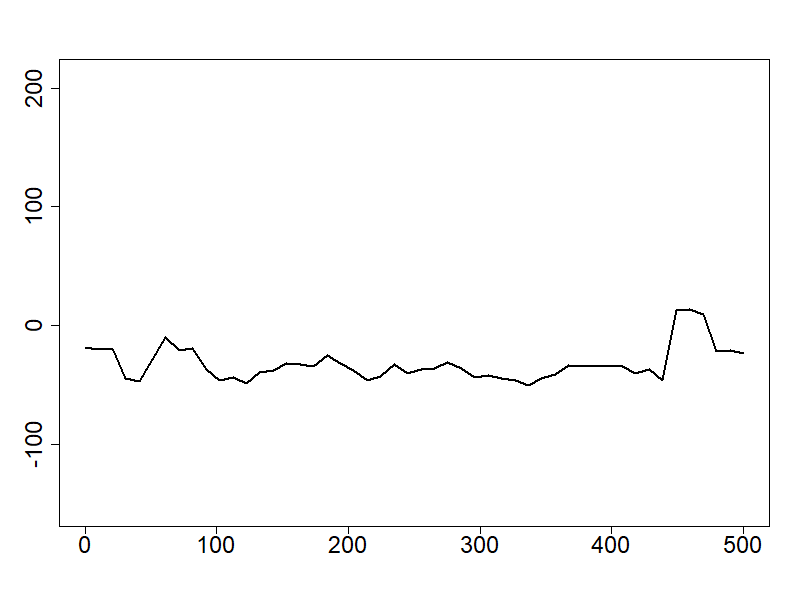}
    \end{tabular}
    \end{tabular}
    \caption{\textit{FingerMovements} data. Each subject is represented by $p=28$ EEG recordings (right). The sensors are {disposed} on the scalp according to the map in the left figure. F, C, P and O stand respectively for the Frontal, Central, Parietal and Occipital regions. } 
    \label{fig0}
\end{figure}

Let $X$ be a functional random variable valued in some {\color{black} Hilbert} space of real-valued functions defined on the time interval $[0, T]$, $T > 0$. {\color{black} Without loss of generality, we assume that this space is the set of squared integrable functions $L_2([0, T])$ \citep{ramsay2008}.} The setting we consider in this paper assumes that $X$ is observed under $p$ different conditions $\{\mathcal{C}_1, \ldots, \mathcal{C}_p\}$, $p\geq 1$.  For instance, these conditions can represent times or/and locations (regions) in some metric space $(\mathcal{S},d)$, typically $(\mathbb{R}^{s}, \| \cdot \|_2)$, for some natural integer $s\geq 1$.  Thus, proximity or grouping structures of conditions can be considered through the distance $\color{black}d(\cdot, \cdot)$ which, depending on the space $\mathcal{S}$, could be the Euclidean distance or some other well-suited distance (for example the great circle distance if $\mathcal{S}$ is a sphere). { An example of this type of data model comes from the field of neuroscience:}  electroencephalography recordings (EEG) \citep{great} represent the brain activity through the electric field intensity over a time interval of $T = 500 ms$ at different regions of the brain, using $p=28$ electrodes/sensors evenly distributed (Figure \ref{fig0}).

Let denote with $X^{(j)}$ the observation of $X$ under the condition $\mathcal{C}_j$, $j=1, \ldots, p$, and with $\mathbf{X}$, the random vector 
$$\mathbf{X}=(X^{(1)}, \ldots, X^{(p)})^{\top}.$$
{ The realizations of $\mathbf{X}$
are known as repeated functional data: the functional random variable $X$ is repeatedly observed $p$ times under different conditions.}  { In this framework, the first major contribution is due to \cite{chen2012} for the development of the principal component analysis (PCA) method.} The authors use a double PCA exploiting the metric structure of the space of conditions $\{\mathcal{C}_1, \ldots, \mathcal{C}_p\}$ belong. In \cite{jacques2014} that structure is ignored and $\mathbf{X}$ is viewed as a $p$-dimensional functional random vector of which principal components are used for { visualization and unsupervised classification}.
\par { To the best of our knowledge, supervised learning with repeated functional data has not been specifically addressed in the literature. In the existing works on supervised models, the repeated feature of data is taken into account as in the classical multivariate functional setting, through the covariance operators of ${\bf X}$ (see e.g \cite{yi2022},
\cite{gorecki2015}, \cite{beyaztas2022}, \cite{moindjie2024classification}). 
These methods ignore that the components of the random vector ${\bf X}$ have a common underlying model ($X$) endowed with some dependence structure given by the conditions $\{\mathcal{C}_1, \ldots, \mathcal{C}_p\}$.} 
\par 
{In this work, we propose methods that can consider, through the (observed) conditions $\{\mathcal{C}_1,
\ldots, \mathcal{C}_p\}$, the
topology of $(\mathcal{S}, d)$ in the estimation procedure of the linear regression model with univariate response $Y$ (scalar or binary random variable) and $\mathbf{X}$ as a predictor. In particular, neighborhood relationships or group membership
among the components of $\mathbf{X}$ are used for the estimation of the regression function.} \par

{ Taking into account the topology of $(\mathcal{S}, d)$ enhances the interpretability of the proposed model, especially in the context of brain activity data: since neighboring brain regions often exhibit correlated activity patterns, the components $X^{(j)}$ that are spatially close are likely to provide similar information in the regression model.  Therefore, considering the spatial proximity of sensors allows for a more accurate and meaningful representation of the underlying neural processes, as adjacent sensors are expected to capture related brain activity. This approach takes into account the inherent spatial structure of brain data to improve the robustness and interpretability of the model's estimates.} As an example, in the { EEG} classification application,
each subject is writing a text and the electric field intensity $X$ is measured simultaneously
at $p=28$ spatial positions (sensors) of the scalp during $T=500ms$. Two groups of subjects
are considered: the right-handed $(Y=0)$ and left-handed $(Y=1)$ writers. The question is
to know and interpret in what measure a person's ability to be left or right-handed is associated with some different activity of the brain. {Our hypothesis is that close
sensors might share/provide similar information {in} this classification problem}. 
\par 
{
The proposed methodologies rely on the standard functional linear regression model}. It assumes that there exist $\beta^{(0)} \in \mathbb{R}$ and { a} regression coefficient function $\boldsymbol{\beta} = (\beta^{(1)}, \ldots, \beta^{(p)})^{\top} \in {{\mathcal{H}^p} = \{L_2([0,T])\}^p}$ such that 

\begin{equation}
\label{ml}
 \mathbb{E}(Y|\mathbf{X}) \approx { \color{black} \color{black}\beta^{(0)} + \sum_{j=1}^{p} \langle X^{(j)}, \beta^{(j)} \rangle_{L_2} }
\end{equation}
where $$\langle X^{(j)}, \beta^{(j)} \rangle_{L_2}= \int_{0}^{T}X^{(j)}(t)\beta^{(j)}(t)dt, $$ for $j=1, \ldots, p$. If $\{(\mathbf{X}_i, Y_i)\}_{i=1:n}$ is an i.i.d. sample of size $n$, $n \geq 1$, drawn from the same distribution as $(\mathbf{X}, Y)$ and $\{(x_i,y_i)\}_{i=1:n}$ is an observation of that sample{, }the estimation of the model (\ref{ml}) is based on the minimization of the mean of the squared errors (MSE), that is,   
 \begin{equation}
 \label{critere}
 (\hat \beta^{(0)}, \hat\beta)=\argmin_{(\psi^{(0)}, \psi)\in \mathbb{R}\times  {\mathcal{H}^p} }\frac{1}{n}\sum_{i=1}^{n} \left( y_i -\left(\psi^{(0)}+ \sum_{j=1}^{p}\langle x_i^{(j)}, \psi^{(j)} \rangle_{L_2}\right) \right)^2. 
 \end{equation}
 
Because of the non-invertibility of the covariance operator, the direct estimation of the coefficient $\boldsymbol{\beta}$ under the minimization of the MSE criterion is an ill-posed problem \citep{cardot1999}. The principal component regression (PCR) and the partial least squares (PLS) have been successful alternatives in this case (see e.g \cite{escabias2005}, \cite{aguilera2006using}, \cite{PLS}, \cite{moindjie2024classification}). {\color{black}However, in these approaches, the estimated coefficient functions are sometimes difficult to interpret: why do two components $X^{(j)}$ and $X^{(j')}$ that are associated to close conditions $\mathcal{C}_j$ and $\mathcal{C}_{j'}$, i.e. the distance $d(\mathcal{C}_{j}, \mathcal{C}_{j'})$ is small, have very different associated coefficient functions $\beta^{(j)}$ and $\beta^{(j')}$?}
This situation occurs especially when $p$ is large. 
In \cite{groupfda}, the authors propose to add the constraint $\mathcal{P}= \sum_{j=1}^{p} ||\psi^{(j)}||_{_{L_2}}$ in the regression model. 
In this case, $\mathcal{P}$ is a generalization to functional multivariate variables of the {group lasso }penalty (GL), originally introduced in the multivariate data case (\cite{group_rl}, \cite{yuan2006model}).\par 
This penalty leads to achieving a trade-off between a minimum number of contributing components $X^{(j)}$ and model fit. Our hypothesis is that closeness between components $X^{(j)}$, in the sense of the distance $d$ between the corresponding conditions $\mathcal{C}_j$, can help for a better interpretation of $\boldsymbol{\beta}$. For this purpose, the {\it fusion penalty} was introduced in the finite multivariate setting in \cite{fusion_variable}. \par Let $v$ be a surjective function,  $v :  \{\mathcal{C}_1, \ldots, \mathcal{C}_p\}\mapsto \{1, \ldots, K\} $, $K\leq p$, and define the 

 fusion penalty (in the functional framework) as
 $$ \mathcal{P}(\boldsymbol{\beta})=\sum_{k=1}^{K} \sqrt{\sum_{j\in \mathcal{I}_k} \left|\left|\beta^{(j)}- \bar{\beta}_{\mathcal{I}_k}\right|\right|^2_{L_2}},   $$
where for each $k= 1:K$, $\mathcal{I}_k=\left\{ j : v(\mathcal{C}_j)=k \right\}$ and $\bar{\beta}_{\mathcal{I}_k}(t) = \frac{1}{|\mathcal{I}_k|}\sum_{j\in \mathcal{I}_k}\beta^{(j)}(t)$ for $t\in [0,T]$ and $|\mathcal{I}_k|$ denotes the cardinal of $\mathcal{I}_k$ for $k=1, \ldots, p$. 

Then, the proximity between conditions $\mathcal{C}_1, \ldots, \mathcal{C}_p$ can be integrated through the function $v$ and the distance $d$ : $v^{-1}(k)$  represents all conditions closest to  $\mathcal{C}_k$. This penalty favors close dimensions of ${\bf X}$ to have similar corresponding dimensions of the regression function (the $\beta^{(j)}$'s functions).
\par To our knowledge, this penalty has not been explored in the case of regression with repeated functional variables (nor multivariate functional variables). In the classical multivariate setting, 
the models that have this penalty are known as the variable fusion model (FU) \citep{fusion_variable} and, when a lasso penalty is added, as the fused lasso method (FL) \citep{fused_lasso}. More recently, the group fusion method introduced in \cite{bleakley2011} extended this penalty from unique conditions to groups of conditions. However, in these cases, the function $v$ was defined as a way to integrate consecutive conditions (or dimensions), i.e. $v$ is defined as $v(\mathcal{C}_j)=j+1$, for $1\leq j \leq p-1$ and $v(\mathcal{C}_p)=p$. Even if this case can be well-suited for the setting $\mathcal{S}\subset \mathbb{R}$ ($s=1$), it limits the number of applications for multivariate locations, i.e. $s\geq 2$. {\par In this setting we define the function $v$ 
through the distance $d$ and the 1-nearest neighbor graph (1-NN). Thus, it extends the
variable fusion method \citep{fusion_variable} to such a spatial structure of components of
${\bf X}$.
We show that under this penalty, after a convenient
reformulation of the optimization problem, the regression parameter functions can be estimated using
the algorithm of the group lasso method (\cite{groupfda}, \cite{group_rl}). Moreover, as the $1$-NN case can be restrictive in practice, we introduce a second penalty that considers a more general
grouping structure of conditions, yielding to what we call a {\em group fusion lasso model}.
This second penalty allows testing the equality among the dimensions of the regression coefficient
function belonging to the same cluster of conditions}.\par 
The paper is organized as follows. Section \ref{method} presents the proposed methodologies and their estimation strategies using basis function expansion techniques. A comparison study of the two methods and the group lasso approach is performed using simulated data in Section \ref{sims}.  A real data application from the EEG classification task is presented in  Section \ref{FM}. The paper ends with a discussion in  Section \ref{dis}. 

 \section{Two new fusion methods for linear regression with multivariate functional data}

Without loss of generality assume that $\mathbf{X}$ and $Y$ are zero mean random variables. {\color{black} Moreover, we consider that $\{(\boldsymbol{x}_i, y_i)\}_{i=1,\ldots, n}$ is an observation of $\{(\boldsymbol{X}_i, Y_i)\}_{i=1, \ldots, n}$, an {i.i.d.} sample of size $n\geq 1 $ drawn from the joint distribution { of } $(\mathbf{X},Y)$.} 
\\ {\color{black} Under the zero mean assumption of $\bf X$ and $Y$, the intercept $\beta^{(0)}$ in (\ref{ml}) vanishes and the mean square criterion  
(\ref{critere}) becomes:}
\begin{equation*}
  \hat{\boldsymbol{\beta}}=\argmin_{ \psi \in  {\mathcal{H}^p} }\frac{1}{n}\sum_{i=1}^{n} \left( y_i -\sum_{j=1}^{p}\langle x_i^{(j)}, \psi^{(j)} \rangle_{L_2} \right)^2.  
 \end{equation*}

\noindent Remind that for each $i  = 1, \ldots, n$, $y_i \in \mathbb{R}$ and $\boldsymbol{x}_i$ is a multivariate function defined on $[0,T]$, 

$$\boldsymbol{x}_i(t) = \left ( x_i^{(1)}(t), \ldots, x_i^{(j)}(t), \ldots, x_i^{(p)}(t)\right)^\top, \ \ \ t\in [0,T], $$
where each dimension $x_i^{(j)}$  is observed under the condition $\mathcal{C}_j$, $j =1, \ldots, p$.

Let us now introduce our first model of the penalty based on the distance among conditions.

\label{method}

\subsection{Fusion method based on the neighbor relationship among conditions} 
\label{FL}
 The basic idea is that if two conditions $\mathcal{C}_j$ and $\mathcal{C}_{j'}$ are close in the space $\mathcal{S}$ (with respect to distance $d$), then the contributions brought by the components $X^{(j)}$ and $X^{(j')}$ in the linear model (\ref{ml}), i.e., $\beta^{(j)}$ and $\beta^{(j')}$,  might be comparable. 
 Allowing for identical coefficients $\beta^{(j)}$ associated with close conditions, the variable fusion methodology is a candidate to obtain a parsimonious model and to compete with existing linear model approaches (\cite{fusion_variable}, \cite{fused_lasso}, \cite{bleakley2011}). 
 \\ When the conditions $\mathcal{C}_j$ belong to $\mathbb{R}^s$ with $s\geq 2$, the distance $d$ defines a neighbor relationship between conditions and thus it can be used to estimate the regression coefficient functions accordingly. 
 More precisely, following the ideas in \cite{fusion_variable}, 
 the \textit{1-NN} variable fusion model (FU) can be formulated as the following optimization problem: 
\begin{equation}
\label{f_lasso}
\begin{split}
    \hat{\boldsymbol{\beta} }_{\lambda}&= \argmin_{\beta \in {\mathcal{H}^p}}  \frac{1}{2} \sum_{i=1}^{n}\left( y_i-\sum_{j=1}^{p} \langle x_i^{(j)}, \beta^{(j)}\rangle_{L_2}   \right)^2+ \lambda \sum_{j=1}^{p}||\beta^{(j)}-\beta^{(v(\mathcal{C}_j))}||_{L_2},  
\end{split}
\end{equation}
where $\lambda \geq 0$, $v: \{\mathcal{C}_1,\ldots, \mathcal{C}_p\}\to \{1, \ldots, p\}$ denotes the neighbor function 
\begin{equation}
v(\mathcal{C}_j)=\argmin_{i \in \{1, \ldots, p\} \backslash \{j\} } d(\mathcal{C}_i, \mathcal{C}_j), \; j = 1,\ldots, p .
\label{v_1}
\end{equation}
The function $v$ helps to integrate into the estimation process of $\boldsymbol{\beta}$ the information brought by the conditions (locations, spatial distributions).
Notice that if the set of $\argmin$ in \eqref{v_1} is not unique, then we choose randomly or experimentally an element of this set.\\ 
For ease of notation, let denote with $\langle\cdot, \cdot\rangle_{\mathcal{H}^p}$ the inner product in ${\mathcal{H}^p}$ defined by:  
$$  \langle  \boldsymbol{f}, \boldsymbol{g} \rangle_{\mathcal{H}^p}= \sum_{i=1}^p \langle f^{(i)}, g^{(i)} \rangle_{L_2}, $$
 for all $ \boldsymbol{f}, \boldsymbol{g} \in {\mathcal{H}^p}$.
 
\begin{figure}[ht]
    \centering
    \includegraphics[scale=.25, align=c]{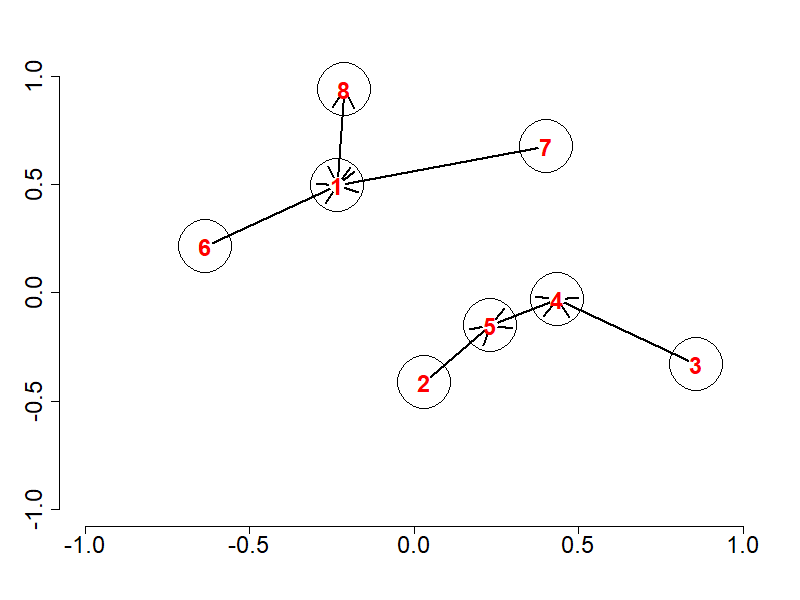}
    \caption{{1-NN} graph: $a \rightarrow b$ means $b$ is the neighbor of $a$.  }
    \label{1NN}
\end{figure}

\noindent The penalty function in \eqref{f_lasso} can then be rewritten as  $$
\displaystyle \sum_{j=1}^{p} ||\beta^{(j)}-\beta^{(v(\mathcal{C}_j))}||_{L_2}= ||\mathbf{L}\boldsymbol{\beta}||_{L_2, 1}, 
$$ where $\mathbf{L}=\mathbf{W}-\mathbb{I}_{p\times p}$ and $\mathbf{W}=( w_{i, j})_{ 1\leq i \leq p  , 1\leq j \leq p } $ is the adjacency matrix with elements
\begin{equation*}
w_{i,j} =\left\{
    \begin{array}{ll}
        1 & \mbox{if  $v(\mathcal{C}_i)=j$ }\\
        0 & \mbox{otherwise.}
    \end{array}
\right. 
\label{w}
\end{equation*}
$\mathbb{I}_{p\times p}$ is the $p \times p$ identity matrix and $||\cdot||_{L_2, 1}$ is the norm on ${\mathcal{H}^p}$ defined as:  
$$||\boldsymbol{f}||_{L_2, 1} =  \sum_{i=1}^{p}||f^{(i)}||_{L_2}, \; \boldsymbol{f}\in {\mathcal{H}^p}.$$
For illustrative purposes, consider the toy example shown in Figure \ref{1NN}, with $\mathcal{S} \subset \mathbb{R}^2$ and $p=8$. It represents $p=8$ points corresponding to the conditions $\mathcal{C}_j \in \mathbb{R}^{2}$, $j=1, \ldots, p$.
The neighborhood relationship among the conditions is given by the following $v$ function: 
$v(\mathcal{C}_1)=8$, $v(\mathcal{C}_2)=5$, $v(\mathcal{C}_3)=4$, $v(\mathcal{C}_4)=5,v(\mathcal{C}_5)=4, v(\mathcal{C}_6)=1, v(\mathcal{C}_7)=1$ and $ v(\mathcal{C}_8)=1$.  \par 
Remark that the rank of the matrix $\mathbf{L}$ is generally lower than $p$, since symmetric relationships are possible (contrary to consecutive conditions case, see e.g \cite{fusion_variable}). For example, Figure \ref{1NN} shows that $\mathcal{C}_1$ is the neighbor of $\mathcal{C}_8$ and $\mathcal{C}_8$ is the neighbor of $\mathcal{C}_1$, the same for the couple $(\mathcal{C}_5,\mathcal{C}_4)$.\\


\begin{lem} \label{lem1} If $r$ is the rank of the matrix $\mathbf{L}$, there exists a $r\times p$ full rank matrix $\mathbf{L}_0$ such as 
\begin{equation}
     ||\mathbf{L}\boldsymbol{f} ||_{L_2, 1} = ||\mathbf{L}_0\boldsymbol{f} ||_{L_2, 1}, \; \boldsymbol{f}\in {\mathcal{H}^p}
     \label{eq_prop}. 
\end{equation} 
\end{lem}
Thus, $\mathbf{L}_0$ avoids redundancy. Its construction consists of finding the couples of rows corresponding to symmetric relations {and}, for each such a couple, {replacing} it with a row representing the double of the replaced ones. The rank of the matrix $\mathbf{L}$ coincides with the number of vertices of the undirected version of the 1-NN graph. 
\\ 
As an illustration, in our toy example (Figure \ref{1NN}), we have the following matrices    
 \begin{equation*}
 \mathbf{L}=\begin{pmatrix}
      -1 & 0 & 0 & 0 & 0 & 0 & 0 & 1 \\ 
  0 & -1 & 0 & 0 & 1 & 0 & 0 & 0 \\ 
   0 & 0 & -1 & 1 & 0 & 0 & 0 & 0 \\ 
   0 & 0 & 0 & -1 & 1 & 0 & 0 & 0 \\
   0 & 0 & 0 & 1 & -1 & 0 & 0 & 0 \\ 
   1 & 0 & 0 & 0 & 0 & -1 & 0 & 0 \\ 
   1 & 0 & 0 & 0 & 0 & 0 & -1 & 0 \\
   1 & 0 & 0 & 0 & 0 & 0 & 0 & -1 
     \end{pmatrix}
 \end{equation*}
 and 
 \begin{equation*}
 \mathbf{L}_0=\begin{pmatrix}
      -2 & 0 & 0 & 0 & 0 & 0 & 0 & 2 \\ 
  0 & -1 & 0 & 0 & 1 & 0 & 0 & 0 \\ 
   0 & 0 & -1 & 1 & 0 & 0 & 0 & 0 \\ 
   0 & 0 & 0 & -2 & 2 & 0 & 0 & 0 \\ 
   1 & 0 & 0 & 0 & 0 & -1 & 0 & 0 \\ 
   1 & 0 & 0 & 0 & 0 & 0 & -1 & 0 
     \end{pmatrix}. 
 \end{equation*}
\par  
Hence, Lemma \ref{lem1} implies that there's an alternative reformulation of \eqref{f_lasso}.  Similarly to the variable fusion methodology \citep{fusion_variable}, Proposition \ref{prop_1} shows that \eqref{f_lasso} can be resolved using a lasso method.

\begin{prop} \label{prop_1}The solution of \eqref{f_lasso} is given by 
$$\hat{\boldsymbol\beta}_\lambda=\mathbf{D}^{-1} \hat{\boldsymbol\psi}_\lambda, $$ where 
\begin{equation}
    \hat{\boldsymbol\psi}_\lambda= \argmin_{\boldsymbol{f}\in {\mathcal{H}^p}} \frac{1}{2} \sum_{i=1}^{n}\left( y_i-\langle (\mathbf{D}^{-1})^{\T}x_i, \boldsymbol{f}\rangle_{\mathcal{H}^p}   \right)^2+ \lambda \sum_{j=1}^{r}||f^{(j)}||_{ L_2}, 
    \label{f_lasso_2}
\end{equation}
 $\mathbf{D}=\begin{pmatrix}
\mathbf{L}_0\\ 
\mathbf{T}
\end{pmatrix}$, $\mathbf{L}_0$ is the $r\times p$ reduced matrix of $\mathbf{L}$ and  $\mathbf{T}$ is a $(p-r)\times p $ matrix which rows form a basis of the null space of $\mathbf{L}_0$,  $\mathbf{L}_0\mathbf{T}^{\T}=\mathbf{0}_{r\times (p-r)}$ and $\mathbf{0}_{r\times (p-r)}$ is the $r\times (p-r)$ matrix of zeros. 
\end{prop}
Note that the estimation of the non-penalized part of $f$ in \eqref{f_lasso_2}, $f^{(r+1)}, \ldots, f^{(p)}$,  might lead (by putting maximum weights on the non-constrained part of $\boldsymbol{\beta}$) to model overfitting, especially in the functional context. This is because the penalty only considers the difference between the coefficients and not the overall norm of $\boldsymbol{\beta}$. To avoid the issue of overfitting, we propose {to constrain the overall norm $\bf D \boldsymbol{\beta} $ by modifying the penalty term in (\ref{f_lasso_2}) as:
    $$||\mathbf{L}\boldsymbol{\beta} ||_{L_2, 1}+\frac{\sqrt{p-r}}{\eta}||\mathbf{T}\boldsymbol{\beta}||_{L_2, 2},$$
where $\eta$ is the Frobenius matrix norm of $\mathbf{T}$, and $||\cdot||_{L_2,2}$ denotes the Frobenius norm of ${\mathcal{H}^p}$: 
$$||\boldsymbol{f}||_{L_2, 2} =  \sqrt{\sum_{i=1}^{p}||f^{(i)}||^2_{L_2}}, $$
for $\boldsymbol{f}\in {\mathcal{H}^p}$. \\
Thus, the  optimization problem \eqref{f_lasso_2} becomes:
\begin{equation}
    \hat{\boldsymbol \psi}_\lambda= \argmin_{\boldsymbol{f}\in {\mathcal{H}^p}} \frac{1}{2} \sum_{i=1}^{n}\left( y_i-\langle (\mathbf{D}^{-1})^{\T}x_i, f\rangle_{\mathcal{H}^p}   \right)^2+ \lambda \left( \sum_{j=1}^{r}||f^{(j)}||_{ L_2}+ \frac{\sqrt{p-r}}{\eta} \left(\sum_{j=r+1}^{p}||f^{(j)}||^2_{ L_2}\right)^{1/2} \right). 
    \label{final_fused}
\end{equation}
In other words, the $r$ first values of $\hat{\psi}_\lambda$ are constrained using the $||\cdot||_{L_2,1}$ penalty and the $p-r$ remaining are considered as a new synthetic group. Hence, this modified penalty favors $\hat{\boldsymbol\beta}_\lambda= {\bf D}^{-1} \hat{\boldsymbol\psi}_{\lambda}$ to not take values arbitrary large. 

\par 

This methodology is based on only one neighbor. In the next section, we introduce a similar methodology based on more than one neighbor, we call it the group fusion lasso. 
\subsection{The group fusion lasso }
\label{CFL}
Let consider the example represented in Figure \ref{grp}. In this example, we assume that the conditions are labeled according to $K=3$ groups: the yellow group $(\mathcal{C}_3, \mathcal{C}_4, \mathcal{C}_7)$,  the red group $(\mathcal{C}_1,\mathcal{C}_6,\mathcal{C}_8)$ and the { blue} group $(\mathcal{C}_2, \mathcal{C}_5)$. For this configuration, more than one neighbor must be considered. Indeed, the following sets $(\mathcal{C}_1,\mathcal{C}_6,\mathcal{C}_8)$ $(\mathcal{C}_3,\mathcal{C}_4,\mathcal{C}_7)$, $(\mathcal{C}_2, \mathcal{C}_5)$ have symmetric neighborhood relations (i.e. $\mathcal{C}_8$ has $(\mathcal{C}_1,\mathcal{C}_6)$ as neighbours, $\mathcal{C}_6$ has $(\mathcal{C}_1,\mathcal{C}_8)$ as neighbours, etc.). Rather than examining the interactions of conditions individually{, we propose in this section to test the resulting group relations}. \par  
\begin{figure}[ht]
    \centering
    \includegraphics[scale=.25, align=c]{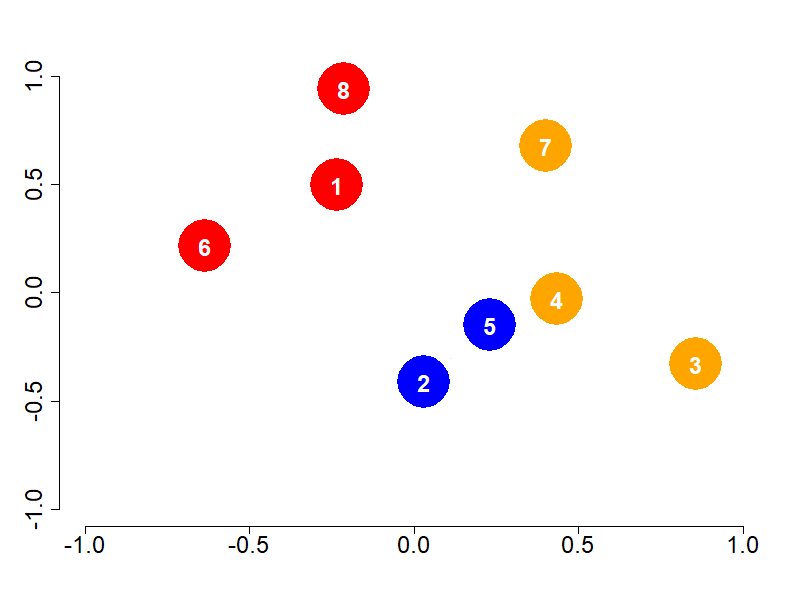}
    \caption{Conditions with grouping structure}
    \label{grp}
\end{figure}
\noindent The grouping structure of conditions is now given by the surjective function $v$: \begin{equation}v: \{\mathcal{C}_1, \ldots, \mathcal{C}_p\} \to \{1, \ldots, K\}, 
\label{eq_grps}
\end{equation}
where $K$ is a number of groups, $K\leq p$.  \\
We recall the definition of the sets of index   
$$
\mathcal{I}_k=\{ j\in \{1, \ldots, p\},\;  v(\mathcal{C}_j)=k \}, \; k=1, \ldots, K. 
$$

\noindent Let denote the size of each group by $$
p_k = |\mathcal{I}_{k}|, \ \ \forall k \in 1, \ldots, K.$$
{The idea behind the group fusion methodology is to introduce criteria that favor similar coefficients for components corresponding to conditions belonging to the same group.} In the example presented in Figure \ref{grp}, $p=8$, $K=3$ and 
\begin{itemize}
\item[-] $v(\mathcal{C}_1) =v(\mathcal{C}_6) =v(\mathcal{C}_8) =1$, the "red" group, 
\item[-] $v(\mathcal{C}_2) =v(\mathcal{C}_5)  =2$, the "{blue}" group
\item[-] $v(\mathcal{C}_3) =v(\mathcal{C}_4) =v(\mathcal{C}_7) =3$ the "yellow" group.
\end{itemize}
Then, as in the lasso regularization framework, this estimation methodology forces the clusters of conditions to have close coefficient functions and, eventually, some of them be the same:
$$\{\beta^{(1)} = \beta^{(6)} =\beta^{(8)}\}\ \ \mbox{and/or} \  \{\beta^{(2)} =\beta^{(5)} \}\ \mbox {and/or} \ \{\beta^{(3)} =\beta^{(4)} = \beta^{(7)}\}.$$ 
For this purpose, let modify the criterion (\ref{f_lasso}) by adding a term penalty for each group $k$ of coefficient functions, $\mathcal{P}_{k}(\cdot)$, $k = 1,\ldots, K$,  as follows: 
\begin{equation}
  \hat{\boldsymbol \beta}_{\lambda} =\argmin_{ \boldsymbol{\beta} \in {\mathcal{H}^p}} \frac{1}{2}\displaystyle\sum_{i=1}^n(y_i- \langle \boldsymbol{x}_i, \boldsymbol{\beta}\rangle_{\mathcal{H}^p})^2 +\lambda \displaystyle\sum_{k=1}^K \mathcal{P}_{k}(\boldsymbol{\beta} ), 
     \label{GFU}
\end{equation}
where $\mathcal{P}_{k}(\boldsymbol{\beta} ) = 
   \sqrt{ p_k } \sqrt{\displaystyle\sum_{i\in \mathcal{I}_k} || \beta^{(i)} - \bar{\beta}_{\mathcal{I}_k }||_{L_2}^2}$    
and  $\bar{\beta}_{\mathcal{I}_k}(t) =  \frac{1}{p_k}\displaystyle\sum_{j\in \mathcal{I}_k}\beta^{(j)}(t)$, $t\in [0,T]$. \\ 
\begin{remark} If for some $k \in 1,\ldots,K$,   $\mathcal{I}_{k}=\{j\}$, then there is no penalty on the $j$-th component (dimension) of the corresponding coefficient function, $\beta^{(j)}$.
\end{remark}
As in the previous criterion (\ref{f_lasso}),  the optimization criterion \eqref{GFU} might lead to model overfitting (see Proposition \ref{prop_2}): the fusion penalties have no control over all terms in the norm of $\boldsymbol{\beta}$. To overcome this difficulty, we introduce the group fusion lasso (GFUL) methodology as a modified version of the elastic-net strategy \citep{zou2005}, that is,    
\begin{equation}
  \hat{\boldsymbol\beta}_{\lambda, \alpha} =\argmin_{ \boldsymbol{\beta} \in {\mathcal{H}^p}} \frac{1}{2}\sum_{i=1}^n(y_i- \langle \boldsymbol{x}_i, \boldsymbol{\beta}\rangle_{\mathcal{H}^p})^2 +\lambda \sum_{k=1}^K \mathcal{P}_{\alpha, k}(\boldsymbol{\beta}), 
     \label{opti_princ}
\end{equation}
with 
\begin{equation*}
  \mathcal{P}_{\alpha, k}(\boldsymbol{\beta}) =(1-\alpha)  \mathcal{P}_{k}(\boldsymbol{\beta}) + \alpha   ||\bar{\beta}_{\mathcal{I}_k }||_{L_2}, \ \ \alpha \in (0,1). 
\end{equation*}

\bigskip

The purpose of GFUL is related to the group lasso methodology where, given some grouping structure of predictor variables, the objective is to force to zero all the coefficients of variables within some group(s) (for more details see \cite{group_rl}, \cite{yuan2006model}). From this perspective, GFUL aims to obtain some {group(s) of conditions} with the same coefficient functions, which is a more general statement. \\
  
\begin{remark}The penalty function is composed of two terms: the first one,  $\mathcal{P}_{k}(\boldsymbol{\beta})$  is of fusion type; $\mathcal{P}_{k}(\boldsymbol{\beta})$ is zero if only if  $ 
 \beta^{(j)}=\beta^{(k)},\; \; \forall j, k \in \mathcal{I}_k $;  the second term, $ 
||\bar{\beta}_{\mathcal{I}_k }||_{L_2} 
 $ is a group-lasso-like penalty. 
\end{remark}

As for  FU methodology (see Proposition \ref{prop_1}), we show now that GFUL estimation reduces to a group-lasso one.

In the GFUL methodology, the membership of conditions to groups is a central notion. Let define the indicator matrix $\mathbf{M}=(m_{k, j})_{ 1\leq k \leq K  , 1\leq j \leq p } $ as 
\begin{equation*}
m_{k,j} =\left\{
    \begin{array}{ll}
        1 & \mbox{if } j \in \mathcal{I}_{k} \\
        0 & \mbox{otherwise.}
    \end{array}
\right. 
\end{equation*}
In the toy example (Figure \ref{grp}), the matrix $\mathbf{M}$ is given by   
$$\displaystyle\begin{pmatrix}
    1 & 0 & 0 & 0 & 0& 1& 0 &1 \\
    0 & 1 & 0 & 0 & 1 & 0 & 0& 0 \\
    0 & 0 & 1 & 1 & 0 & 0 & 1 & 0
    \end{pmatrix}.
$$\\
In a general case, up to a permutation of columns, $\mathbf{M}$ can be written as  
\begin{equation*}
    \mathbf{M} = \begin{pmatrix}  \Vec{1}_{p_1}^{\T} & \Vec{0}_{p_2}^{\T} & ... & 0 \\ 
    \Vec{0}_{p_1}^{\T}  & \Vec{1}_{p_2}^{\T} & ... & 0 \\
    ...  &  &  &  ...\\ 
    \Vec{0}_{p_1}^{\T}  & \Vec{0}_{p_2}^{\T}  & ...& \Vec{1}_{p_K}^{\T} 
    \end{pmatrix}, 
\end{equation*}
where $\Vec{1}_{p_k}$,$ \Vec{0}_{p_k}$  are respectively the $p_k$ column vector of ones and the $p_k$ column vector of zeros. \\ 
Let denote by $\bar{\mathbf{M}}$ the standardized version of $\mathbf{M}$, i.e. $\bar{\mathbf{M}}=\text{diag}(1/p_1, 1/p_2, \ldots, 1/p_K) \mathbf{M}$. \\ 
Then, similarly as in Lemma \ref{lem1}, the following result holds.
\begin{lem}
\label{lemma2} Let $f  \in {\mathcal{H}^p}$, $\alpha \in (0,1)$ and $p_k\geq 2$, for $k=1, \ldots, K$. Consider $2K$  synthetic groups $\{\tilde{\mathcal{I}}_k\}_{k=1}^{2K},$ defined as 
\begin{align*}
    \tilde{\mathcal{I}}_k= \left\{\begin{array}{c c}
        \left\{j\in \{1, \ldots, p\}\left| 1+\sum_{l=1}^{k-1}(p_l-1)\leq j \leq \sum_{l=1}^{k}(p_l-1)\right \} \right. & k=1, \ldots, K \\
        \\ 
        \left\{ k+p-2K\right\} & k=K+1, \ldots, 2K.  
    \end{array} \right.
\end{align*}
Up to a permutation of dimensions, the penalty function of GFUL can be written as   
\begin{equation}
\label{eq_prop_2}
    \sum_{k=1}^K \mathcal{P}_{\alpha, k}(\boldsymbol{f})= \sum_{k=1}^{2K} \sqrt{\sum_{i\in \tilde{\mathcal{I}}_k} || (\mathbf{G}_{\alpha}\boldsymbol{f})^{(i)}||_{L_2 }^2}, \ \boldsymbol{f}\in {\mathcal{H}^p}
\end{equation}
    where $ \mathbf{G}_{\alpha}$ is the $p\times p$ non-singular matrix given by: $$\mathbf{G}_{\alpha}= \begin{pmatrix}(1-\alpha)\mathbf{R} \\ 
\alpha \bar{\mathbf{M}} \end{pmatrix}, $$ with  $\mathbf{R}$ is the block diagonal matrix composed of the following elements $\sqrt{p_1}\mathbf{R}_1, \ldots, \sqrt{p_K}\mathbf{R}_K$, and for $k=1, \ldots, K$, $\mathbf{R}_k$ is the upper triangular $(p_k-1) \times p_k$ matrix obtained from the reduced rank QR decomposition of 
 $\mathbf{P}_k = \mathbb{I}_{p_k \times p_k}-\displaystyle\frac{1}{p_k} \mathbf{1}_{p_k\times p_k}$; here $\mathbf{1}_{p_k \times p_k }$ denotes the $p_k\times p_k$ matrix of ones. 
\end{lem}
Using the non-singularity of $\mathbf{G}_\alpha$, the following proposition provides a way to estimate GFUL using a simpler model. 
\begin{prop} \label{prop_2} For $\alpha \in (0, 1)$, the solution of \eqref{opti_princ}, holds $\hat{\boldsymbol\beta}_{\alpha, \lambda}=\mathbf{G}^{-1}_{\alpha}\hat{\boldsymbol\psi}_{\lambda}$, where 
\begin{equation}
    \hat{\boldsymbol\psi}_{\lambda}=\argmin_{ \boldsymbol{f} \in {\mathcal{H}^p}} \frac{1}{2}\sum_{i=1}^n(y_i- \langle (\mathbf{G}_{\alpha}^{-1})^{\T} \boldsymbol{x}_i, \boldsymbol{f}\rangle_{\mathcal{H}^p})^2 + \lambda \sum_{k=1}^{2K} \sqrt{\sum_{i\in \tilde{\mathcal{I}}_k } || f^{(j)}||_{L_2}^2} 
    \label{g_lasso}, 
\end{equation}
\end{prop}
\noindent {\color{black}The proof of this proposition follows as a direct consequence of the non-singularity of $\mathbf{G}_\alpha$ and Lemma \ref{lemma2}. } 
\begin{remark}
    The case of $\alpha=1$ or $\alpha=0$ can be resolved using the same technique as in Proposition \ref{prop_1}. Indeed, the non-null part of $\mathbf{G}_{\alpha}$ is full rank for $\alpha \in \{0, 1\}$ 
\end{remark}

The direct estimation of $\boldsymbol\beta$, under least squares regression, is generally an ill-posed inverse problem (\cite{cardot1999}, \cite{aguilera2006using}). The basis expansion technique, a well-known dimension reduction technique as an alternative to solve this problem, is presented in the next section.  
\subsection{Computational aspect: Basis expansion}
The basis expansion technique assumes that there exists a set of linearly independent functions $\{\phi_k\}^M_{k=1}$, such as, for each $i=1, \ldots, n$, $x_i$ can be written as 
\begin{equation}
    x_i^{(j)}(t)= \sum_{k=1}^{M} a_{i,k}^{(j)} \phi_k(t)= (\boldsymbol{a}_i^{(j)})^\top\boldsymbol{\phi}(t), \   t\in [0, T]
    \label{star}
\end{equation}
where $a_{i,k}^{(j)} \in \mathbb{R}$ for $i=1, \ldots, n$, $j=1, \ldots, p$ and 
\begin{itemize}
    \item $\boldsymbol{a}_i^{(j)}= \begin{pmatrix}
    a_{i, 1}^{(j)} & \ldots & a_{i, M}^{(j)}
\end{pmatrix}^\top $, 
\item $\boldsymbol{\phi}= \begin{pmatrix} \phi_1 & \ldots & \phi_M  \end{pmatrix}^{\T}$ is the vector of functions. {\color{black} The most common choices of $\boldsymbol{\phi}$ are Fourier or B-splines functions, depending on the periodicity of $\bf X$ \citep{ramsay2008}. }
\end{itemize}
Note that for each $\boldsymbol{x}_i$, we have that 
$$
\boldsymbol{x}_i=\begin{pmatrix}
    x_i^{(1)} \\ 
    \vdots \\ 
    x_i^{(p)}
\end{pmatrix} = \Phi \boldsymbol{a}_i $$
where  
\begin{align*}
\boldsymbol{a}_i= \begin{pmatrix}
    \boldsymbol{a}_i^{(1)} \\ 
    \vdots \\ 
    \boldsymbol{a}_i^{(p)}
\end{pmatrix} 
\text{ and }
    \Phi=\begin{pmatrix} \phi_1&  \ldots & \phi_M & 0 &\ldots &0 &\ldots & 0&\ldots & 0 \\
    0 & \ldots & 0 &  \phi_1 &\ldots &  \phi_M & \ldots& 0&\ldots& 0  \\
    \vdots &   &   &  & &   &   & & &\vdots \\ 
      0 & \ldots & 0 &  0 &\ldots &  0&  \ldots&\phi_1 &\ldots &  \phi_M 
    \end{pmatrix}. 
\end{align*}

Notice that in the expression in \eqref{star}, we use the same basis $\boldsymbol \phi$ for all dimensions of $\bf X$. This seems realistic since $X^{(1)}, \ldots, X^{(p)}$ measure the same parameter $X$. However that is not mandatory, each dimension $X^{(j)}$ can be expressed on its own basis.\\  
We assume that the coefficient function $\boldsymbol{\beta}$ can also be expressed as 
\begin{equation*}
 \boldsymbol{\beta}(t) =\Phi(t)\boldsymbol{b},\; t \in [0,T]  
\end{equation*}
where
$$
\boldsymbol{b}=\begin{pmatrix}
    \boldsymbol{b}^{(1)} \\ 
    \vdots \\ 
    \boldsymbol{b}^{(p)} 
\end{pmatrix}, \text{ with } \boldsymbol{b}^{(j)} \in \mathbb{R}^{M}. 
$$
Remark that the predictors $\boldsymbol{x}_i$ and $\boldsymbol \beta$ admit also the equivalent matrix notations     
\begin{equation}
    \boldsymbol{\beta}(t)=\mathbf{B}\boldsymbol{\phi}(t) \text{ and } \boldsymbol{x}_i(t)= \mathbf{A}_i \boldsymbol{\phi}(t) 
    \label{b_ex}
\end{equation}
where $\mathbf{A}_i$ and $\mathbf{B}$ are the following matrices of size $p\times M$, \newline   
 $\mathbf{B}= \begin{pmatrix}
    \boldsymbol{b}^{(1)} & \ldots & \boldsymbol{b}^{(p)}
\end{pmatrix}^\top$ and   $\mathbf{A}_i=  \begin{pmatrix}\boldsymbol{a}_i^{(1)} &  
     \boldsymbol{a}_i^{(2)} & 
     ... & 
     \boldsymbol{
     a}_i^{(p)}
     \end{pmatrix}^{\T}
 $, for all $i=1, \ldots, n$. 

\begin{prop} The following statements hold 
\begin{enumerate} 
    \item $ ||\boldsymbol{\beta}||_{L_2, 1} \doteq \displaystyle\sum_{j=1}^{p} || \beta^{(j)} ||_{L_2}  = ||\mathbf{B}\mathbf{F}_{\boldsymbol{\phi}}^{1/2}||_{2,1}  $,
    where $||\cdot||_{2,1}$ is the $(2,1)$ matrix norm and $\mathbf{F}^{1/2}_{\boldsymbol{\phi}}$ is the square root matrix of $\mathbf{F}_{\boldsymbol{\phi}}$= $\{ \langle \phi_i, \phi_j \rangle \}_{i, j}$.
    
    \item Let $k$ be an integer in $\{0, \ldots, p-1\}$ and $\bf Z$ be a matrix of size $(p-k)\times p$. Define $\boldsymbol{\beta}_0(t)=\mathbf{Z}\boldsymbol{\beta}(t)$, for all  $t \in [0, T]$. Then $$\boldsymbol{\beta}_0(t)= \boldsymbol{b}_0\Phi(t), \ \ \forall t\in [0,T], $$ where $\boldsymbol{b}_0=(\mathbf{Z} \otimes \mathbb{I}_{M \times M}) \boldsymbol{b} $ and $\otimes$ denotes the Kronecker product. 
\end{enumerate}
\label{prop3}
\end{prop}
\noindent The first point states that the norm of $\boldsymbol\beta$ depends on the vector $\boldsymbol{b}$ and the basis $\{\phi_k\}^M_{k=1}$ via the matrix $\mathbf{F}_{\boldsymbol{\phi}}$. As an example, {consider} the following group lasso problem (each dimension represents a group):
\begin{align}
    \hat{\boldsymbol \beta}_\lambda& = \argmin_{\boldsymbol \beta \in {\mathcal{H}^p}} \frac{1}{2} \sum_{i=1}^{n} \left( y_i- \langle \boldsymbol{x}_i, \boldsymbol{\beta} \rangle_{\mathcal{H}^p} \right)^2+ \lambda \sum_{j=1}^{p} || \beta^{(j)} ||_{L_2}. 
    \label{grp_func}
\end{align}
Since $\langle x_i^{(j)}, \beta^{(j)} \rangle = (\boldsymbol{a}_i^{(j)})^\top \mathbf{F}_{\boldsymbol{\phi}} \boldsymbol{b}^{(j)}$ and $||\beta^{(j)}||_{L_2}= \left((\boldsymbol{b}^{(j)})^\top \mathbf{F}_{\boldsymbol{\phi}} \boldsymbol{b}^{(j)} \right)^{\frac{1}{2}}$, the problem in \eqref{grp_func}  is equivalent to the one of finding the vector $$\hat{\boldsymbol b}_\lambda= \begin{pmatrix}
    \hat{\boldsymbol b}^{(1)}_\lambda \\
    \hat{\boldsymbol b}^{(2)}_\lambda \\ 
    \vdots \\
    \hat{\boldsymbol b}^{(p)}_\lambda
\end{pmatrix}$$ such that 
\begin{align}
    \hat{\boldsymbol b}_\lambda= \argmin_{ \boldsymbol{b} \in \mathbb{R}^{pM} } \frac{1}{2} \sum_{i=1}^n \left( y_i - \sum_{j=1}^{p} (\boldsymbol{a}_i^{(j)})^\top \mathbf{F}_{\boldsymbol{\phi}} \boldsymbol{b}^{(j)} \right)+ \lambda \sum_{j=1}^{p} ||\mathbf{F}_{\boldsymbol{\phi}} ^{1/2}\boldsymbol{b}^{(j)}||_2. 
\end{align}
Let denote by $\hat{\boldsymbol \gamma}^{(j)}= \mathbf{F}^{1/2}_{\boldsymbol{\phi}}\boldsymbol{b}^{(j)}$. Then, obtaining $\hat{\boldsymbol \gamma}_\lambda$ as solution of  
\begin{align*}
    \hat{\boldsymbol \gamma}_\lambda= \argmin_{\boldsymbol{\gamma} \in \mathbb{R}^{pM} } \frac{1}{2} \sum_{i=1}^n \left( y_i - \sum_{j=1}^{p} (\boldsymbol{a}_i^{(j)})^\top \mathbf{F}^{1/2}_{\boldsymbol{\phi}}\boldsymbol{\gamma}^{(j)} \right)+ \lambda \sum_{j=1}^{p} ||\gamma^{(j)}||_2 
    \label{grp_lasso_func_2}
\end{align*}
therefore allows to estimate $\boldsymbol{b}^{(j)}$ as  \begin{equation*}
    \hat{\boldsymbol b}_\lambda^{(j)}=(\mathbf{F}^{1/2}_{\boldsymbol{\phi}})^{-1} \hat{\boldsymbol \gamma}^{(j)}_\lambda, \ \ j = 1, \ldots, p.
\end{equation*}
The problem \eqref{grp_func} is studied in \cite{groupfda} using principal component analysis to avoid multicollinearity and high-dimension issues (\cite{aguilera2006using}, \cite{escabias2005}). \par 
The second statement in Proposition \ref{prop3} shows the correspondence (relationship) between expansion coefficients in the basis $\boldsymbol{\phi}$ after linear transformation of a function in ${\mathcal{H}^p}$, in particular for the coefficient function $\boldsymbol{\beta}$. In the next section, this relationship helps to estimate the coefficient regression function under the FU methodology by reducing the problem to a group-lasso-like, as in \eqref{grp_lasso_func}. 
\subsubsection{FU estimation}
\label{FU_est}
To obtain the solution $\hat{\boldsymbol\beta}_\lambda$ of the FU criterion \eqref{final_fused}, we use the second statement in Proposition \ref{prop3} and Proposition \ref{prop_1}. Let $\hat{\boldsymbol\gamma}_\lambda$ be the solution of the minimization problem 
\begin{align}
    \hat{\boldsymbol \gamma}_\lambda  = \argmin_{\boldsymbol{\gamma} \in \mathbb{R}^{pM} } \frac{1}{2} \sum_{i=1}^n \left( y_i - \boldsymbol{a}_i^\top (\mathbf{D} \otimes \mathbb{I}_{ M\times M})^{-1} \mathbf{F}^{1/2}\boldsymbol{\gamma} \right)^2 + \lambda \left( \sum_{j=1}^{r} ||\gamma^{(j)}||_2 + \frac{\sqrt{p-r}}{\eta} \sqrt{\sum_{j=r+1}^{p} ||\gamma^{(j)} ||_2^2 } \right),  
    \label{grp_lasso_func_fl}
\end{align}
where $\mathbf{F}^{1/2}=  \begin{pmatrix}
    \mathbf{F}^{1/2}_{\boldsymbol{\phi}} & 0 & \ldots & 0 \\ 
    0 & \mathbf{F}^{1/2}_{\boldsymbol{\phi}} & \ldots & 0 \\ 
    \vdots &  &   & \vdots  \\ 
    0 & 0 & \ldots & \mathbf{F}^{1/2}_{\boldsymbol{\phi}}   
\end{pmatrix}.$ \\ 

\noindent Then, the coefficient function $\hat{\boldsymbol \beta}_\lambda$ is given by  $$\hat{\boldsymbol \beta}_\lambda =  \Phi(\mathbf{D} \otimes \mathbb{I}_{M\times M}) (\mathbf{F}^{1/2})^{-1}\hat{\boldsymbol\gamma}_\lambda.$$ 

\subsubsection{GFUL estimation }

We use a similar procedure as in Section \ref{FU_est} for the estimation of $\hat{\boldsymbol\beta}_{\lambda, \alpha}$. \par 

Let define the sets $\mathcal{G}_1, \ldots, \mathcal{G}_{2K}$ as follows. 
\begin{align*}
\mathcal{G}_k&=\left\{ j \left|  1+M\sum_{l=1}^{k-1}(p_l-1) \leq j \leq  M\sum_{l=1}^{k}(p_l-1) \right. \right\} &k=1, \ldots, K, \\ \mathcal{G}_k&=\left\{ j \left|   p'+M(k-1-K)+1 \leq j \leq p'+M(k-K) \right. \right\} & k=K+1, \ldots, 2K,  
\end{align*}
where $p'=M(p-K)$, and $K$ is the number of groups in GFUL. Note that $\{\mathcal{G}_k\}_{k=1}^{2K}$ correspond to $\{\tilde{\mathcal{I}}_k\}_{k=1}^{2K}$ (see Lemma \ref{lemma2}) under the basis expansion hypothesis, i.e when each $\beta^{(j)}$ is represented by $M$ expansion coefficients.  

For convenient notation, let define the permutation matrix $\mathbf{S}= \left(s_{u,v} \in \{0,1\}\right)_{(u, v) \in \{1, \ldots, p\}^2}$,  such that   
\begin{equation*}
    \mathbf{S}\boldsymbol{\beta}= \begin{pmatrix}\boldsymbol{\beta}_{\mathcal{I}_1} \\ 
    \boldsymbol{\beta}_{\mathcal{I}_2}\\ 
    \ldots \\ 
    \boldsymbol{\beta}_{\mathcal{I}_K} 
    \end{pmatrix},  
\end{equation*}
where $\boldsymbol{\beta}_{ \mathcal{I}_k}$ is the vector of components of $\boldsymbol{\beta}$ corresponding to the set of indexes $\mathcal{I}_k$, $k=1, \ldots, K$.  \par 

Then, the group fusion lasso problem reduces to determine $\hat{\boldsymbol \gamma}_{\lambda, \alpha}$ as solution of the following problem:
\begin{align}
    \hat{\boldsymbol \gamma}_{\lambda , \alpha} & = \argmin_{ \boldsymbol{\gamma} \in \mathbb{R}^{pM} } \frac{1}{2} \sum_{i=1}^n \left( y_i - \boldsymbol{a}_i^\top (\mathbf{G}_\alpha\mathbf{S} \otimes \mathbb{I}_{M\times M})^{-1} \mathbf{F}^{1/2}\boldsymbol{\gamma} \right)^2 + \lambda \sum_{k=1}^{2K}  ||\boldsymbol{\gamma}_{\mathcal{G}_k}||_2.
    \label{grp_lasso_func}
\end{align}
Therefore,  $\hat{\boldsymbol \beta}_{\lambda, \alpha}$ is given by $$\hat{\boldsymbol \beta}_{\lambda, \alpha} =  \Phi(\mathbf{G}_\alpha\mathbf{S} \otimes \mathbb{I}_{M\times M}) (\mathbf{F}^{1/2})^{-1}\hat{\boldsymbol \gamma}_{\lambda, \alpha}.$$ 

\begin{remark}
    \label{bin}
{The binary response case} can be naturally { considered} in our proposed methodologies. More precisely, as in \cite{group_rl}, the MSE criterion is replaced by the likelihood one (multiplied by -1) whereas
the penalized terms are the same. In this case, the optimization problem in \eqref{f_lasso} becomes:
  \begin{equation}
  \hat{\boldsymbol \beta}_\lambda= \argmin_{{\boldsymbol\beta} \in {\mathcal{H}^p}} - \sum_{i=1}^{n}\left(y_i\langle \boldsymbol{x}_i, \beta \rangle_{\mathcal{H}^p} - \log( 1+   \langle \boldsymbol{x}_i, \beta \rangle_{\mathcal{H}^p}) \right)+ \lambda \sum_{j=1}^{p} || \beta^{(j)}- \beta^{(v(\mathcal{C}_j))} ||_2.
  \end{equation}
\end{remark}
\begin{remark}
\label{remak_post}
 As GFUL and FU, under the basis expansion hypothesis, can be reduced to {a classical} group lasso optimization problems, the use of post-interference techniques (in particular the work of \cite{post_inference}) allows testing the statistical significance of the equality among coefficients. More specifically, the null hypotheses in the group lasso (which state that coefficients among some groups are null) have the following correspondences in our methods.  \begin{itemize}
        \item For FU, the null hypotheses ($H_{0, j}$) are   $$H_{0, j}: \ \beta^{(j)}= \beta^{(v(\mathcal{C}_j))}, \ j=1, \ldots, p. $$ 
        \item  For GFUL, the null hypothesis $H_{0, k}$  are given by   
$$\begin{array}{ccc c}
       H_{0,k}:& (l,m) \in \mathcal{I}_k^2,\  &\beta^{(l)}= \beta^{(m)} &, k=1, \ldots, K.
 \end{array}
 $$
 
    \end{itemize}
   
\end{remark}
\section{Numerical experiments}
\subsection{Simulations}
\label{sims}
We present a simulation study that compares the performance of the proposed methods, FU and GFUL, with competitor lasso methods. Notice that all our models are estimated by using \cite{grplasso} R package. The R code sources of our simulations are available at \url{https://github.com/imoindjie/GFUL-FU}.

\subsubsection{The simulation setting}
The setting of the simulation is as follows. To show the efficiency of taking into account the grouping structure of conditions, we consider two scenarios. In the first one, the number of conditions is fixed to $p=12$ and we show that all the methods perform {similarly} in terms of MSE criteria. In the second one, we increase the number of conditions to $p=80$ and then, we show the efficiency of our methodology with respect to the others. In both scenarios, the number of groups is $K=4$ and the number of conditions in each group is $p_1 = p_2 = \ldots = p_K = \displaystyle\frac{p}{K} \doteq \kappa$.\\ 
Next, we present the construction of our simulation study. 
\begin{itemize}
    \item[(a)] the conditions and the grouping structure, 
    
    \item[(b)]  the theoretical regression coefficient functions,
    \item[(c)] the definition of the predictor and the response variables, 
    \item[(d)] the two simulation settings,
    \item[(e)] the competitor methods, 
    \item[(f)] the goodness of fit and homogeneity among the coefficient regression functions 
\end{itemize}

\paragraph{\it (a) The conditions and the grouping structure.} \ \\ 
\noindent Let consider the $p$ conditions $\mathcal{C}_j$, $j=1, \ldots, p$,   as points in $\mathbb{R}^2$ and their group structure defined as follows: 
 \begin{align*}
\mbox{Group 1:}  \ \  \mathcal{C}_{j}&= \zeta_j+ c_1,  & j=1, \ldots, \kappa,  \\
\mbox{Group 2:} \ \   \mathcal{C}_{j}&=\zeta_j +c_2,
& j=\kappa+1, \ldots, 2\kappa, \\ 
\mbox{Group 3:} \ \ \mathcal{C}_{j}&= \zeta_j+c_3, & j=2\kappa+1, \ldots, 3\kappa,\\ 
\mbox{Group 4:} \ \ \mathcal{C}_{j}&= \zeta_j+c_4, & j=3\kappa+1, \ldots, p,
\end{align*}
where $$\zeta_j= \left(
     \cos(2\pi \frac{j \text{mod} \kappa}{\kappa}), 
     \sin(2\pi \frac{j \text{mod} \kappa}{\kappa})
\right)^\top,$$ and $c_1= \left(0, 0\right)^\top$, $c_2=\left(3, 3\right)^\top$, $c_3=2c_2$, $c_4=3c_2$ are the "centers" of the groups. Figure \ref{expmpl_condtions} presents the conditions for $p=12$ and $p=80$. \\ One can imagine that these conditions correspond to the position of $p$ points in a $10\times 10$ squared metal piece where one observes in each point $\mathcal{C}_j, j = 1,\ldots, p,$ the temperature $X^{(j)}$ over the time interval $[0,1]$.
\begin{figure}[H]
    \centering
    \begin{tabular}{c c}
    \textbf{(a)} & \textbf{(b)} \\
\includegraphics[scale=.35]{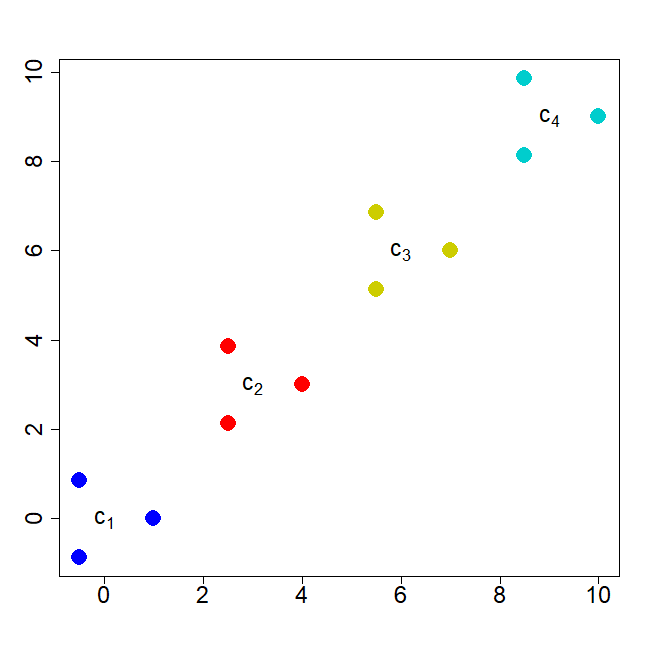}  &  
\includegraphics[scale=.35]{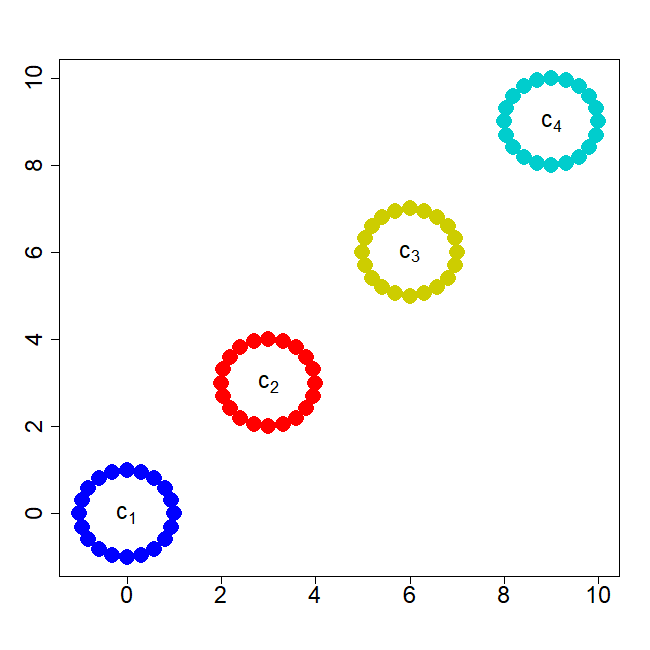} \\
    \end{tabular}
    \caption{Conditions when $p=12$ \textbf{(a)} and $p=80$ \textbf{(b)}. The colors are associated with each group of conditions. }
    \label{expmpl_condtions}
\end{figure}
\paragraph{\it (b)  The theoretical regression coefficient functions.} \ \\ 
\noindent The theoretical coefficient regression function $\boldsymbol{\beta} = (\beta^{(1)}, \ldots, \beta^{(p)})^\top$ is defined as follows:  
\begin{align*}
   \mbox{Groupe 1:} \ \ \beta^{(j)}&= 0,   & j=1, \ldots, \kappa,  \\
     \mbox{Groupe 2:} \ \ \beta^{(j)}&= \sqrt{2}\sum_{k=1}^3\Delta_k, & j=\kappa+1, \ldots, 2\kappa, \\ 
    \mbox{Groupe 3:} \ \ \beta^{(j)}&= b_j\sum_{k=1}^9 \Delta_k, & j=2\kappa+1, \ldots, 3\kappa,\\ 
    \mbox{Groupe 4:} \ \ \beta^{(j)}&= -\sqrt{2}\sum_{k=1}^3 \Delta_k, & j=3\kappa+1, \ldots, p,
\end{align*}
where $b_j=(-1)^{j}\frac{1+j\text{mod} \kappa}{\kappa}$,  the functions $\Delta_1, \ldots, \Delta_9$ denote  the set of functions defined by: $$\Delta_s(t)=(1-0.2(10t-s)^2 )_+, $$ where $(.)_+$ is the positive part function. In this setting, only the third group has different coefficient functions. 


\paragraph{ \it (c) The predictor and the response variables.} \ \\
For $j= 1,\ldots,p$,  $X^{(j)}$ is generated as
$$
X^{(j)}(t)= \sum_{s=1}^{9}a_s\Delta_s(t),
$$
where $a_s \sim \mathcal{N}(0, 1), s= 1, \ldots, 9$ and $t\in [0,1]$.\\ 
The response variable $Y$ is given by  
$$ \displaystyle
Y= \langle \mathbf{X}, \beta \rangle_{\mathcal{H}^p}+ \epsilon,
$$ where $\epsilon \sim \mathcal{N}(0, \sigma^2_\epsilon)$. The values of $\sigma^2_\epsilon$ are set such that the noise to signal ratio, $\displaystyle\frac{\sigma^2_\epsilon}{var(Y)}$, is of about $ 10\%$.
\paragraph{ \it (d) The two simulation settings.} \ \\ 
Two scenarios are presented according to the size of groups, $\kappa$:  
\begin{itemize}
    \item[\textbf{(S1)}] $\kappa=3$, $\sigma_\epsilon=1.6$ 
    \item[\textbf{(S2)}] $\kappa=20$, $\sigma_\epsilon=3.6$ 
\end{itemize}

The theoretical coefficient regression functions $\beta^{(j)}$, $j=1,\ldots, p$ for the two scenarios  are presented in Figure \ref{t_beta}.

\begin{figure}[H]
    \centering

       \begin{tabular}{cccccc}
& & \begin{tabular}[c]{@{}c@{}}Group 1 \\ ${\scriptstyle \beta^{(1)}= \beta^{(2)}= \beta^{(3)}}$\end{tabular} & \begin{tabular}[c]{@{}c@{}}Group 2 \\ ${\scriptstyle \beta^{(4)}= \beta^{(5)}= \beta^{(6)}}$\end{tabular} & \begin{tabular}[c]{@{}c@{}}Group 3 \\${\scriptstyle \beta^{(7)} \neq \beta^{(8)} \neq \beta^{(9)}}$\end{tabular} & \begin{tabular}[c]{@{}c@{}}Group 4 \\ ${\scriptstyle \beta^{(10)} = \beta^{(11)} = \beta^{(12)}}$\end{tabular}  \\
\textbf{S1} & $\beta^{(j)}\mbox{'s}$ & \includegraphics[scale=.1, align=c]{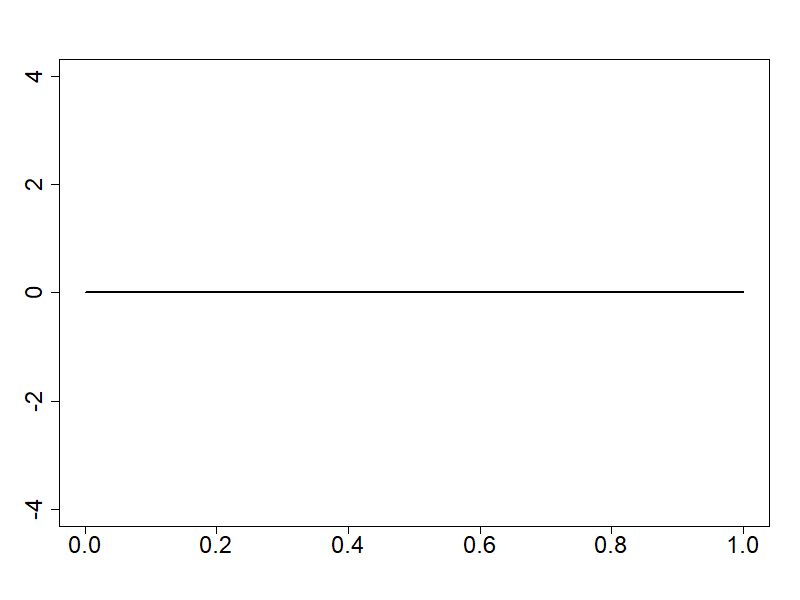}                                   & \includegraphics[scale=.1, align=c]{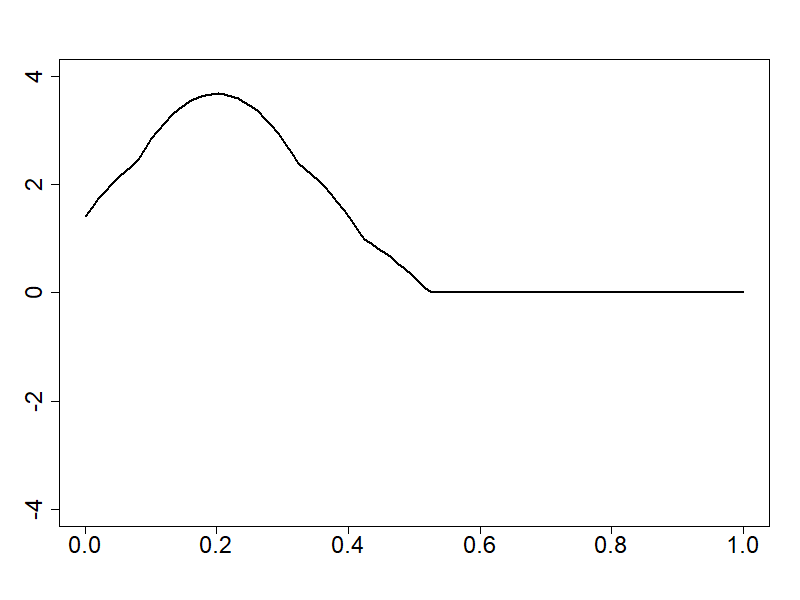}                                   & \includegraphics[scale=.1, align=c]{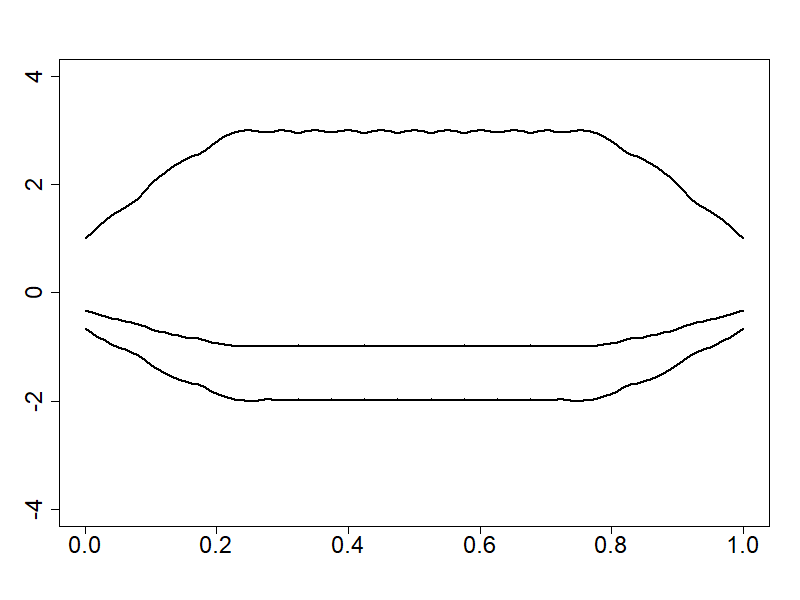}                                           & \includegraphics[scale=.1, align=c]{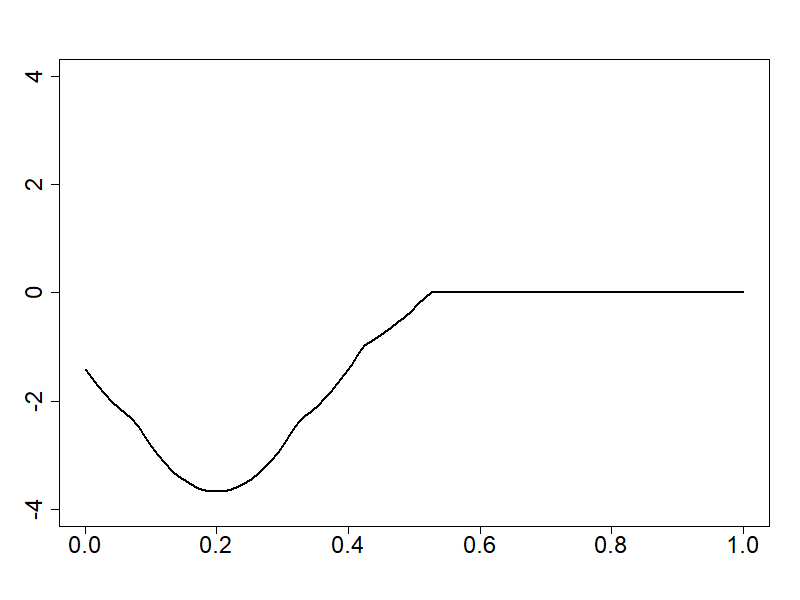}         \\ 
\hline & & \begin{tabular}[c]{@{}c@{}}
${\scriptstyle \beta^{(1)}= \beta^{(2)}= \ldots =\beta^{(20)}}$\end{tabular} & \begin{tabular}[c]{@{}c@{}}
${\scriptstyle \beta^{(21)}= \beta^{(22)}= \ldots =\beta^{(40)}}$\end{tabular} & \begin{tabular}[c]{@{}c@{}}
${\scriptstyle \beta^{(41)} \neq \beta^{(42)} \neq \ldots \neq \beta^{(60)}}$\end{tabular} & \begin{tabular}[c]{@{}c@{}}
${\scriptstyle \beta^{(61)} = \beta^{(62)} = \ldots = \beta^{(80)}}$\end{tabular}  \\

 \textbf{S2} & $\beta^{(j)}\mbox{'s}$ & \includegraphics[scale=.1, align=c]{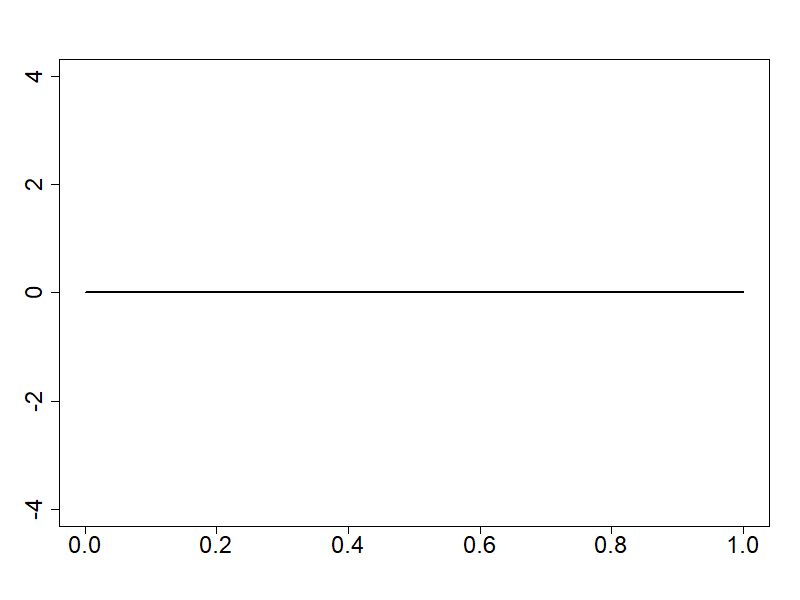}                                   & \includegraphics[scale=.1, align=c]{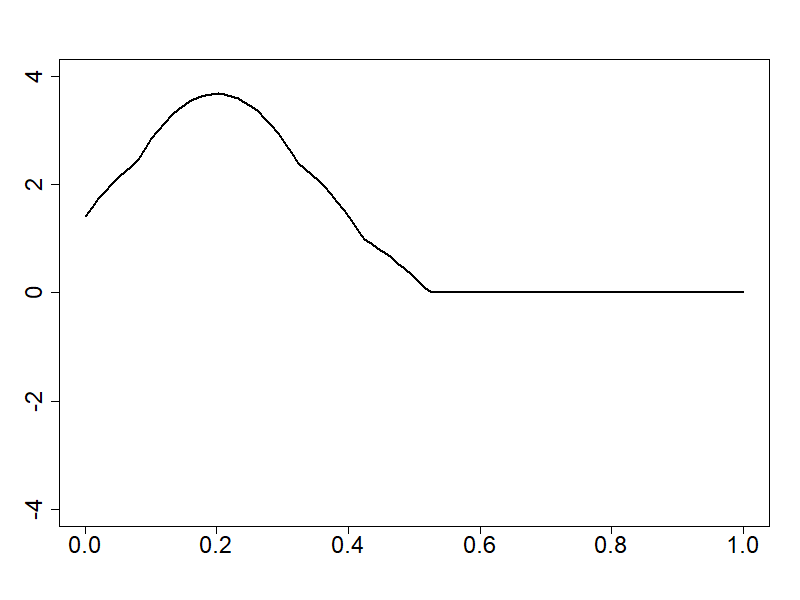}                                   & \includegraphics[scale=.1, align=c]{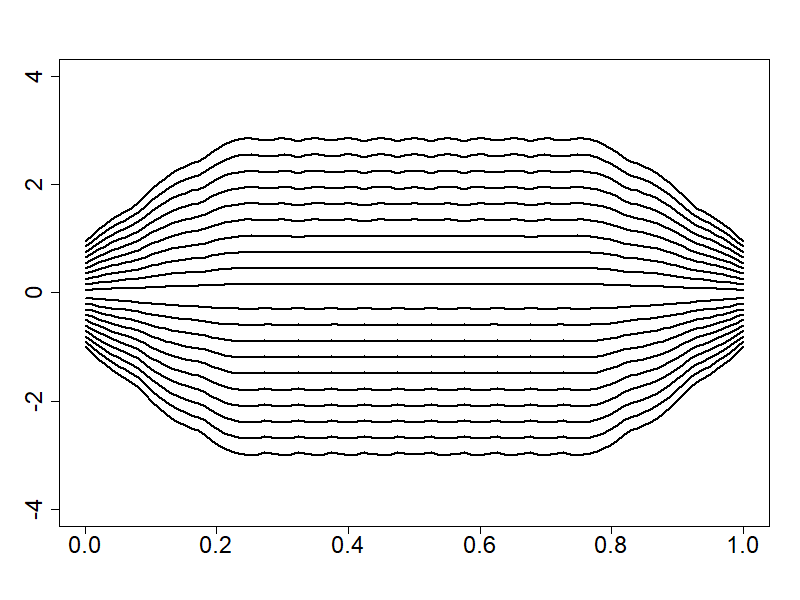}                                           & \includegraphics[scale=.1, align=c]{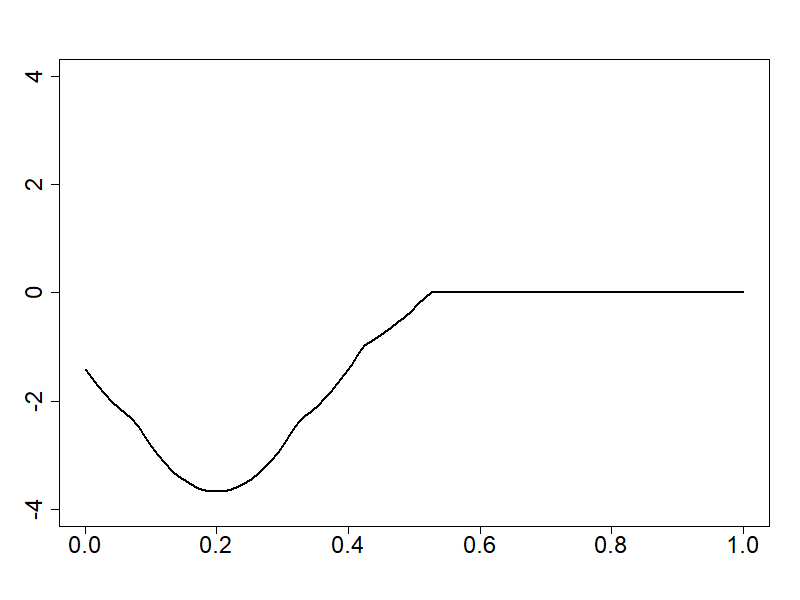}   
\end{tabular}
\caption{Theoretical regression coefficient function $\boldsymbol{\beta}$ for the two scenario}
\label{t_beta}
\end{figure}



\noindent The { predictor function} $\bf X$ is observed on $100$ equidistant sampling time points in the interval $[0,1]$. For all dimensions of $\bf X$, $X^{(j)}$ $j=1, \ldots, p$, we use as an approximation their expansion into a cubic B-splines basis of size $M=20$. To assess model performances, a random training sample of $80\%$ of the data is considered 
 and the remaining $20\%$ is used for prediction. This experiment is repeated $I = 100$ times.
 
\paragraph{\it (e) The competitor methods.} \ \\ 
The variable fusion methodology is employed using the 1-NN relationship among conditions whereas the grouping structure is used for the group fusion lasso.{
To evaluate} their performances, FU and GFUL are compared with the partial least square regression (MFPLS) \citep{moindjie2024classification}, the principal component regression  (MFPCR) \citep{aguilera2006using} and two group lasso methods \citep{groupfda}. The first one, denoted by GL1 ("Group Lasso 1"), uses each dimension $X^{(j)}$ of ${\bf X}$ as a group, as in the classical lasso setting. The second one,  denoted by GL2 (Group Lasso 2), uses the same group definitions as in GFUL (see equation \eqref{eq_grps} ).
\\ In addition to these methods, we propose also the regression model HG (Homogeneous Groups) resuming all the conditions within a group $\mathcal{I}_{k}$ by their mean function, 

$$ m^{(k)} = \frac{1 }{p_k} \sum_{j \in \mathcal{I}_{k}}X^{(j)}, $$
and then fit a multivariate functional linear model 
$$ Y = \sum_{k = 1}^{K}\int_{0}^{T}m^{(k)}(t)\gamma^{(k)}(t)dt + \epsilon, $$ using principal component regression methodology (\cite{aguilera2006using}).
\\ The idea behind this method is to obtain the same coefficient regression function for all conditions within a group, and that for all the groups: $\forall k = 1, \ldots, K$,
$$ \beta^{(j)} = (1/p_{k})\gamma^{(k)}, \ \ \ \forall j \in \mathcal{I}_{k}.$$
The difference with GFUL is that the latter allows only for some groups to have identical coefficient functions, whereas HG imposes it for all groups. 

 \par  Except for the model HG which doesn't have a penalty term, the hyperparameters ($\alpha, \lambda$)  in (\ref{opti_princ}) are tuned by 10-fold cross-validation: $\lambda$ is chosen from the set $$\lambda \in \left \{0.96^{i}\lambda_{\max}, i = 0,1, \ldots, 148\right\} \cup \left\{0\right\}$$ and $$\alpha \in \{0.1,0.2, \ldots, 1\},$$  with $\lambda_{\max}$ is determined as in \cite{grplasso}. { \par For MFPLS and MFPCR methods, the retained numbers of components are chosen by cross-validation. The considered grids comprise $150$ equidistant integers from $1$ to $p(M -1)$. }   
 
 \paragraph{ \it (f) the goodness of fit and homogeneity among the coefficient regression functions } \ \\
 For each method, the goodness of fit is assessed by the mean squared error (MSE) computed on the test set. Their ability to recover the true equality among coefficient functions  $\beta^{(j)}$ is measured by "sensitivity" (Sens) and "specificity" (Spec) metrics. They are defined as follows. For each pair $\left(\beta^{(j)}, \beta^{(k)}\right)$, $j,k  = 1,\ldots, p$,   we define 

$$ Sens(j,k) =  \mathbb{P}\left(\hat{\beta}^{(j)}=\hat{\beta}^{(k)} | \beta^{(j)}=\beta^{(k)}\right), $$  and 

$$ Spec(j,k) =  \mathbb{P}\left(\hat{\beta}^{(j)}\neq \hat{\beta}^{(k)} | \beta^{(j)}\neq\beta^{(k)}\right).  $$  

\noindent Thus, $ Sens(j,k)$ measures the capability of the method to obtain identical estimated coefficient functions $\hat{\beta}^{(j)}=\hat{\beta}^{(k)}$ when the theoretical ones verify that equality, $\beta^{(j)}=\beta^{(k)}$.

Then, as global measures, let define 
\begin{align*}
    \text{Sens}& = \frac{2}{p(p-1)} \sum_{j=1}^{p}\sum_{ k < j} Sens(j,k),  \\ 
    \text{Spec}& = \frac{2}{p(p-1)} \sum_{j=1}^{p}\sum_{k <j} Spec(j,k). 
\end{align*}
  
\par 

\subsubsection{Results}
\paragraph{Scenario 1}Recall that in this scenario $p=12$ and $\kappa=3$. 
The summary of the obtained metrics in  \textbf{S1} is presented Table \ref{res_sc1}.\par 
{In this scenario}, all models with penalty give close results. {The HG model gives the highest MSE and the estimation of the coefficient functions is inconsistent (see Figure \ref{scenario_1})}. Thus, the naive hypothesis that "dimensions in the same group share the same regression coefficient function" might lead to inconsistent results. Table \ref{res_sc1} shows that the variable fusion and the group fusion lasso methods reach the highest scores of sensibility and specificity. This demonstrates the ability of these methodologies to find true equalities among coefficients compared to the group lasso methods (GL1 and GL2). However, their MSEs are higher than those of MPFLS and MFPCR. Although these two methods are the best performers in terms of MSE, they lack interpretability: all their estimated coefficients have different values. 


\begin{table}[ht!]
\centering
\begin{tabular}{rlll}
  \hline
 & MSE & Sens & Spec \\ 
  \hline
GL1 & 6.2(1.59) & 0.22(0.13) & 1(0.01) \\ 
  GL2 & 6.07(1.45) & 0.29(0.12) & 1(0) \\ \hdashline
  FU & 5.97(1.52) & 0.82(0.22) & 0.99(0.01) \\ 
  GFUL & 5.21(1.81) & 0.92(0.22) & 1(0) \\ \hdashline
  HG & 14.85(3.25) & 1(0) & 0.95(0) \\
  \hdashline 
 MFPLS &3.47 (0.85) &0(0) & 1(0) \\ 
MFPCR & 4.10 (1.60) & 0(0) & 1(0) \\
  \hline
\end{tabular}
\caption{Scenario S1: MSE mean and standard error (in parentheses), sensibility and specificity obtained metrics with $I=100$ experiments. }
\label{res_sc1}
\end{table}
\newpage
\begin{figure}[ht]
    \centering
    \begin{tabular}{ccccc}
& \begin{tabular}[c]{@{}c@{}}Group 1 \\ ${\scriptstyle \beta^{(1)}= \beta^{(2)}= \beta^{(3)}}$\end{tabular} & \begin{tabular}[c]{@{}c@{}}Group 2 \\ ${\scriptstyle \beta^{(4)}= \beta^{(5)}= \beta^{(6)}}$\end{tabular} & \begin{tabular}[c]{@{}c@{}}Group 3 \\${\scriptstyle \beta^{(7)} \neq \beta^{(8)} \neq \beta^{(9)}}$\end{tabular} & \begin{tabular}[c]{@{}c@{}}Group 4 \\ ${\scriptstyle \beta^{(10)} = \beta^{(11)} = \beta^{(12)}}$\end{tabular}  \\
\textbf{GFUL}        & \includegraphics[scale=.1, align=c]{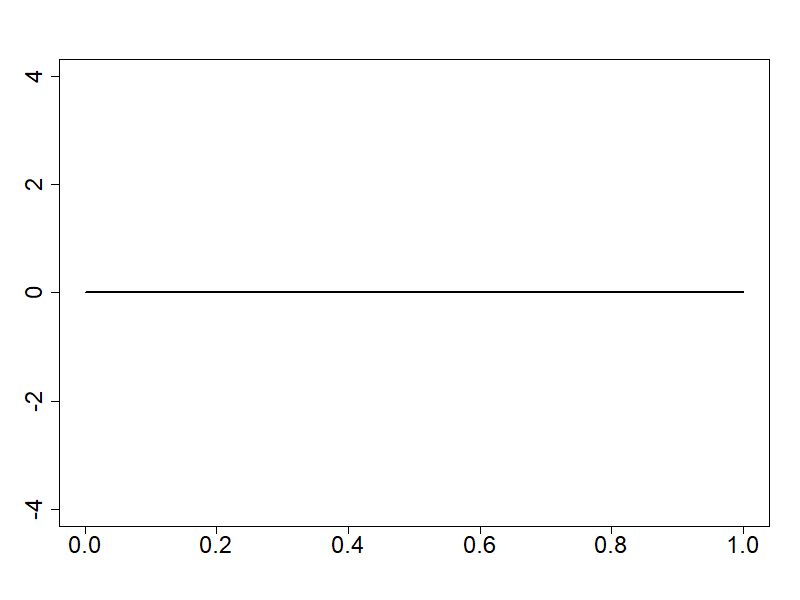}                                     & \includegraphics[scale=.1, align=c]{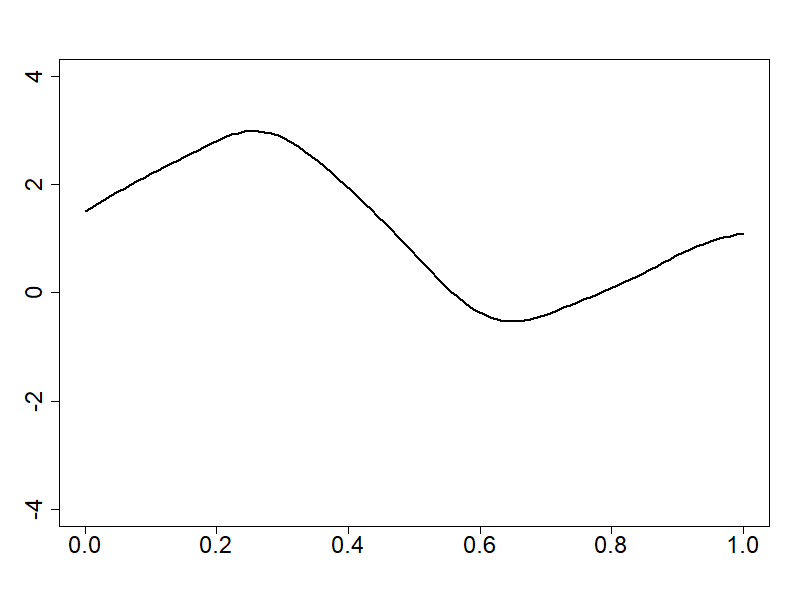}                                     & \includegraphics[scale=.1, align=c]{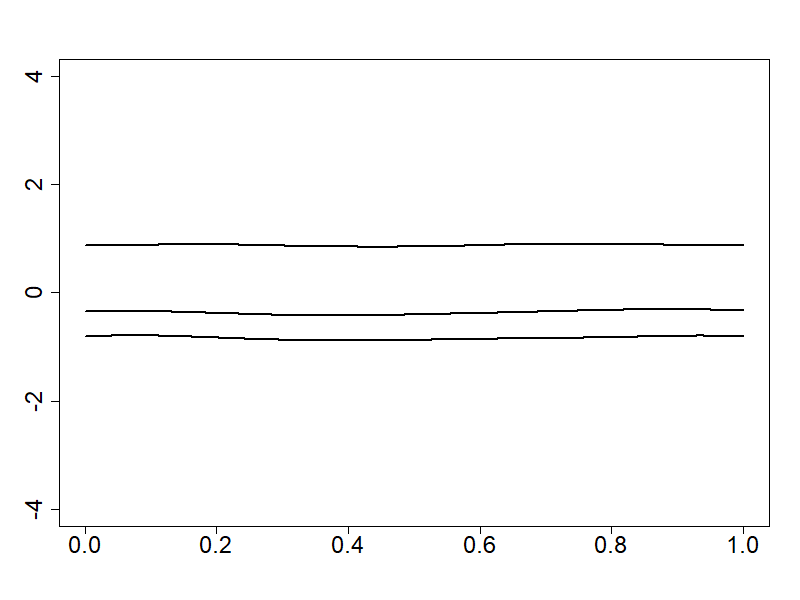}                                             & \includegraphics[scale=.1, align=c]{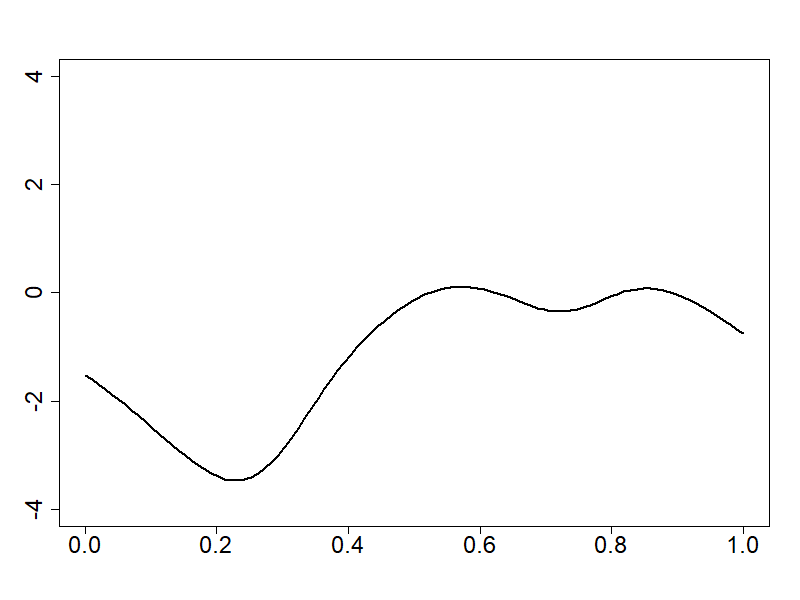}                                         \\
\textbf{FU}          & \includegraphics[scale=.1, align=c]{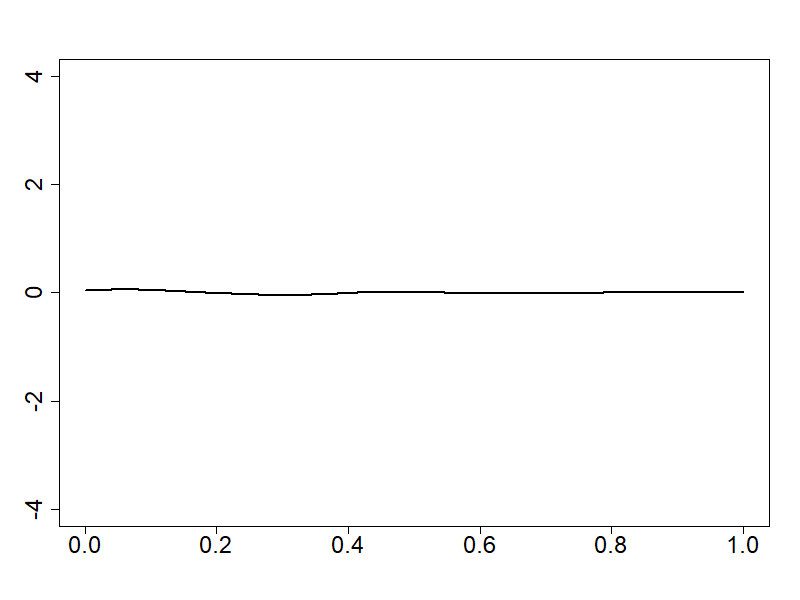}                                     & \includegraphics[scale=.1, align=c]{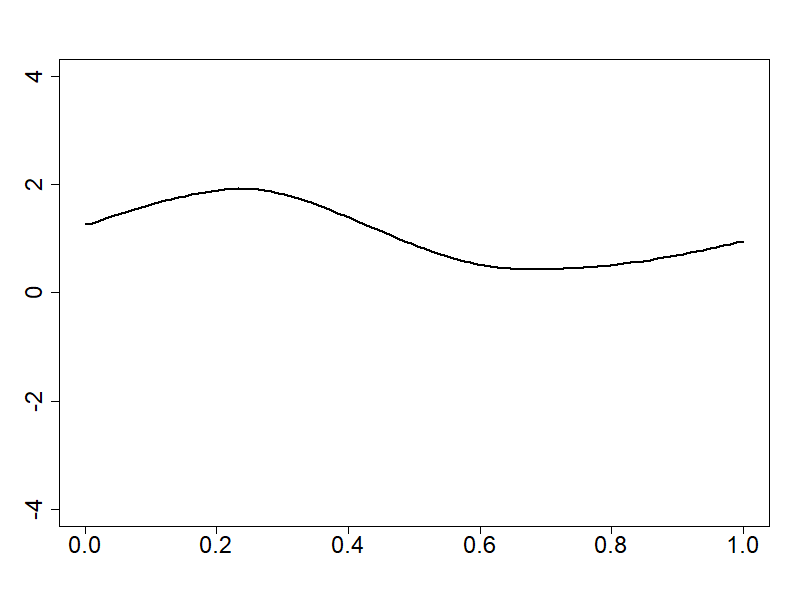}                                     & \includegraphics[scale=.1, align=c]{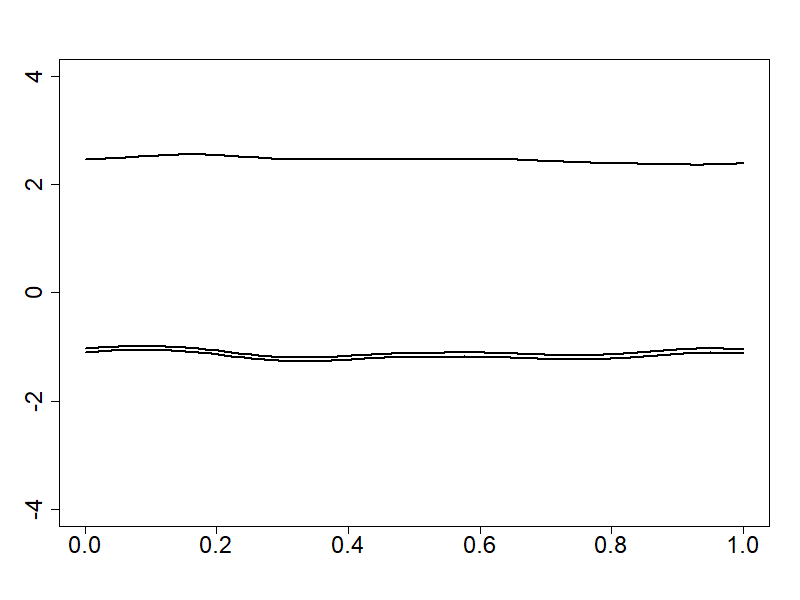}                                             & \includegraphics[scale=.1, align=c]{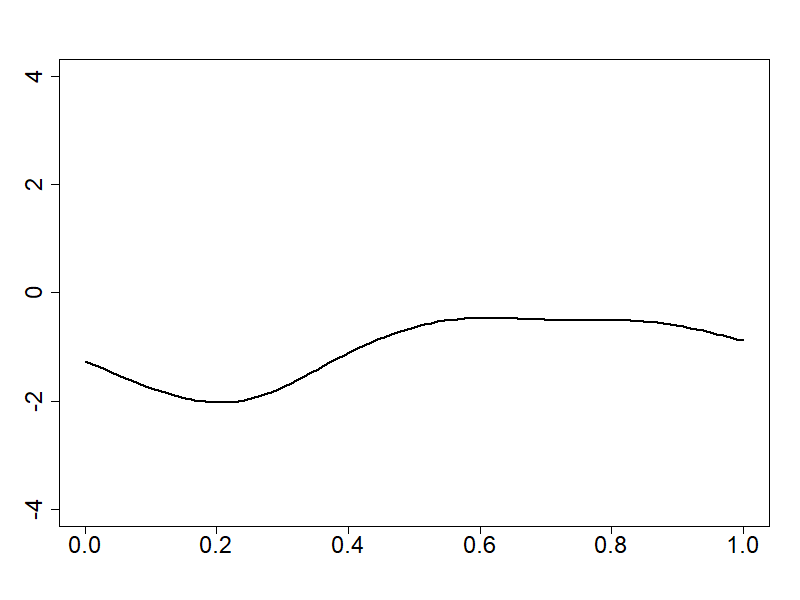}                                         \\
\textbf{GL1}          & \includegraphics[scale=.1, align=c]{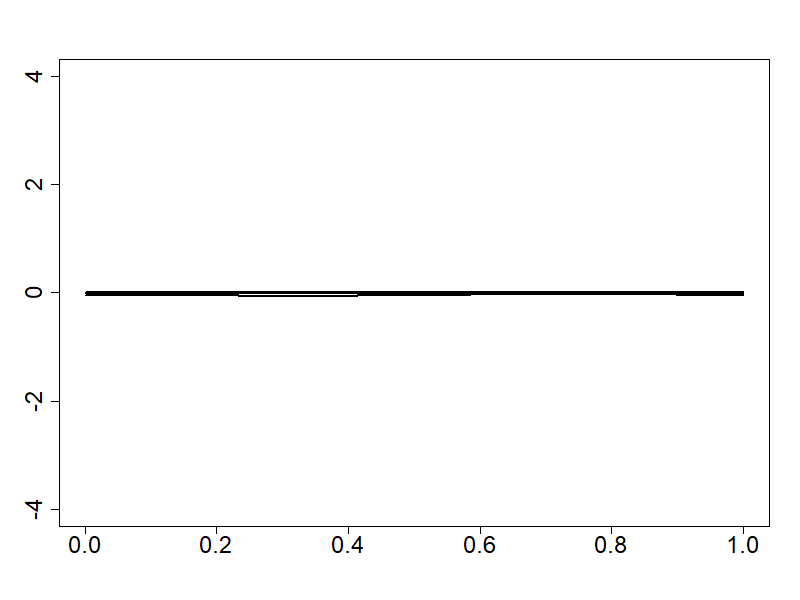}                                     & \includegraphics[scale=.1, align=c]{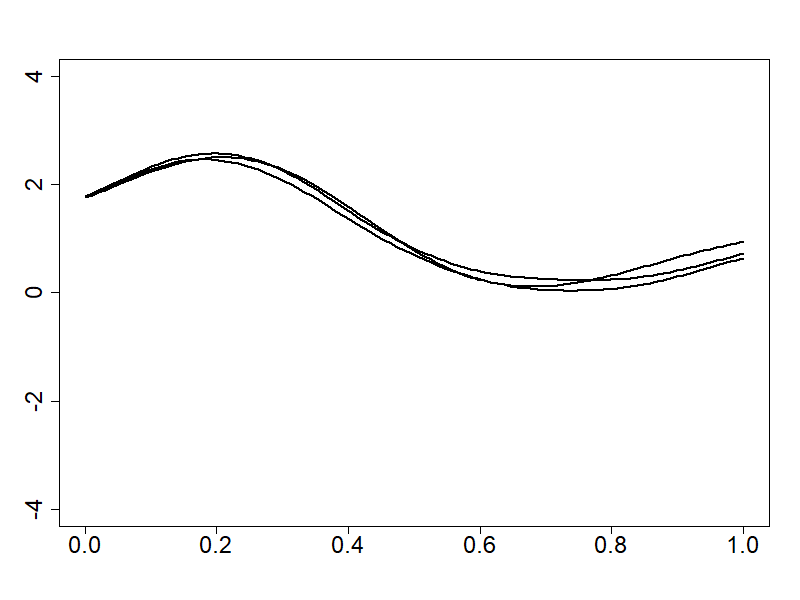}                                     & \includegraphics[scale=.1, align=c]{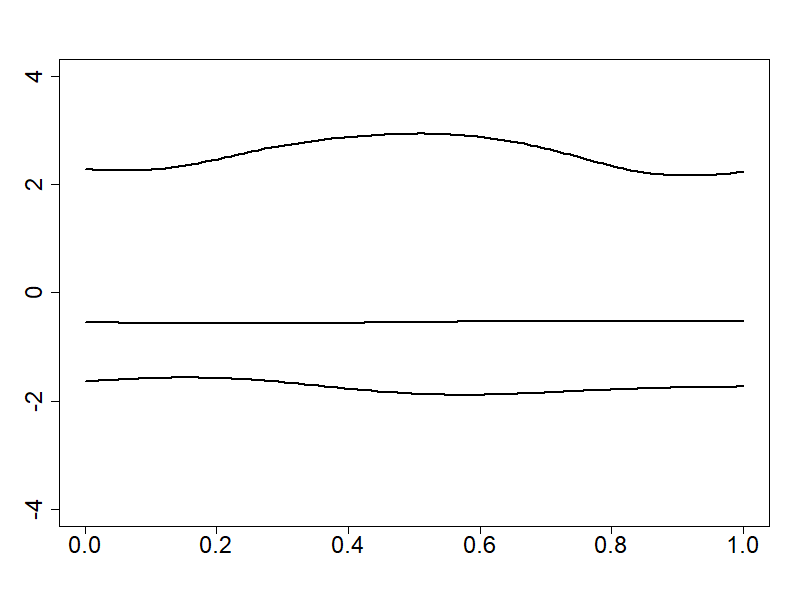}                                             & \includegraphics[scale=.1, align=c]{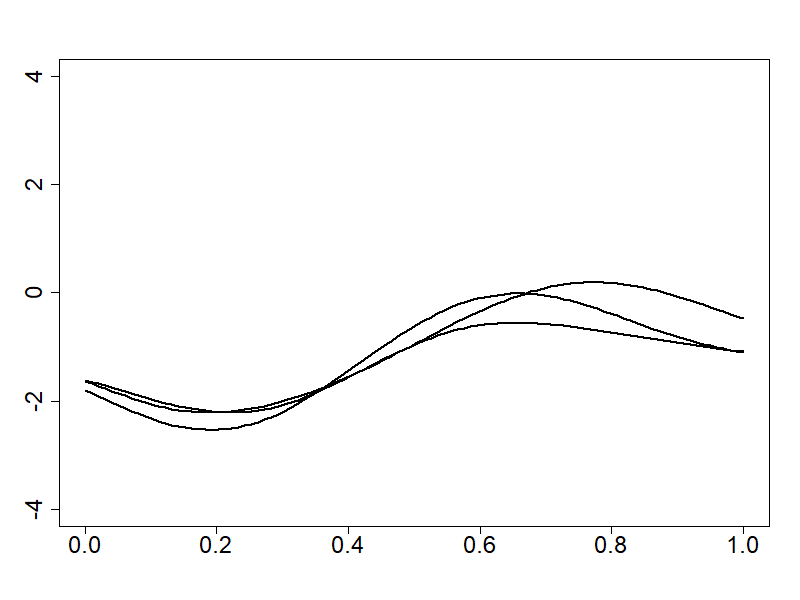}                                         \\
\textbf{GL2}          & \includegraphics[scale=.1, align=c]{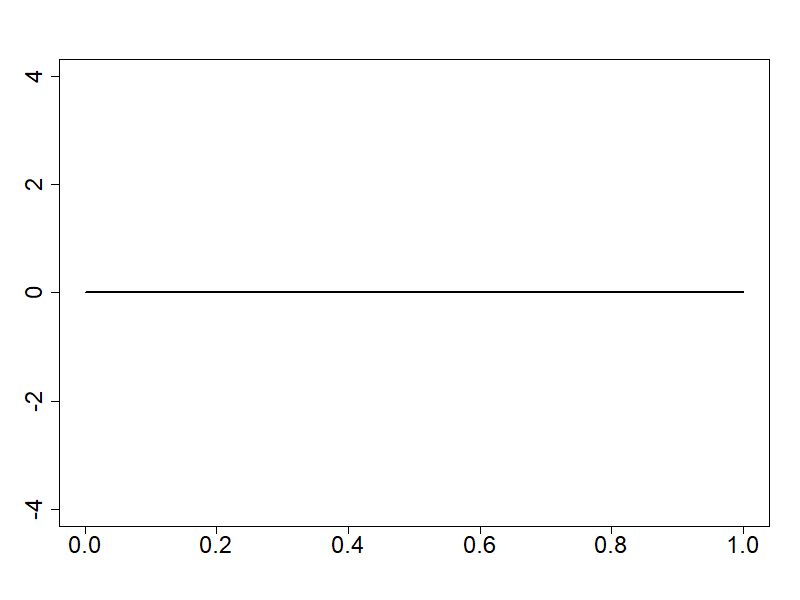}                                     & \includegraphics[scale=.1, align=c]{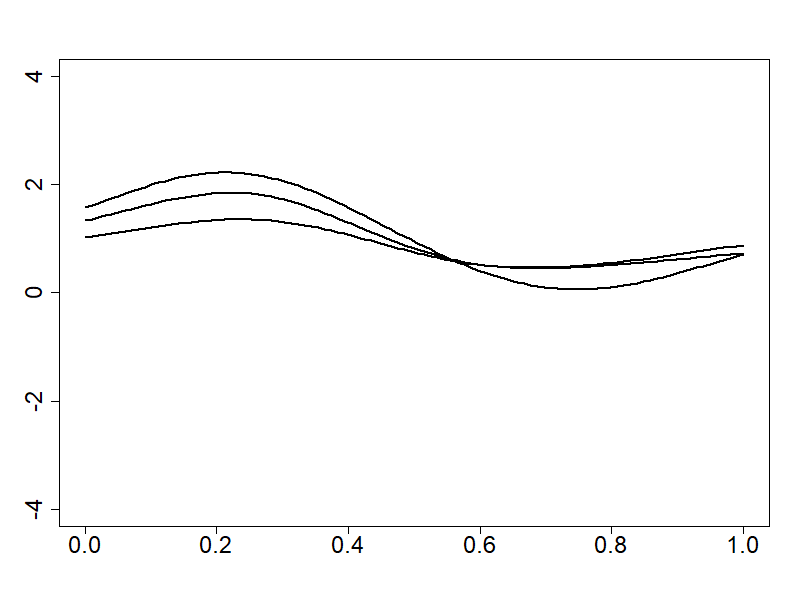}                                     & \includegraphics[scale=.1, align=c]{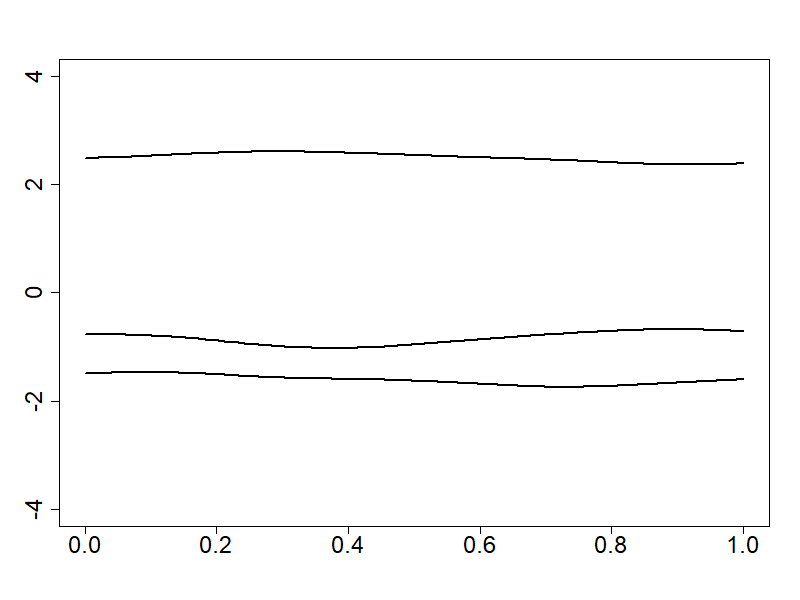}                                             & \includegraphics[scale=.1, align=c]{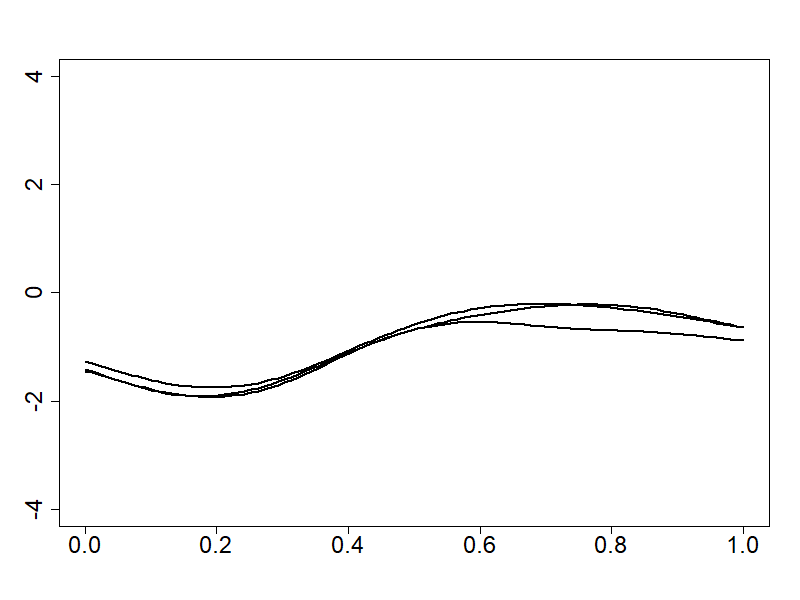}                                         \\

\textbf{HG}          & \includegraphics[scale=.1, align=c]{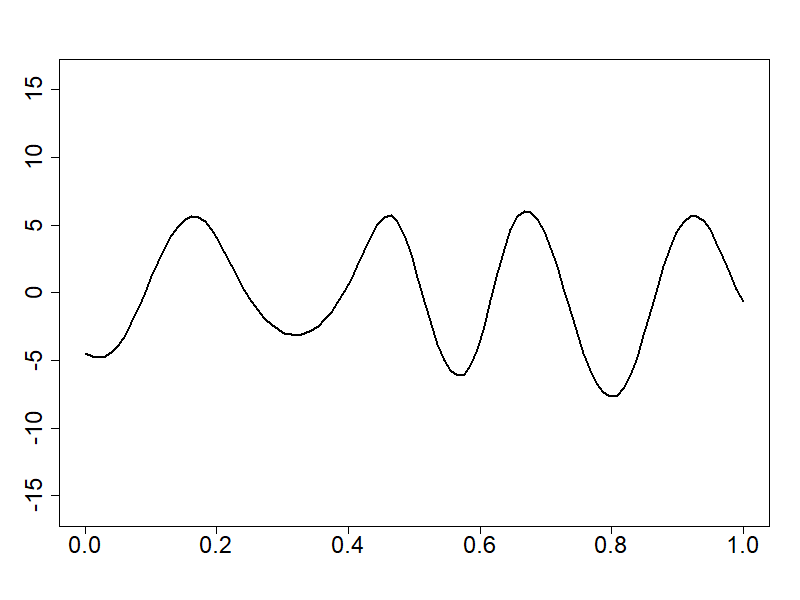}                                     & \includegraphics[scale=.1, align=c]{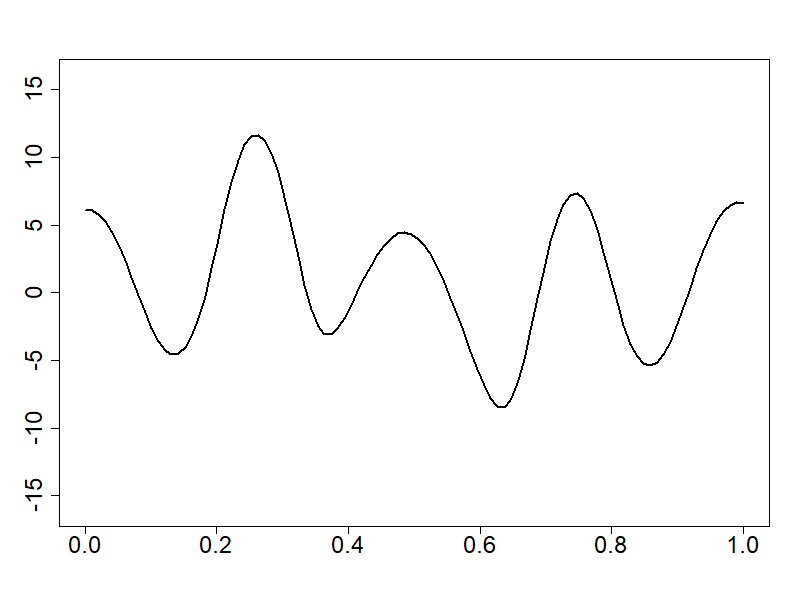}                                     & \includegraphics[scale=.1, align=c]{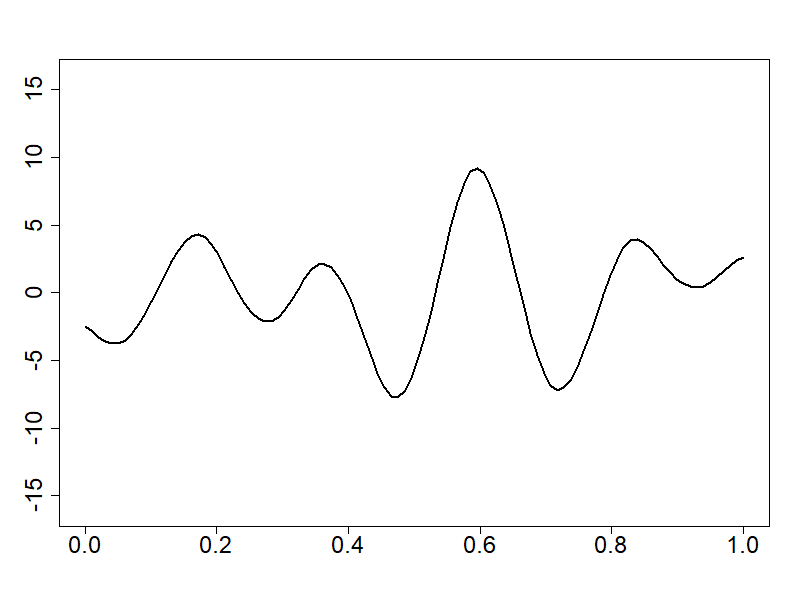}                                             & \includegraphics[scale=.1, align=c]{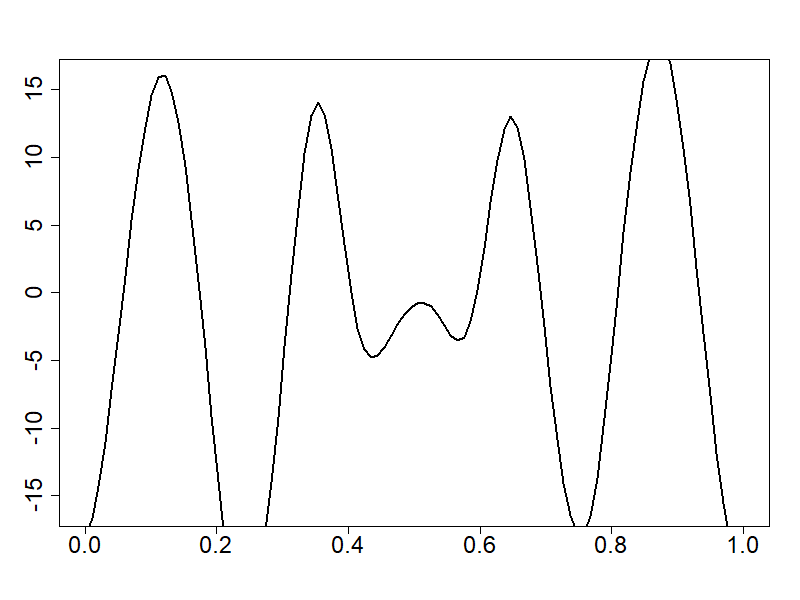}  \\ 

\textbf{MFPLS}          & \includegraphics[scale=.1, align=c]{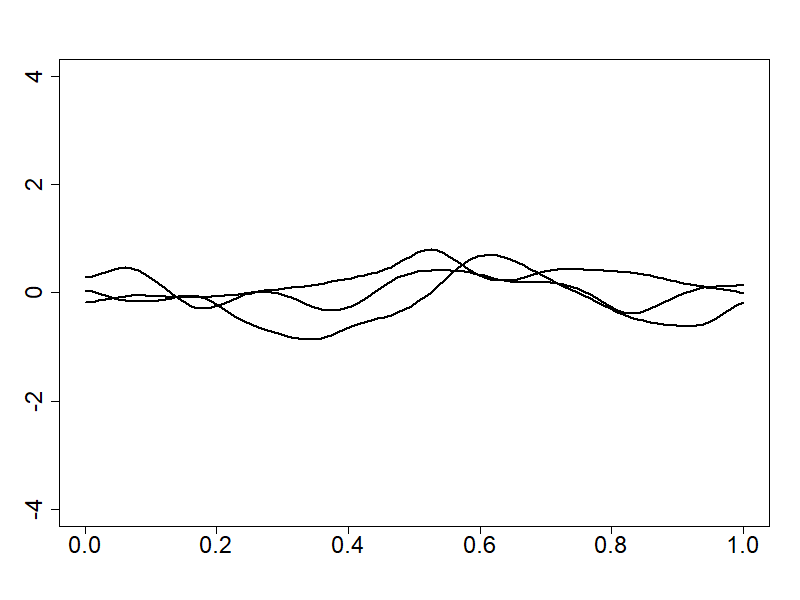}                                     & \includegraphics[scale=.1, align=c]{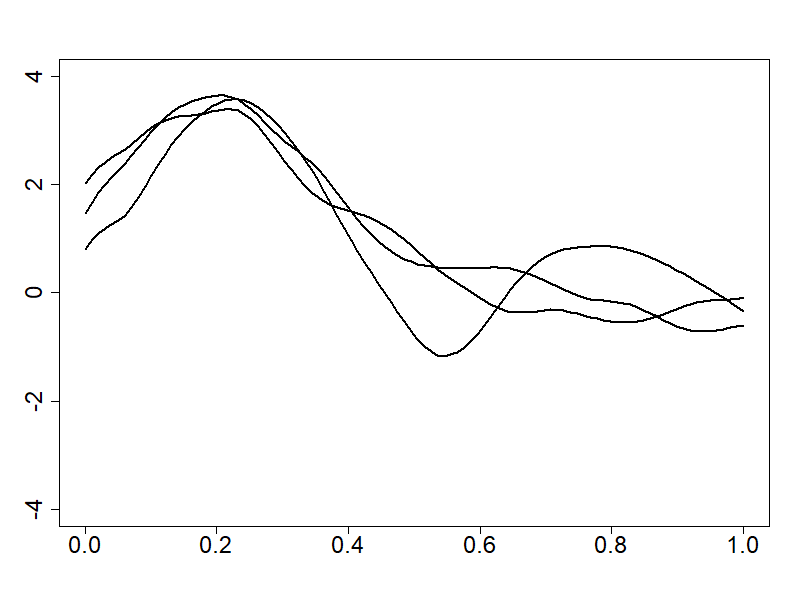}                                     & \includegraphics[scale=.1, align=c]{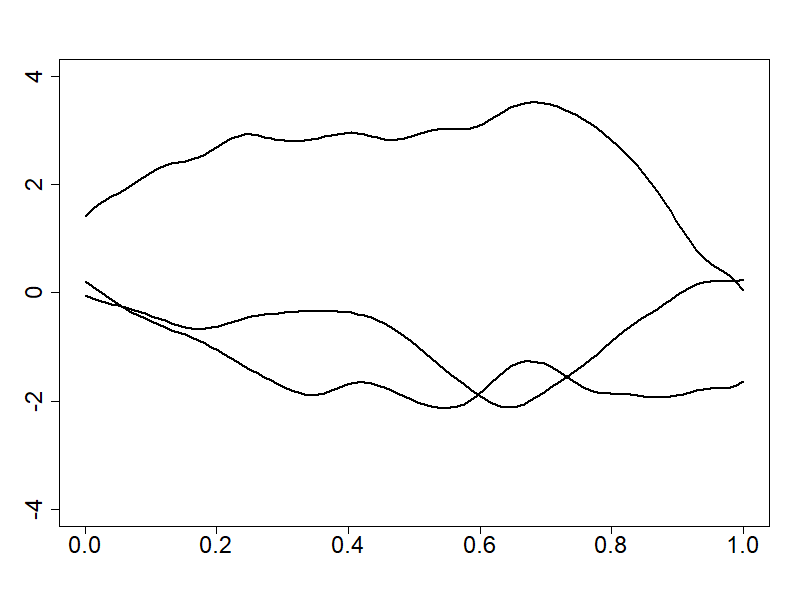}                                             & \includegraphics[scale=.1, align=c]{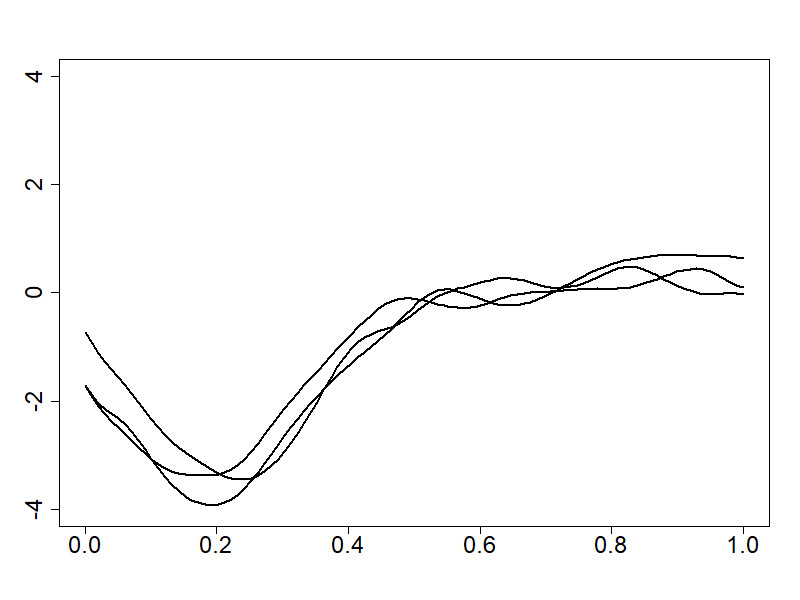}  \\ 
\textbf{MFPCR}          & \includegraphics[scale=.1, align=c]{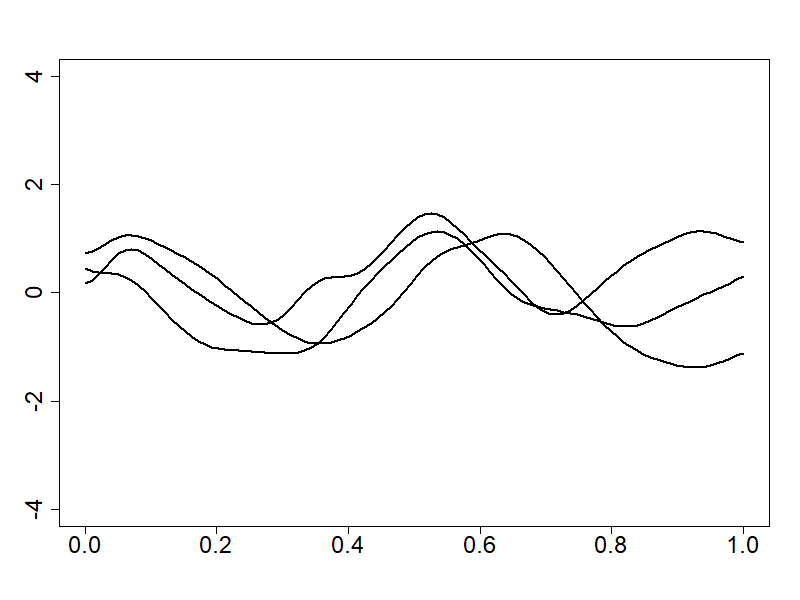}                                     & \includegraphics[scale=.1, align=c]{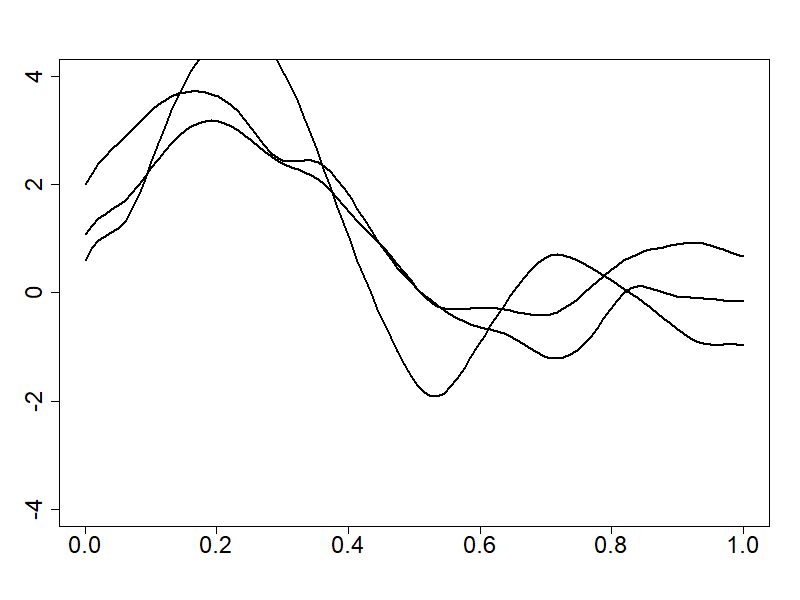}                                     & \includegraphics[scale=.1, align=c]{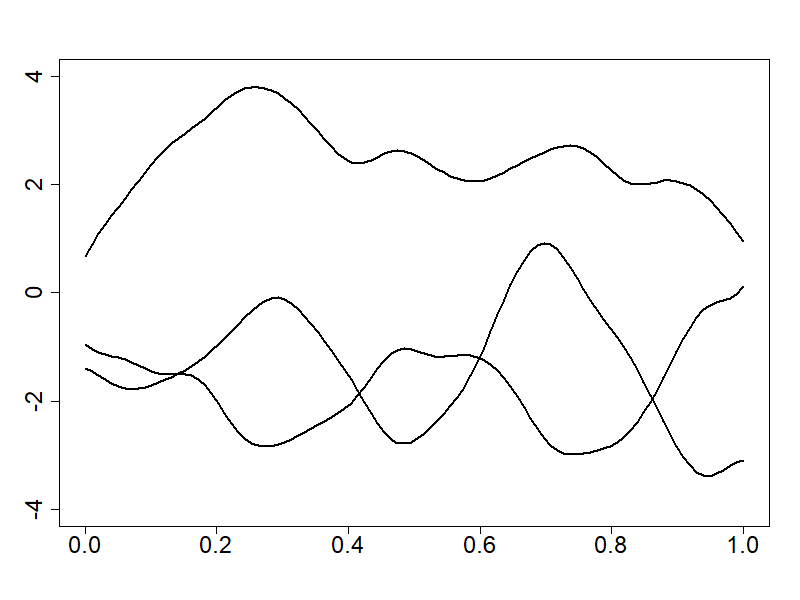}                                             & \includegraphics[scale=.1, align=c]{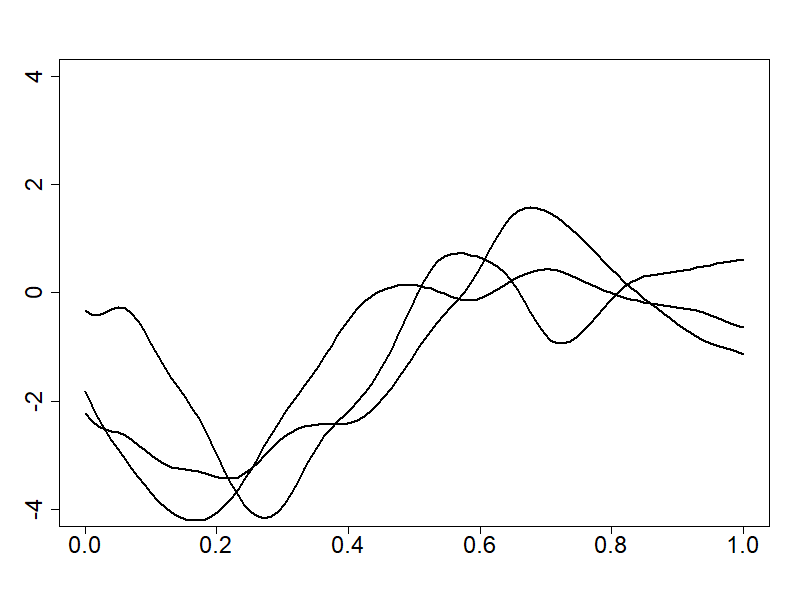}  
\end{tabular}
    \caption{Scenario 1- The estimations of $\boldsymbol{\beta}$ by the different methods (first simulation).}
    \label{scenario_1}
\end{figure}

\newpage 

\paragraph{Scenario 2: }
{In this scenario $p=80$ and $\kappa=20$. Table \ref{res_s2} shows that GFUL is the best methodology for the MSE metric}. The high specificity and the low sensitivity of the competitor methods indicate that they ignore that some groups share the same regression coefficient functions (see Figure \ref{scenario_2}). This is also reflected in the MSE criteria. \par 
Let observe that all the other methods, including FU,   provide quite bad results with respect to GFUL. This can be explained by the fact that these methods are clearly not adapted to consider the grouping structure of conditions. 
 
\begin{table}[ht]
\centering
\begin{tabular}{rlll}
  \hline
 & MSE & Sens & Spec \\ 
  \hline
GL1 & 88.74(23.75) & 0.2(0.19) & 0.85(0.18) \\ 
  GL2 & 64.29(17.22) & 0.08(0.14) & 1(0) \\ \hdashline
  FU & 70.58(18.53) & 0.07(0.01) & 1(0) \\ 
  GFUL & 31.68(16.33) & 0.73(0.4) & 1(0) \\ \hdashline 
  HG & 69.37(15.61) & 1(0) & 0.93(0) \\
  \hdashline 
MFPLS & 70.28(17.81)    & 0(0)   &1(0)  \\  
  MFPCR &  75.76(19.93)  & 0(0) &  1(0)\\ 
   \hline
\end{tabular}
\caption{Scenario S2: MSE mean and standard error (in parentheses), Sensibility and specificity obtained metrics with $I=100$ experiments.}
\label{res_s2}
\end{table}


\newpage 
\begin{figure}[ht]
    \centering
      \begin{tabular}{ccccc}
& \begin{tabular}[c]{@{}c@{}}Group 1 \\ ${\scriptstyle \beta^{(1)}= \beta^{(2)}= \ldots =\beta^{(20)}}$\end{tabular} & \begin{tabular}[c]{@{}c@{}}Group 2 \\ ${\scriptstyle \beta^{(21)}= \beta^{(22)}= \ldots =\beta^{(40)}}$\end{tabular} & \begin{tabular}[c]{@{}c@{}}Group 3 \\${\scriptstyle \beta^{(41)} \neq \beta^{(42)} \neq \ldots \neq \beta^{(60)}}$\end{tabular} & \begin{tabular}[c]{@{}c@{}}Group 4 \\ ${\scriptstyle \beta^{(61)} = \beta^{(62)} = \ldots = \beta^{(80)}}$\end{tabular}  \\
\textbf{GFUL}        & \includegraphics[scale=.1, align=c]{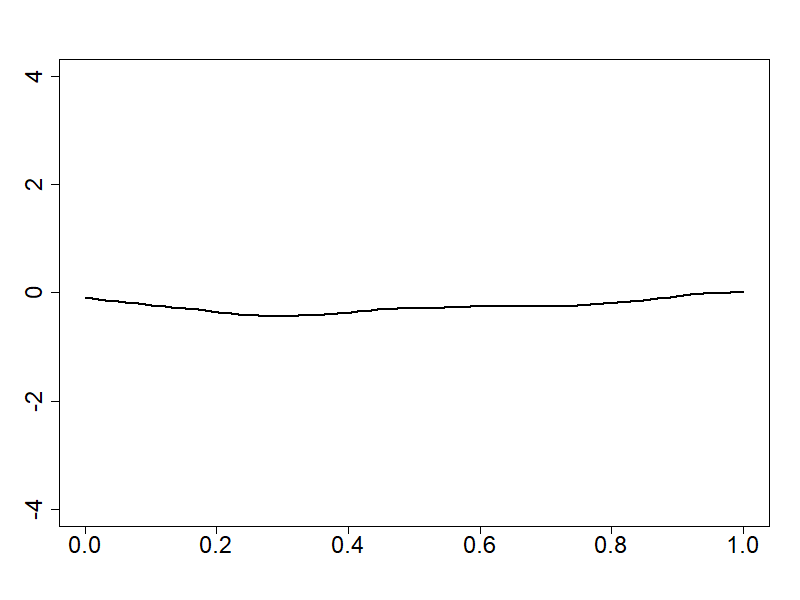}                                     & \includegraphics[scale=.1, align=c]{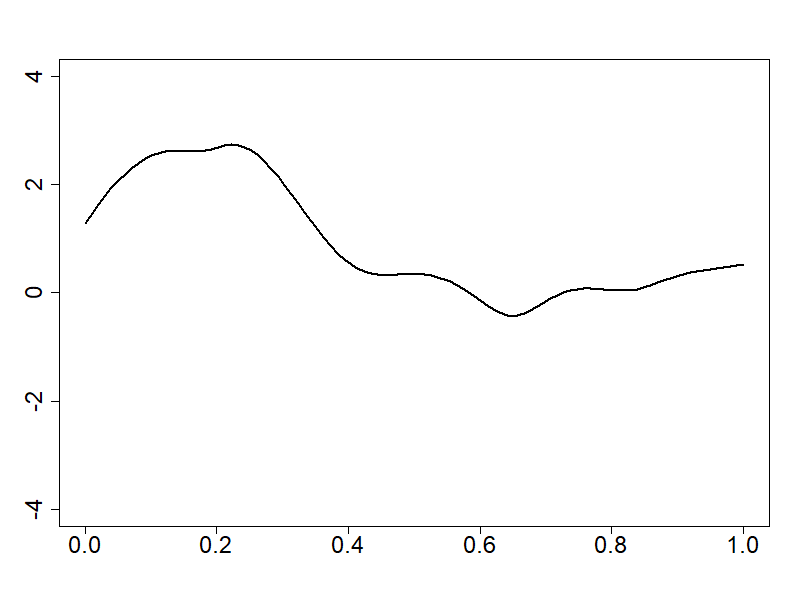}                                     & \includegraphics[scale=.1, align=c]{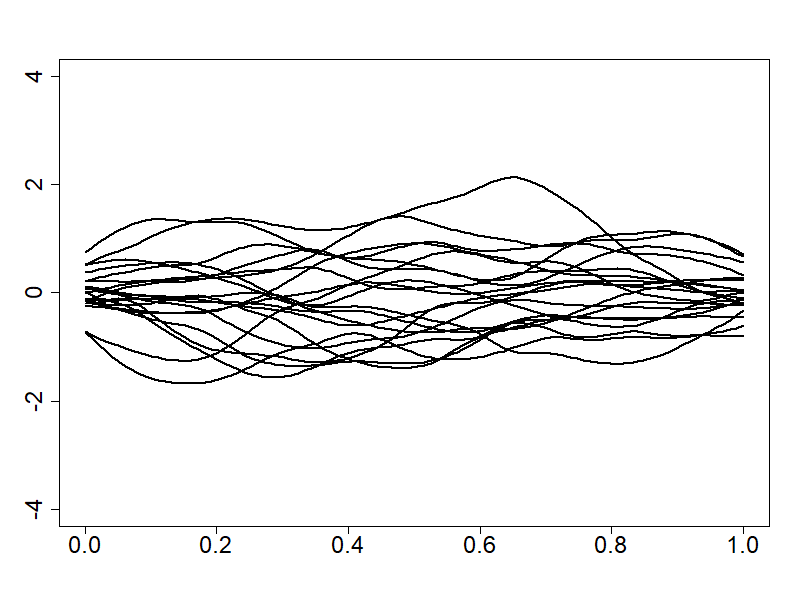}                                             & \includegraphics[scale=.1, align=c]{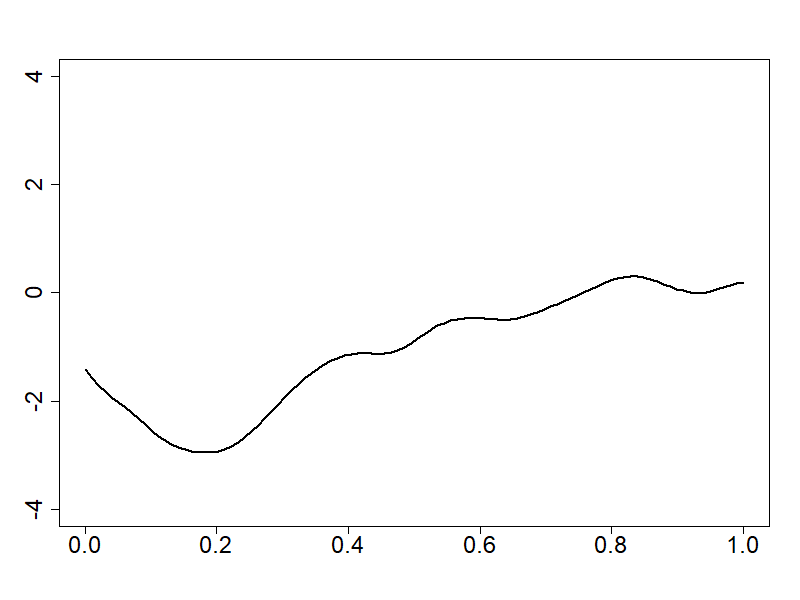}                                         \\
\textbf{FU}          & \includegraphics[scale=.1, align=c]{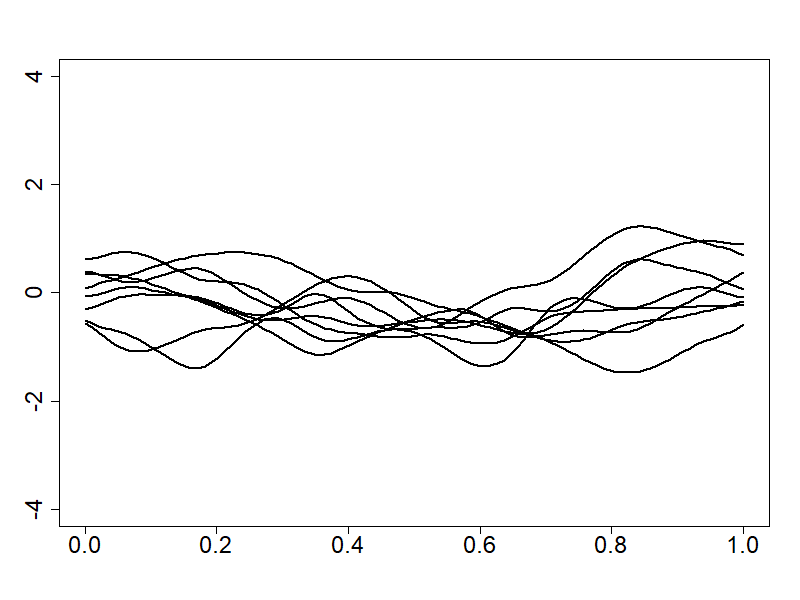}                                     & \includegraphics[scale=.1, align=c]{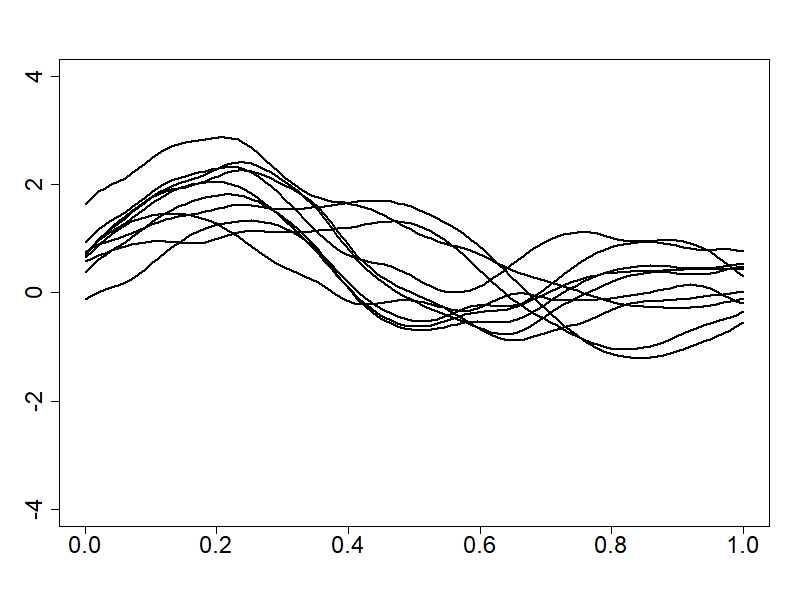}                                     & \includegraphics[scale=.1, align=c]{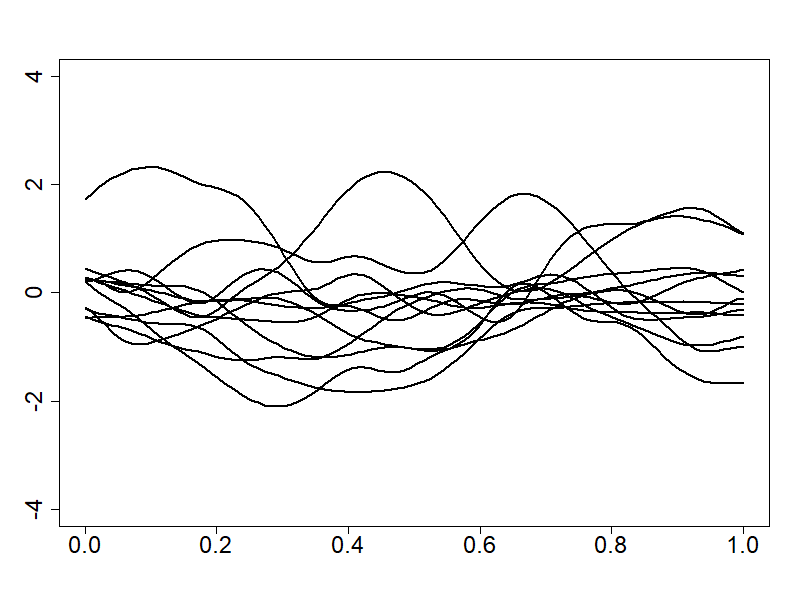}                                             & \includegraphics[scale=.1, align=c]{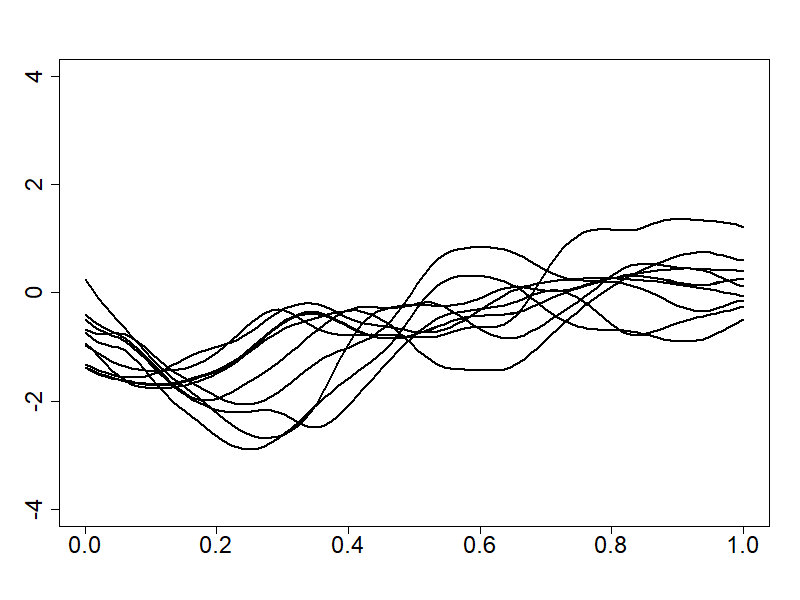}                                         \\

\textbf{GL1}          & \includegraphics[scale=.1, align=c]{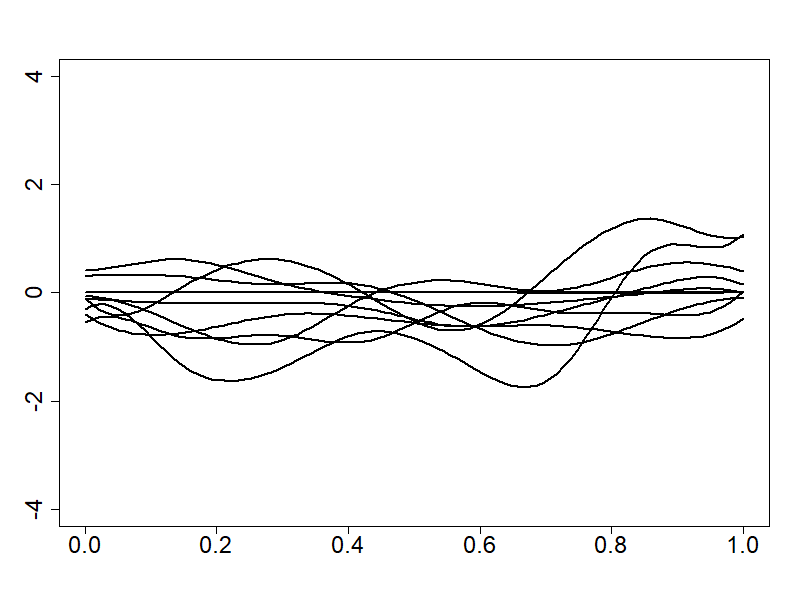}                                     & \includegraphics[scale=.1, align=c]{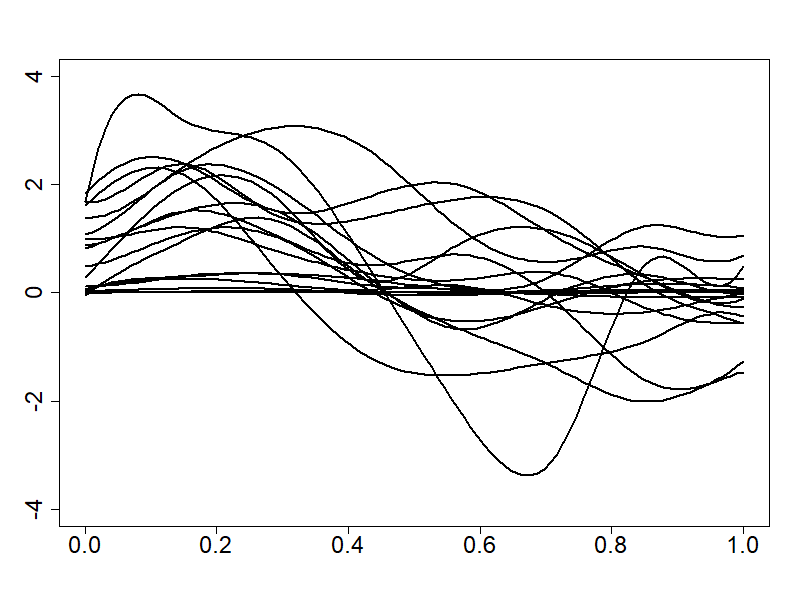}                                     & \includegraphics[scale=.1, align=c]{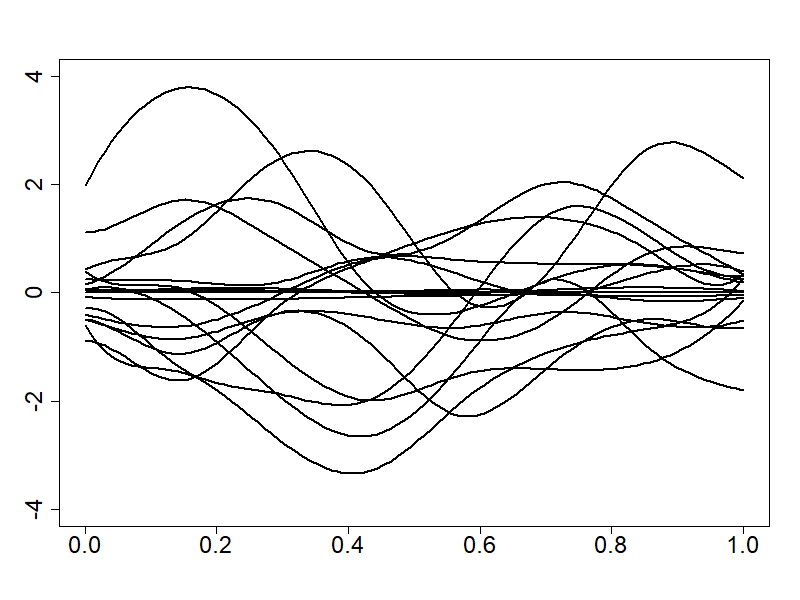}                                             & \includegraphics[scale=.1, align=c]{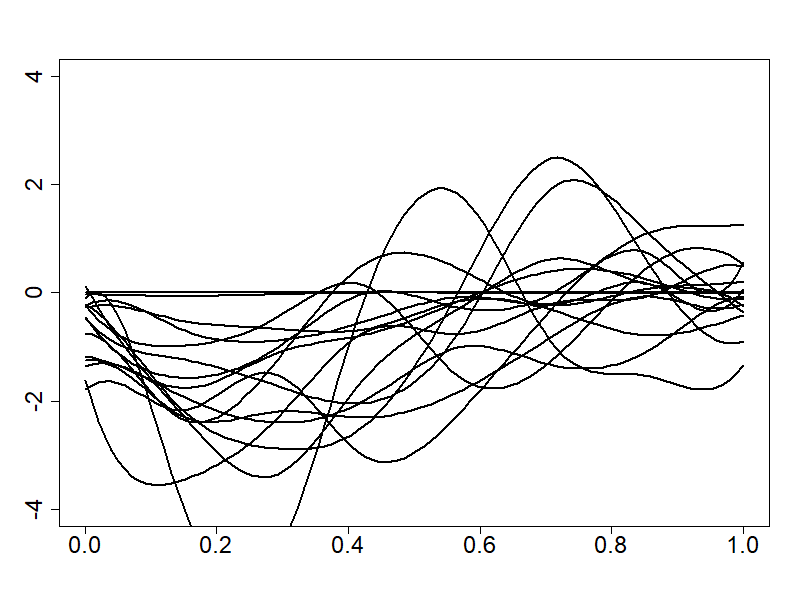}                                         \\

\textbf{GL2}          & \includegraphics[scale=.1, align=c]{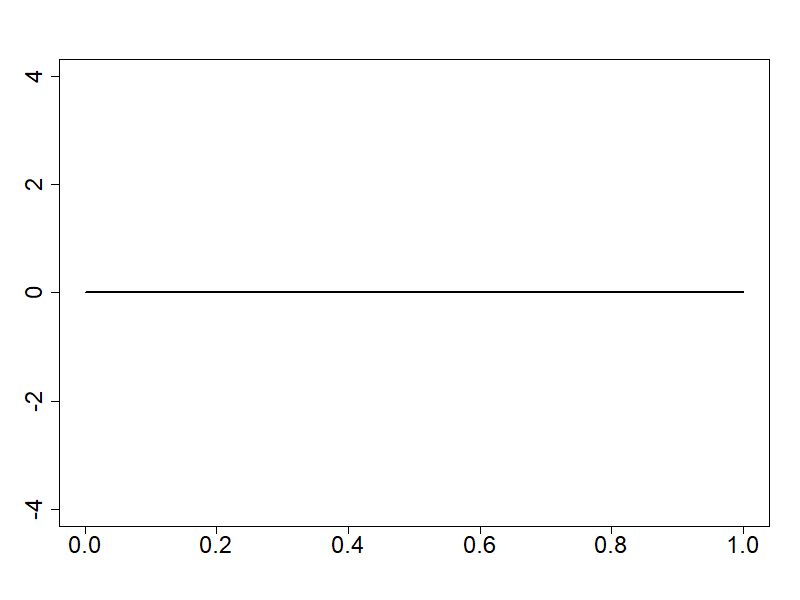}                                     & \includegraphics[scale=.1, align=c]{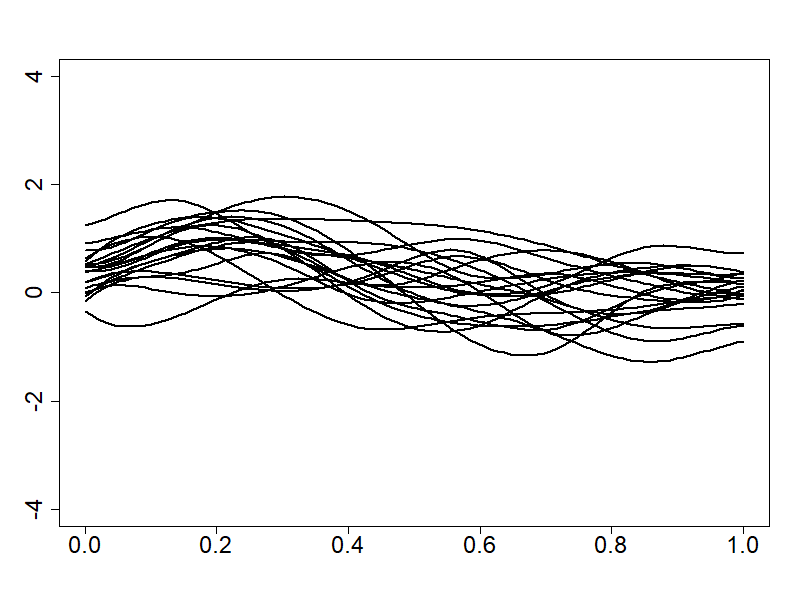}                                     & \includegraphics[scale=.1, align=c]{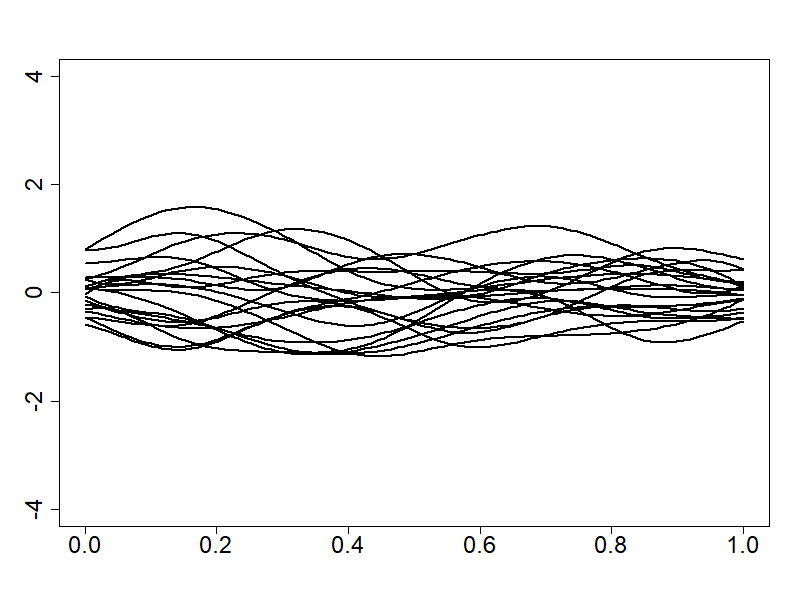}                                             & \includegraphics[scale=.1, align=c]{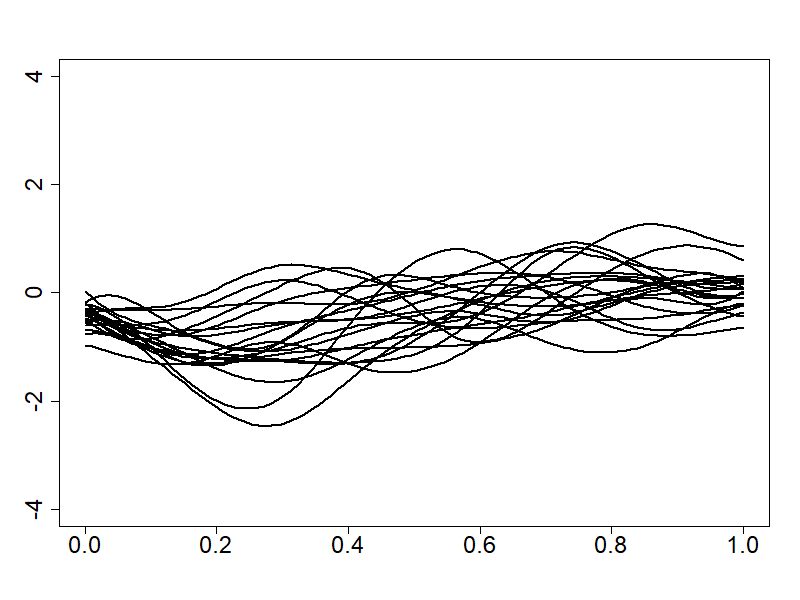}                                         \\
\textbf{HG}          & \includegraphics[scale=.1, align=c]{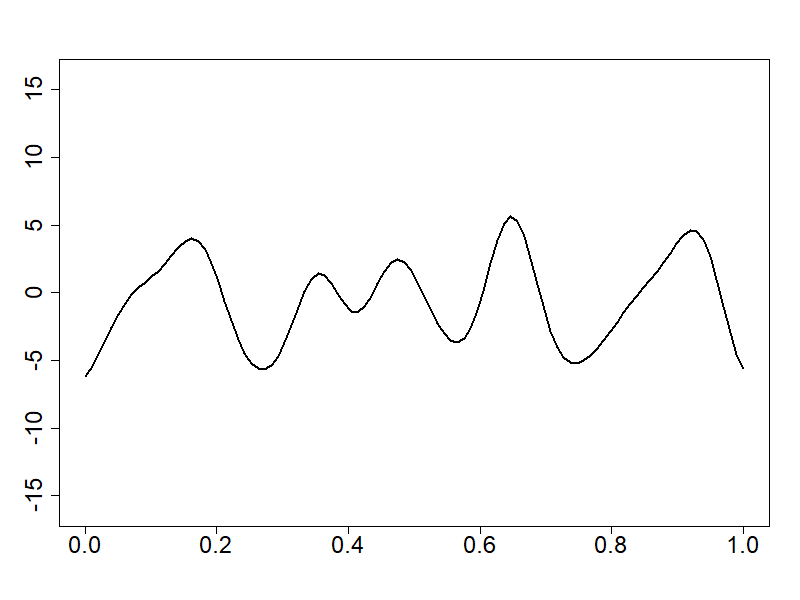}                                     & \includegraphics[scale=.1, align=c]{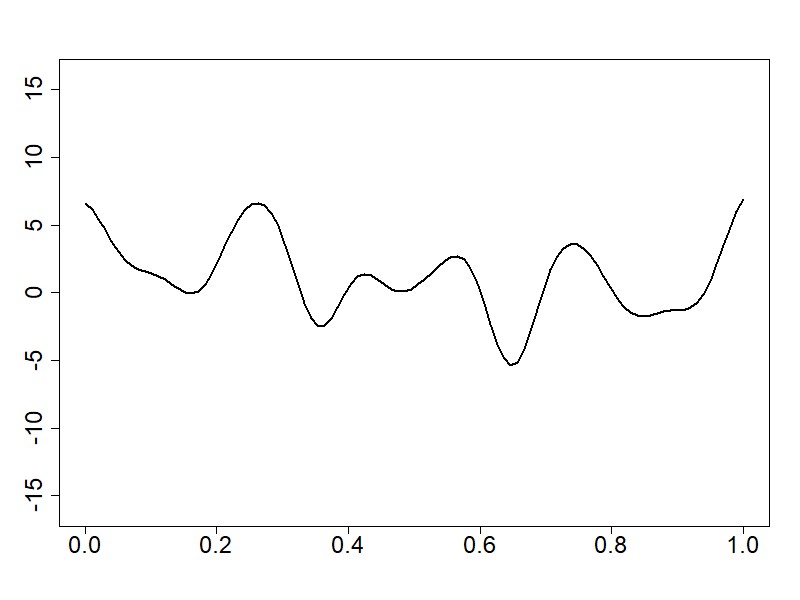}                                     & \includegraphics[scale=.1, align=c]{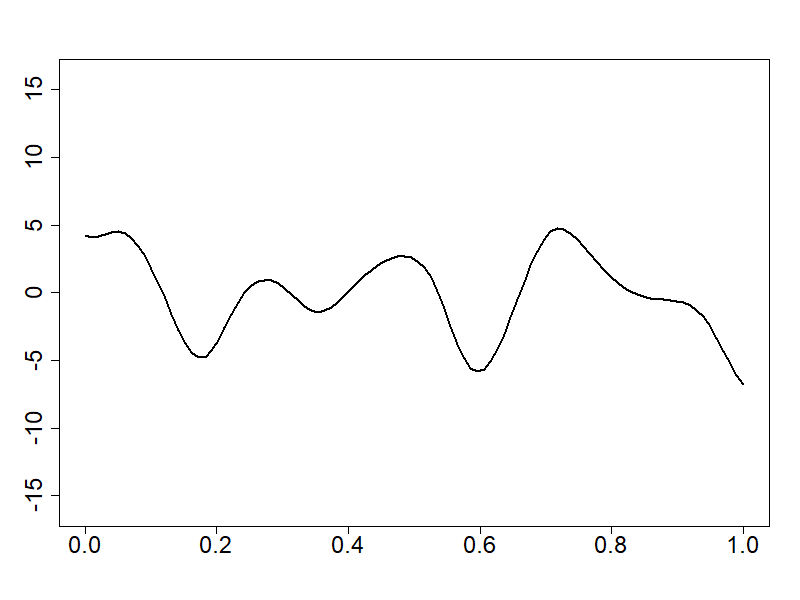}                                             & \includegraphics[scale=.1, align=c]{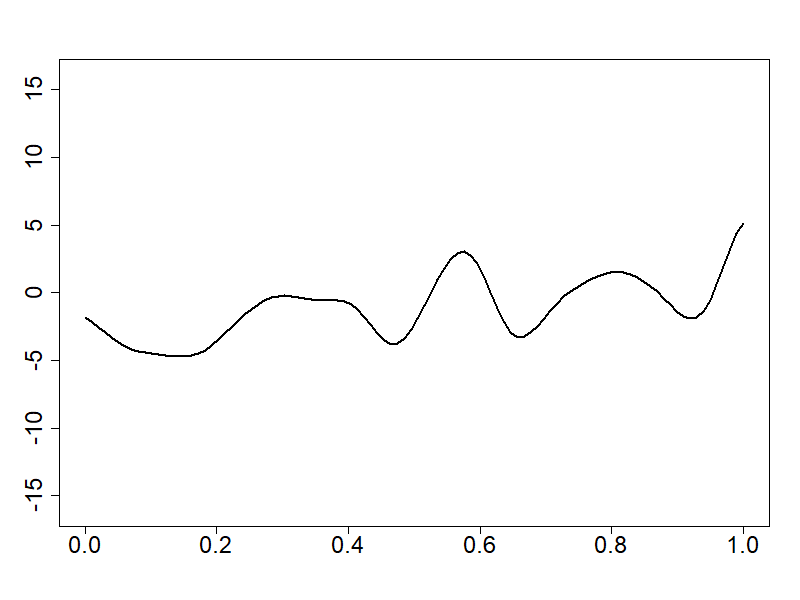} \\ 
 \textbf{MFPLS}          & \includegraphics[scale=.1, align=c]{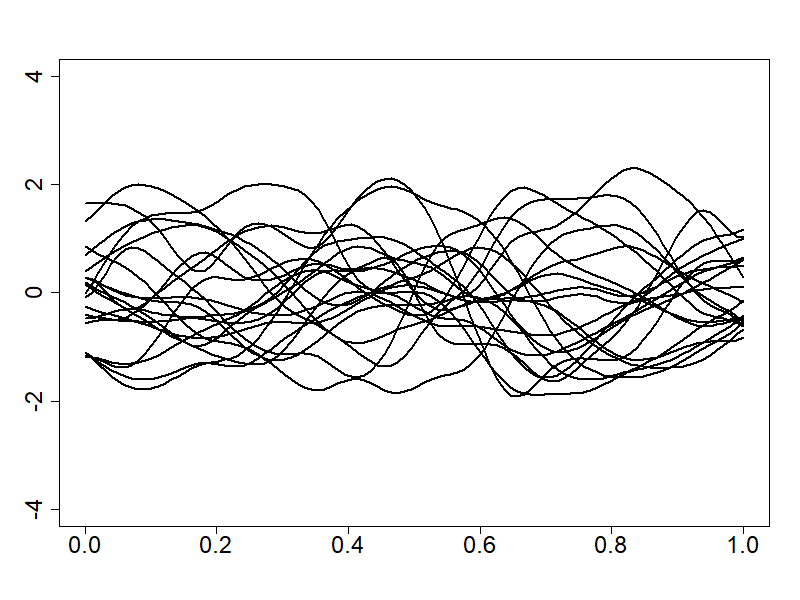}                                     & \includegraphics[scale=.1, align=c]{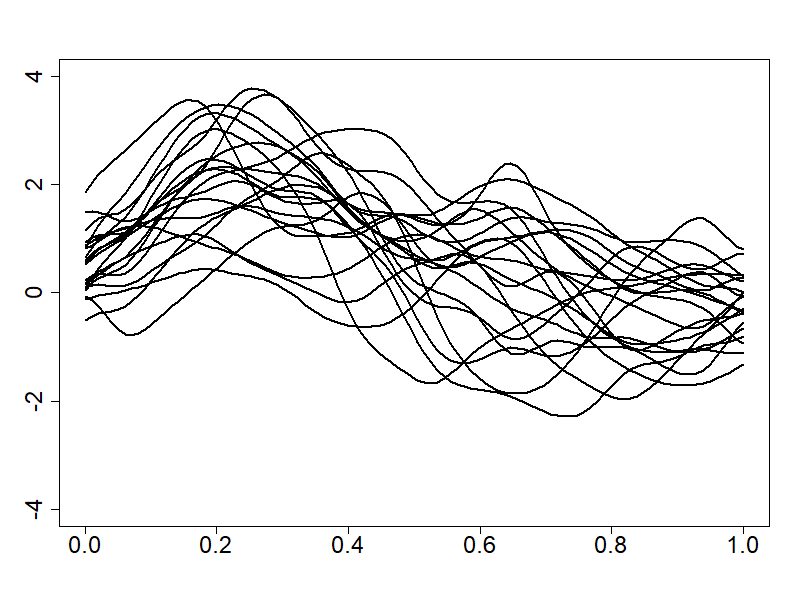}                                     & \includegraphics[scale=.1, align=c]{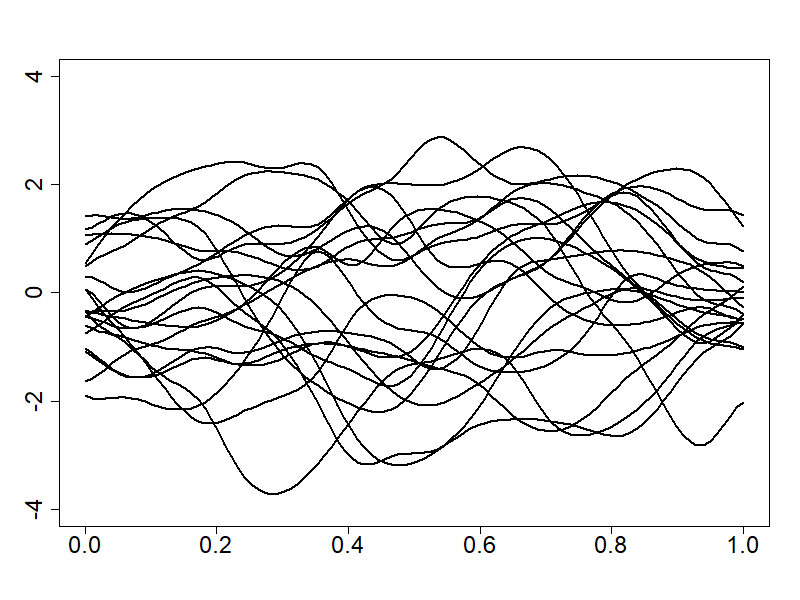}                                             & \includegraphics[scale=.1, align=c]{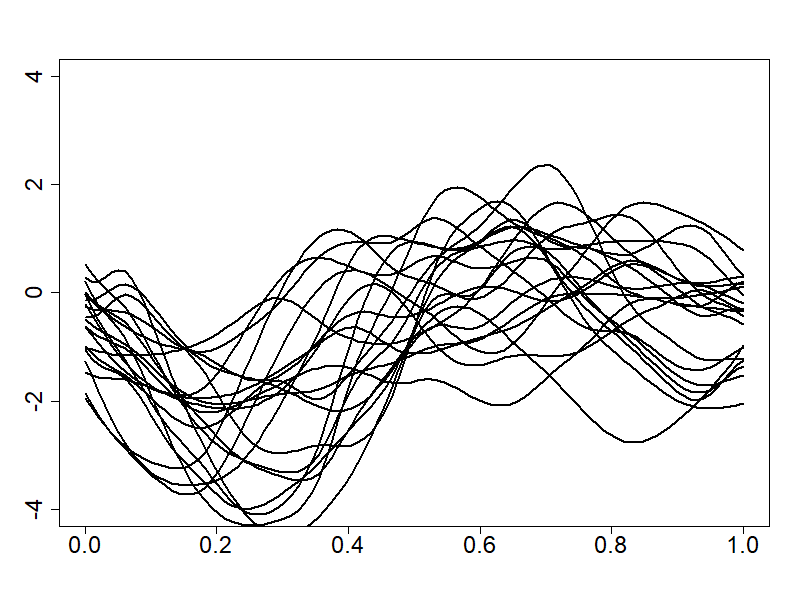}   \\ 
 \textbf{MFPCR}          & \includegraphics[scale=.1, align=c]{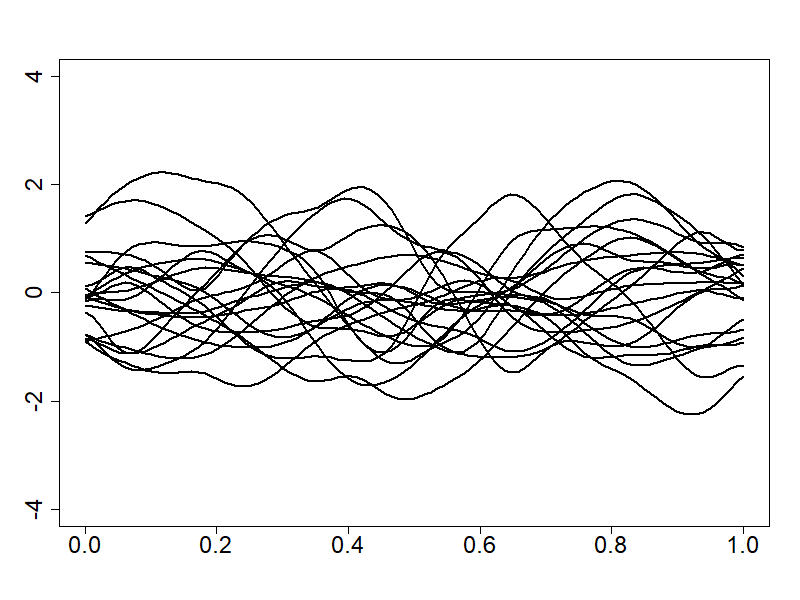}                                     & \includegraphics[scale=.1, align=c]{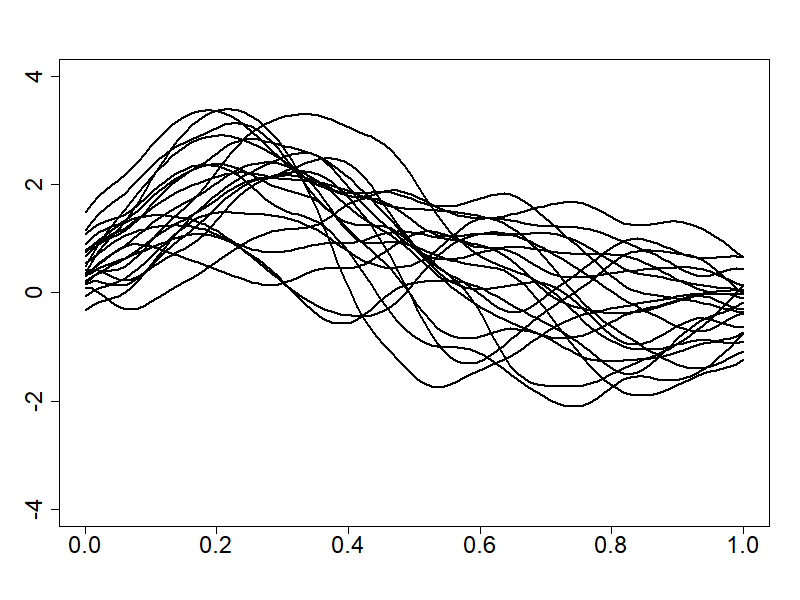}                                     & \includegraphics[scale=.1, align=c]{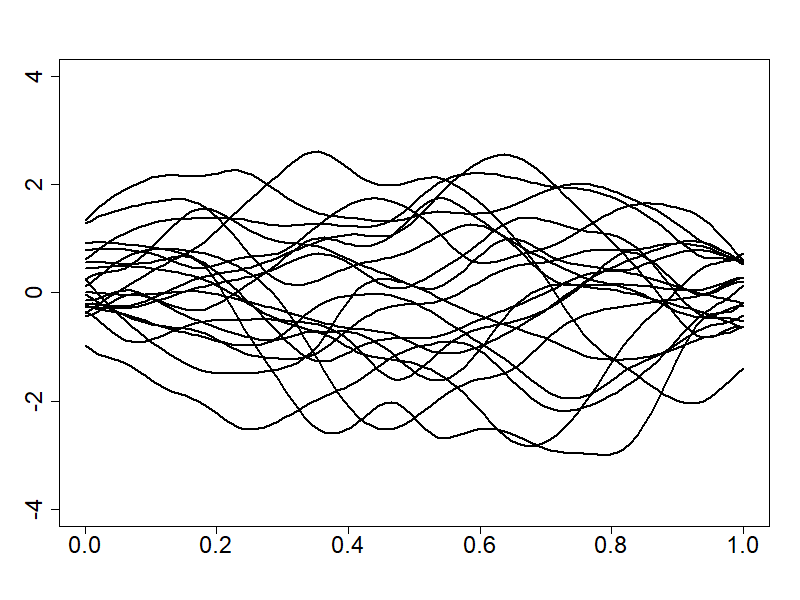}                                             & \includegraphics[scale=.1, align=c]{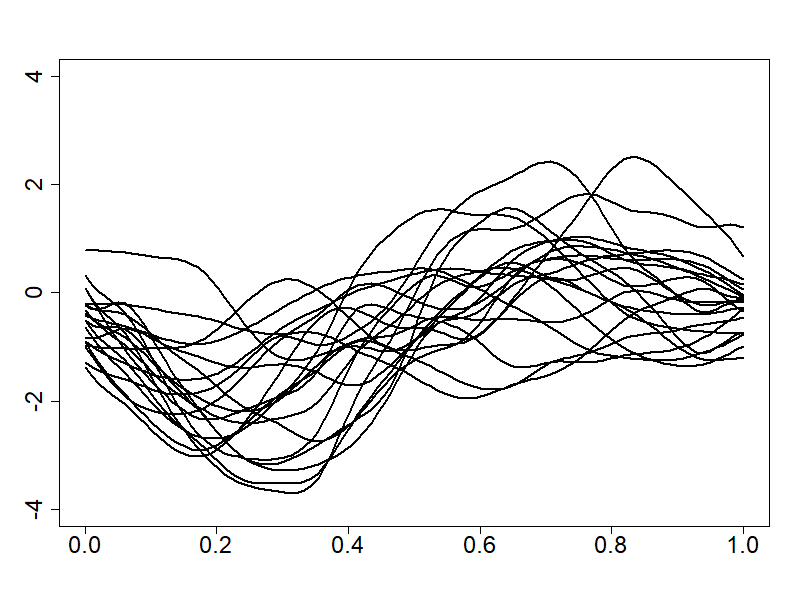} 

\end{tabular}
    \caption{Scenario 2- The estimations of $\boldsymbol{\beta}$ by the different methods (first simulation).}
    \label{scenario_2}
\end{figure}

\newpage
\subsection{Application: FingerMovements}
\label{FM}
In this section, we are interested in a supervised binary classification problem for  FingerMovements\footnote{\url{https://www.timeseriesclassification.com/description.php?Dataset=FingerMovements}} dataset. These data come from the brain-computer interface domain and are used for binary classification as a benchmark.
More precisely, a subject has been asked to type characters using only the index and the pinky fingers of the right ($Y=0$) and the left ($Y=1$) hands. The challenge is to determine, based on their electroencephalography (EEG) recording ($\bf X$), the hand that has been used. The EEG signal is recorded {during $500$ ms} by $p=28$ sensors located on the scalp. Thus, for any subject, $p=28$ curves are available. Each curve is summarized by $50$ equidistant times points in the interval $[0, 500ms]$. Figure \ref{fig0} (see the Introduction section) presents the curves registered by a sample of 6 sensors (named F1,F2, F3, F4,O1,O2), for a given subject.
The dataset comprises $N=416$ subjects and is split into a training set of $n=316$ units and a test set of $100$ units. 
\par This dataset has been used in \cite{great}. The authors showed that the Inception Time (IT) model \citep{inception} provides the best predictions among the state-of-the-art models. 

\par In this section we compare the results obtained by our methodologies (FU and GFUL) with the competitors, i.e. GL1, GL2, and IT.  The relationship between linear discriminant analysis and regression enables our methods, GL1 and GL2 to perform binary classification. The latter is based on a convenient re-coding of the response variable (see \cite{moindjie2024classification} for more details). Using this relationship, instead of logistic regression, is intended to allow the application of the post-inference method proposed in \cite{post_inference} in the latent group lasso problems.  
 
\par 
For the estimation of FU and GFUL methods, two distances are used:  the Euclidean ($d_e(., .)$) and the great circle distance ($d_c(.,.)$). The distance $d_e(., .)$ is computed using the spatial location of sensors in the 3-D space $\mathcal{C}_j \in \mathbb{R}^3$, $j=1,\ldots, 28$. For the calculation of $d_c(., .)$, which is the shortest distance between two points on the surface of a sphere, we used the sensors' spherical coordinates and assumed that the patient's head could be considered as a sphere. 

\par  The two computed FU methods (FU$_\text{Euclidean}$ and FU$_\text{Circle}$) are based on the 1-NN graph built with respectively $d_e(., .)$ and $d_c(.,.)$ distances. For the GFUL methods (GFUL$_\text{Euclidean}$ and GFUL$_\text{Circle}$), the dissimilarity matrices obtained using the Euclidean distance and the great circle distance are used for clustering sensors based on their locations. For $d_e(.,.)$, $K=10$ clusters of conditions were obtained from the k-means clustering algorithm applied to sensor locations. The $K=10$ groups correspond to well-defined scalp regions (group $1$ = frontal left, group $2$ = frontal right, etc). We use an agglomerative hierarchical clustering approach with average linkage for clustering based on $d_c(.,.)$. Using the majority rule, we determine $K=5$ clusters. Figure \ref{fig_8} presents the groups and 1-NN graphs associated with $d_e(.,)$ and $d_c(.,.)$. 
\begin{figure}[H]
    \centering
\resizebox{\textwidth}{!}{ 
\begin{tabular}{c |c |c }
 \textbf{Conditions} & \textbf{1-NN} & \textbf{Groups} \\ 
    \includegraphics[scale=.5, align=c ]{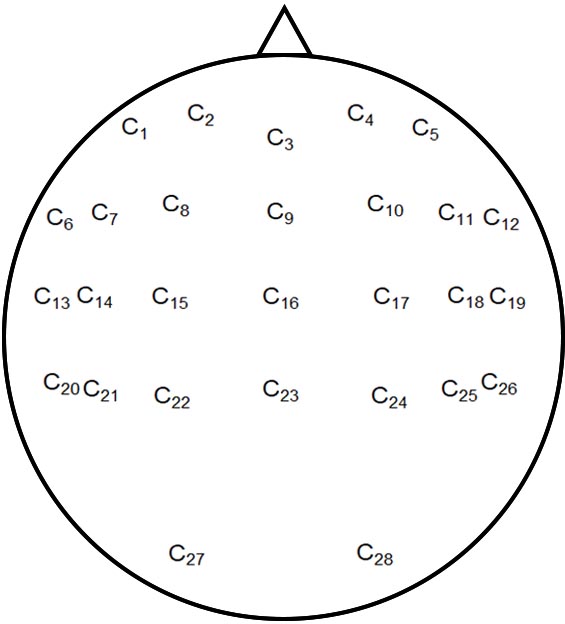} & \begin{tabular}{c c}
        \includegraphics[scale=.5, align=c]{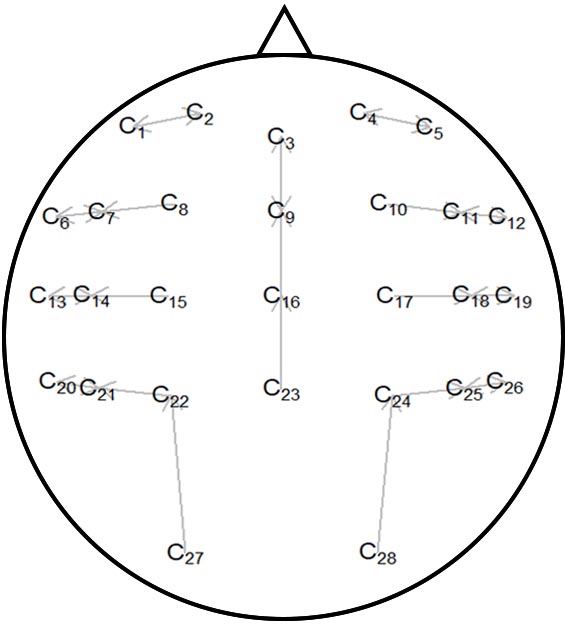} & \includegraphics[scale=.5, align=c ]{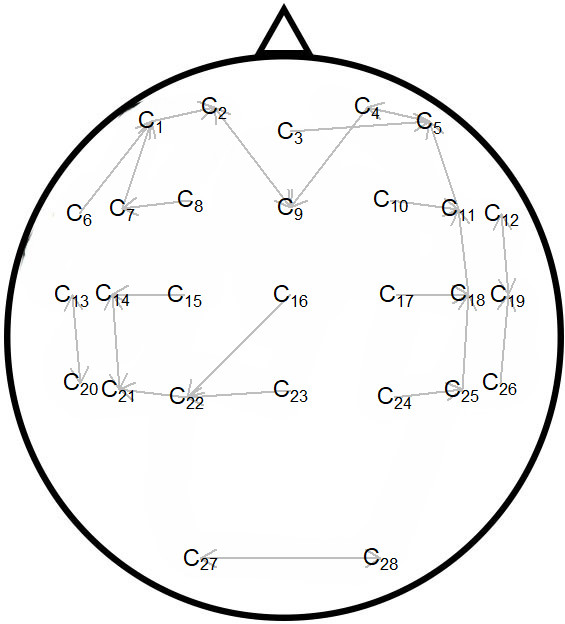} \\ 
        Euclidean distance ($d_e$) & Great circle distance ($d_c$)
    \end{tabular} & \begin{tabular}{c c}
        \includegraphics[scale=.5, align=c ]{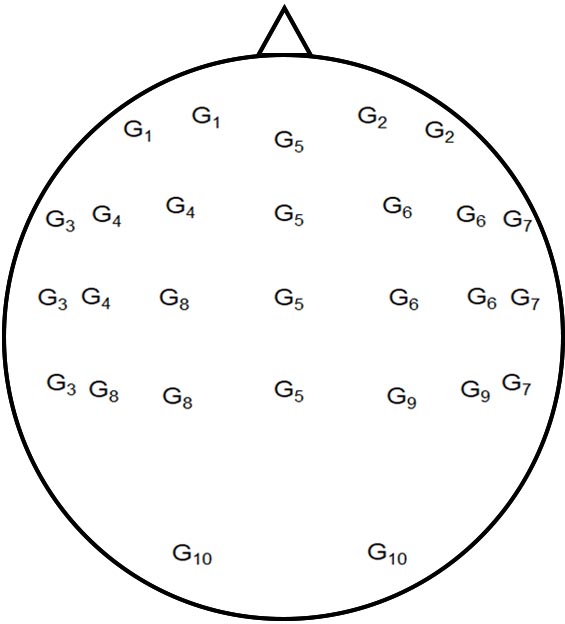}   &  \includegraphics[scale=.5, align=c ]{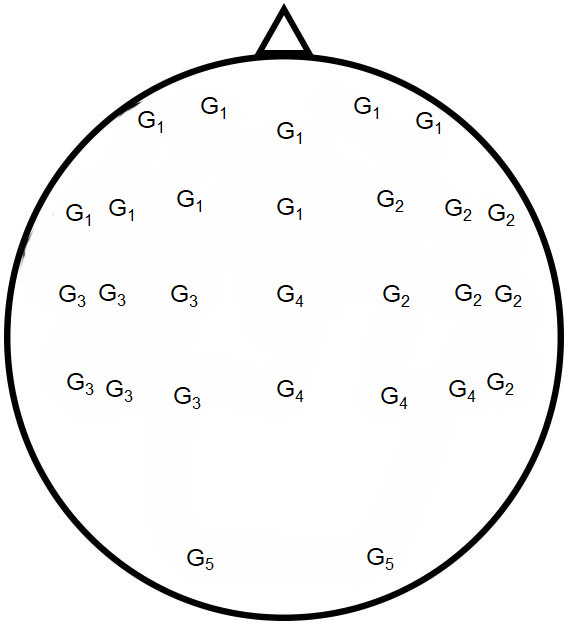} \\ 
        Euclidean distance ($d_e$) & Great circle distance ($d_c$)
        
    \end{tabular}  \\
\end{tabular}
}
\caption{Conditions and groups for the FingerMovements Dataset}
\label{fig_8}
\end{figure}

Three group lasso models are also fitted: GL1$_\text{Euclidean}$, GL1$_\text{Circle}$ and GL2. Similarly to the simulation study, the GL1 method uses each dimension as a group whereas GL1$_\text{Euclidean}$ uses the same grouping structure as GFUL$_\text{Euclidean}$ and GL1$_\text{Circle}$ uses the same grouping structure as GFUL$_\text{Circle}$. \par  
 For all dimensions $X^{(j)}$, $j=1, \ldots, 28$, a basis of $M = 30$ B-splines is used to reconstruct the functional form of the predictors. The hyperparameters $\lambda$ and $\alpha$ are tuned by a 10-fold cross-validation procedure, on the following grids $$\lambda \in \left \{0.96^{i}\lambda_{\max}, i = 0,1, \ldots, 148\right\} \cup \left\{0\right\}$$ and $$\alpha \in \{0, 0.1,0.2, \ldots, 1\},$$
  where $\lambda_{\max}$ is the minimum value such that the penalty term vanishes ($\mathcal{P}(\hat{\beta}_{\lambda, \alpha})= 0$).  
\subsection{Results}
\begin{table}[H]
\centering
\begin{tabular}{l c  }
     \textbf{Methods}& \textbf{Accuracy}  \\ \hline 
    $\textbf{FU}_\text{Euclidean}$ &62\% \\ 
    $\textbf{GFUL}_\text{Euclidean}$ & 68\% \\ 
    $\textbf{FU}_\text{Circle}$ & 59\% \\ 
     $\textbf{GFUL}_\text{Circle}$ & 68\%\\ \hdashline
    \textbf{IT} & 56.7\% \\ \hdashline
    \textbf{GL1} & 63\% 
    \\ 
    $\textbf{GL2}_\text{Euclidean}$ &  62\% \\ 
    $\textbf{GL2}_\text{Circle}$ &60\% \\ 
\end{tabular}
\caption{Accuracy obtained on the test dataset}
\label{res_FM}
\end{table}

Table \ref{res_FM} shows that our proposed methodologies are competitive or perform better than most competitors (GL2 and IT) in terms of accuracy (well-classified rate) estimated on the test sample. { In particular, the GFUL models perform best}. \\ Figure \ref{fig_FM} shows the grouping structure of the estimated regression coefficient functions obtained with FU and GFUL. Hence, those results provide information about the importance of sensors and their locations (also through the grouping structure) for predicting the response.\\\  
 Table \ref{p_inf} presents the p-values associated with post-inference tests for the GFUL methods. {Note that for GFUL$_\text{Euclidean}$ the p-value of Group 2 is missing as dimensions belonging to this group were identified as sharing the same coefficient}. \par {Although they were computed using the methodology proposed in \cite{post_inference} for group lasso problems, the null hypotheses are not the same in the GFUL case (see Remark \ref{remak_post})}. These p-values give valuable information on the groups' pertinence. For example, Table \ref{p_inf} shows that the first three groups in GFUL$_\text{Circle}$, and {Group 5 and Group 6} for GFUL$_\text{Euclidean}$, have significant (with 5\% of level significance) {different coefficients}. This means that conditions among these groups shouldn't be considered {as equivalent contributors} in the regression model.  \par  
 For readability purposes, the associated p-values for FU, GL1, GL2, and all estimated coefficient functions are presented in the appendix \ref{fig_A}. 
\begin{figure}[H]
\centering
\begin{tabular}{c c c} 
$\textbf{FU}_\text{Euclidean}$ & $\textbf{GFUL}_\text{Euclidean}$ & $\textbf{FU}_\text{Circle}$ \\
\includegraphics[scale=.5]{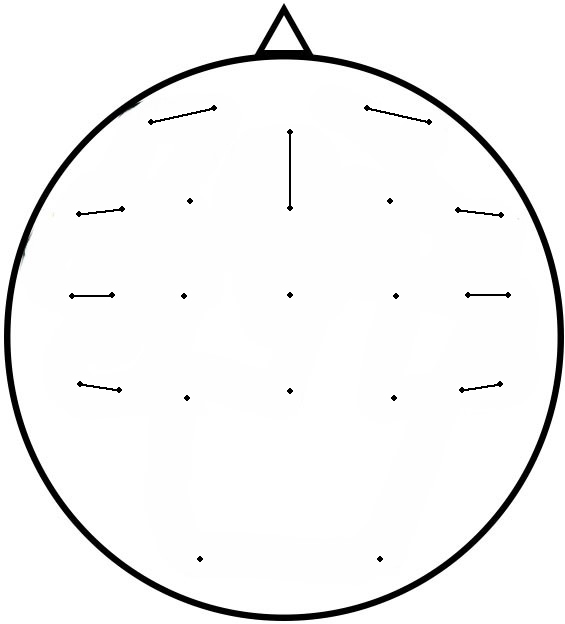} &   \includegraphics[scale=.5]{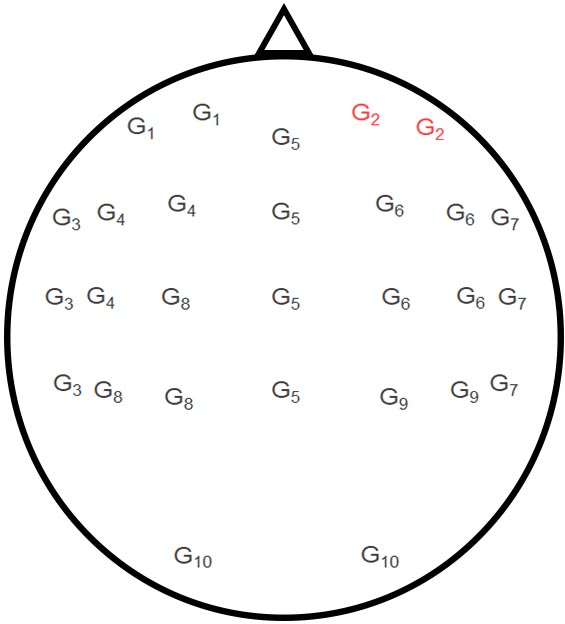} & 
\includegraphics[scale=.5]{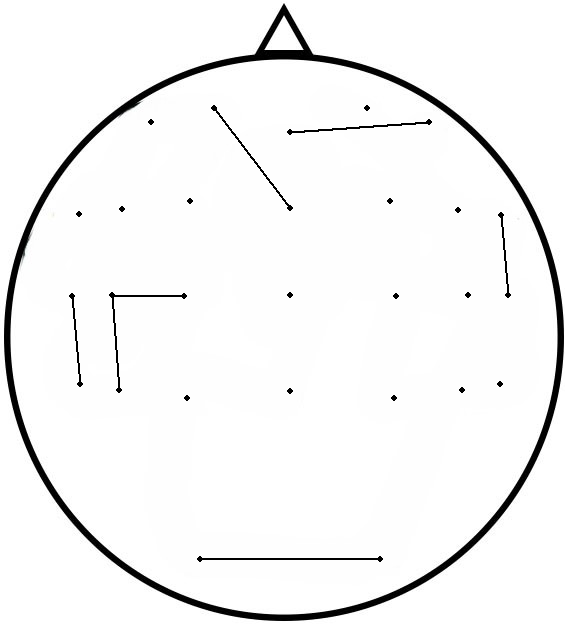 }
 \end{tabular}

 \caption{Estimated structures. \textbf{FU}: connected points share the same coefficients, \textbf{GFUL}: groups in red share the same coefficients. The estimated structure of GFUL$_\text{Circle}$  is not displayed as the selected model by this method didn't identify groups with the same coefficient values.
}
 \label{fig_FM}
 \end{figure}
 
\begin{table}[H]
\centering
\begin{tabular}{c c c}
    &\bf GFUL$_\text{Euclidean}$ & \bf  GFUL$_\text{Circle}$ \\
    \textbf{Group 1}& 0.9968 &0.0000  \\ 
    \textbf{Group 2}& --- & 0.0003 \\ 
    \textbf{Group 3}& 0.3085 & 0.0001 \\ 
    \textbf{Group 4}&0.2205  &0.8789\\ 
    \textbf{Group 5}&0.0070  & 1.0000 \\ 
    \textbf{Group 6}& 0.0024  & ---\\ 
    \textbf{Group 7}&0.1715 & ---\\ 
    \textbf{Group 8}& 0.3454 & ---\\ 
    \textbf{Group 9} &  0.9376 & ---\\ 
    \textbf{Group 10} & 0.9887 &  ---
\end{tabular}
\caption{P-values for GFUL methods}
\label{p_inf}
\end{table}

\section{Discussion}
\label{dis}
In this paper, we introduced two new criteria for estimating a linear regression model with the predictor represented by a functional random variable observed under different conditions (eventually spatially distributed). We called that data repeated functional {data}. When some grouping {or neighborhood} structure of conditions is present, our methods can integrate it into the fitting process through specific penalties: fusion and group fusion-lasso. 

The numerical simulation study, as well as the application to Finger movements data, confirm the efficacy of { approaches integrating grouping structures of conditions.} { Our hypothesis that close sensors might bring similar information helps to obtain competitive models, especially in the Finger movements dataset as our proposed methodologies give similar results or outperform the lasso method competitors.}\par 
The GFUL method can be seen as a generalization of {FU} to more than one neighbor. {It relies on the assumption of $K$ known groups. When this is not the case, we suggest to use clustering algorithms (K-Means, Gaussian Mixture Models, etc.) to obtain meaningful clusters of conditions.\par Notice that FU only requires neighborhood structure in $\mathcal{S}$ and this is suitable for the case where pairs of closest conditions might have the same associated contribution in the linear model. In the presence of group structure among conditions,  GFUL performs better than FU.}
{ Unlike FU, GFUL tests group membership simultaneously instead of testing one-on-one interactions.} This is quite a strong hypothesis, as it assumes that equality relations (among regression coefficient functions) in a group can be either all true or all false. The use of smaller overlaps between groups could be an alternative model. In this setting, the solution is related to the group lasso with overlap, which is more challenging \citep{yuan2011_2}. An extensive study of adapted optimization problems should be done. One can also explore the model group lasso proposed in \cite{jacob2009}. Yet, it seems that using this approach leads to losing the diffusion between overlapped groups, the penalty is no longer defined on the $2,1$ norm (See \cite{jacob2009} for details).   
The integration of sparsity conditions and the study of other types of neighborhood structures in the fusion method can be some future promising developments. 
\subsection*{Acknowledgement}
I-A Moindjié thanks Dr.Dewez and Dr.Grimonprez (DiagRAMS Technologies) for the interesting discussions.  
\begin{appendices}
\section{Additional figures: FingerMovements}
 \label{fig_A}

\begin{figure}[H]
    \centering
    \resizebox*{!}{0.8\textheight}{%
    \begin{tabular}{c c c c  c}
    & 
     $\hat{\beta}^{(1)}=\hat{\beta}^{(2)}$&$\hat{\beta}^{(2)}=\hat{\beta}^{(1)}$ &$\hat{\beta}^{(3)}=\hat{\beta}^{(9)}$&  $\hat{\beta}^{(4)}=\hat{\beta}^{(5)}$ \\ 
    & \includegraphics[scale=.09, align=c]{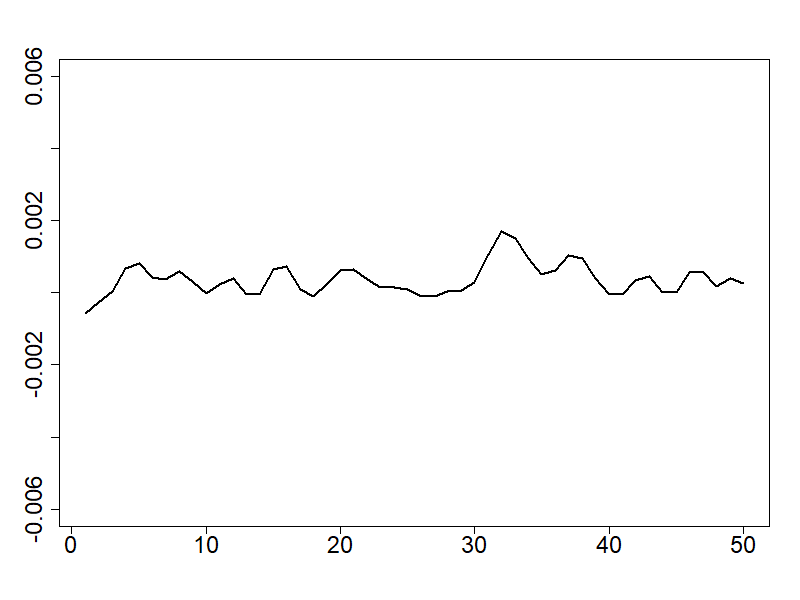}  &  \includegraphics[scale=.09, align=c]{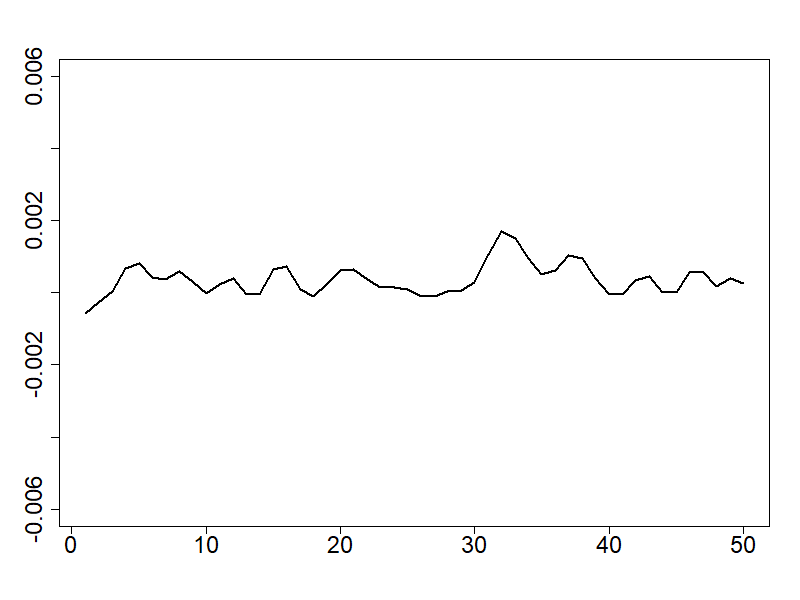} & \includegraphics[scale=.09, align=c]{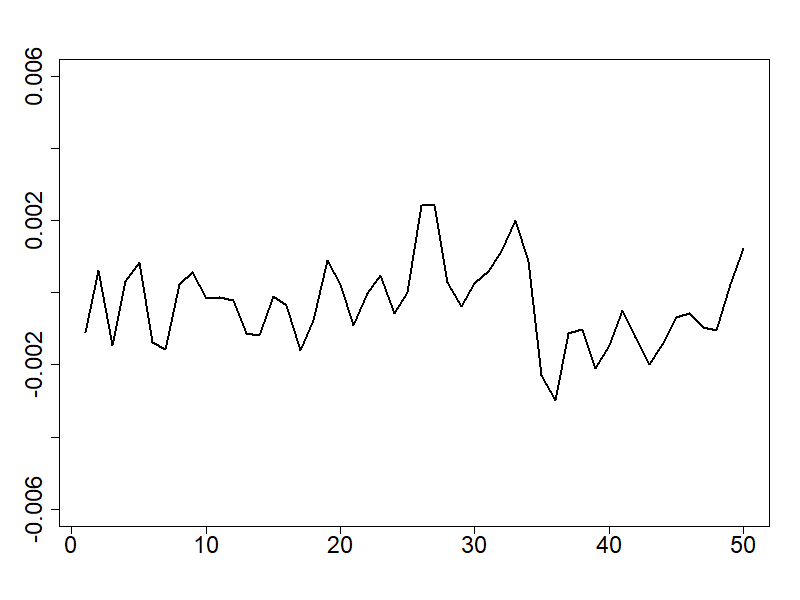} & \includegraphics[scale=.09, align=c]{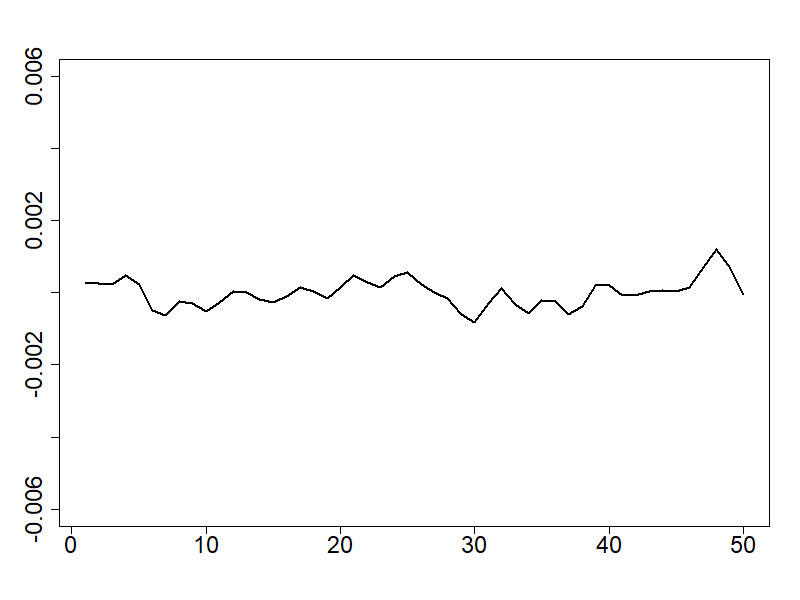} \\ 
     & $\hat{\beta}^{(5)}=\hat{\beta}^{(4)}$&$\hat{\beta}^{(6)}=\hat{\beta}^{(7)}$ &$\hat{\beta}^{(7)}=\hat{\beta}^{(6)}$& \begin{tabular}{c}
          $\hat{\beta}^{(8)}$\\
          $H_0: \beta^{(8)}=\beta^{(7)}$ \\
          (p=0.997)
     \end{tabular}  \\ 
    &\includegraphics[scale=.09, align=c]{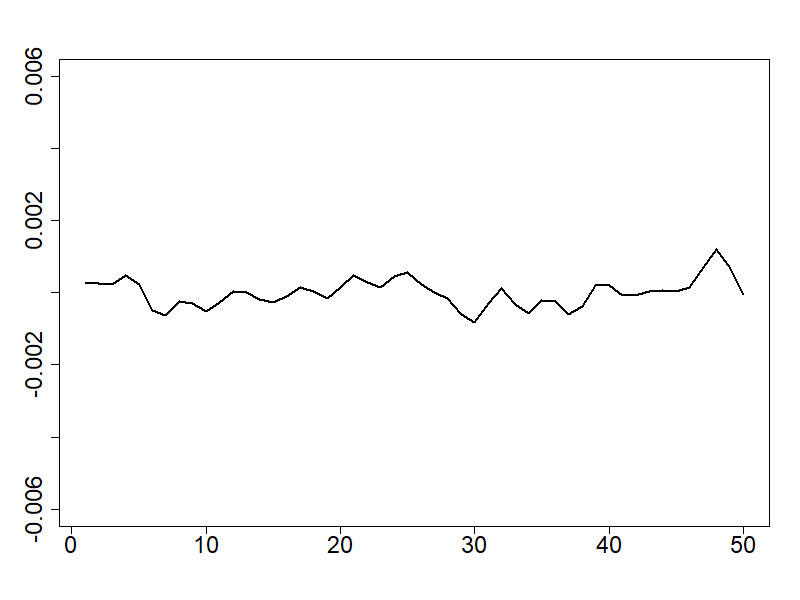} &  \includegraphics[scale=.09, align=c]{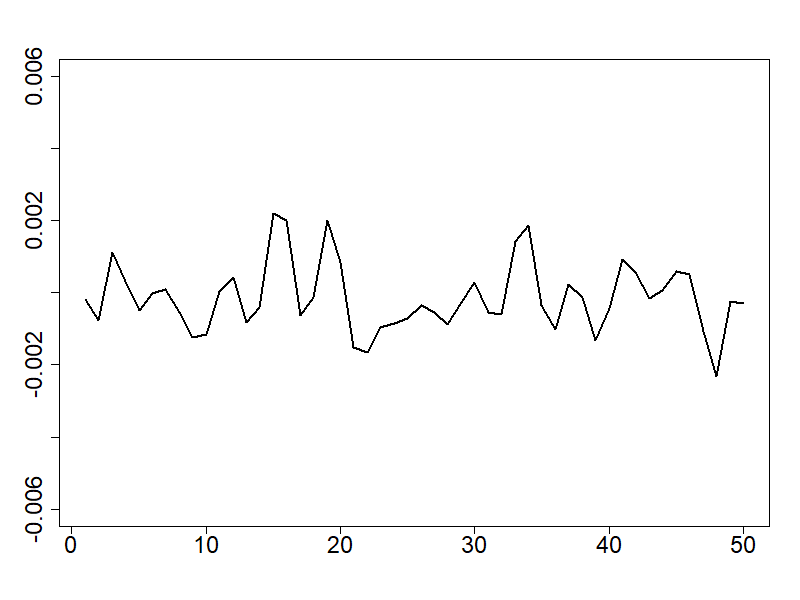} &  \includegraphics[scale=.09, align=c]{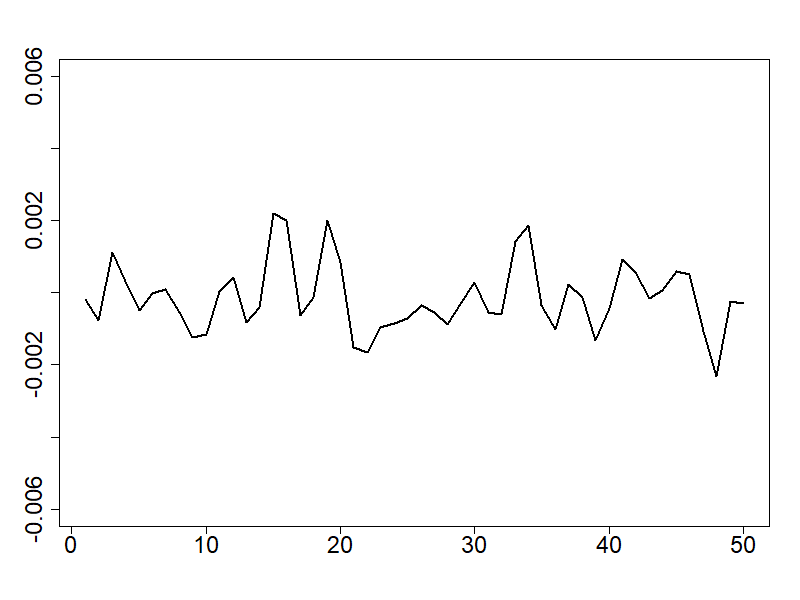} &  \includegraphics[scale=.09, align=c]{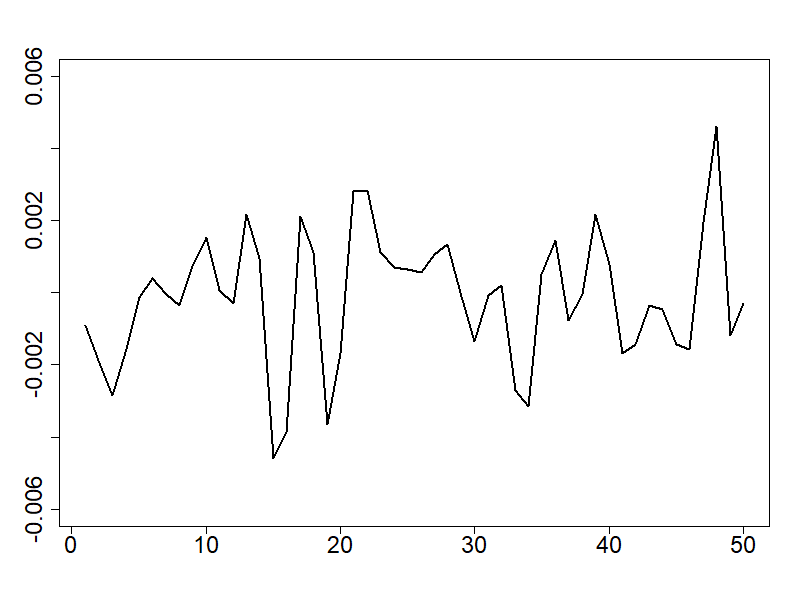} \\ 
      &  $\hat{\beta}^{(9)}=\hat{\beta}^{(3)}$&\begin{tabular}{c}
          $\hat{\beta}^{(10)}$   \\
           $H_0: \beta^{(10)}=\beta^{(11)}$ \\ $\small (p=1)$
      \end{tabular}&$\hat{\beta}^{(11)}=\hat{\beta}^{(12)}$&  $\hat{\beta}^{(12)}=\hat{\beta}^{(11)}$\\ 
     & \includegraphics[scale=.09, align=c]{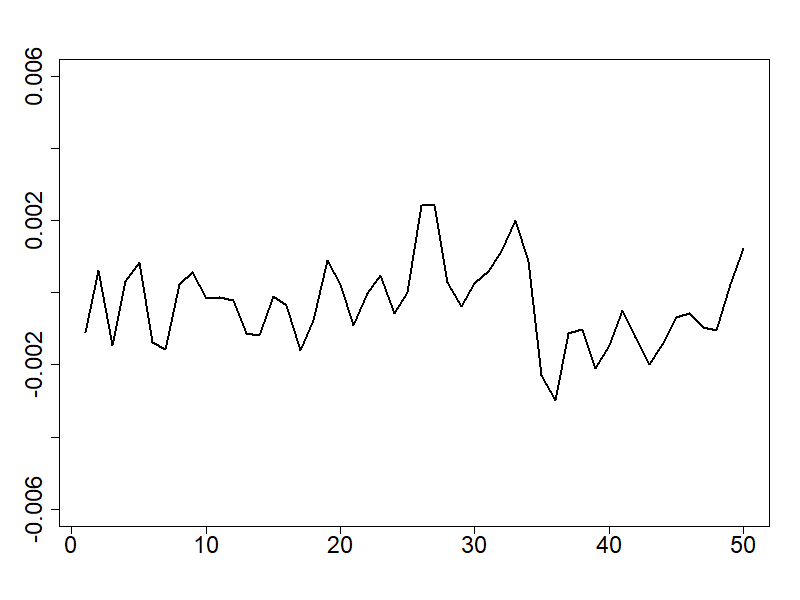} & \includegraphics[scale=.09, align=c]{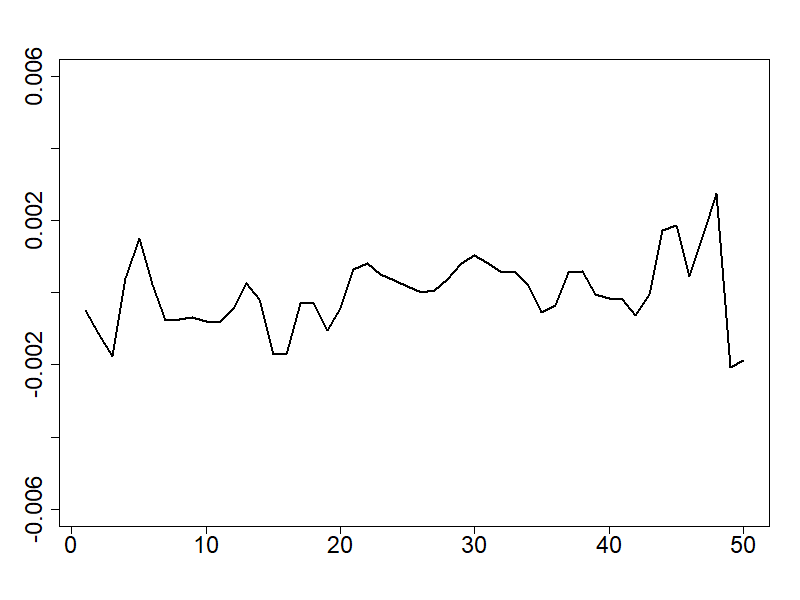} & 
    \includegraphics[scale=.09, align=c]{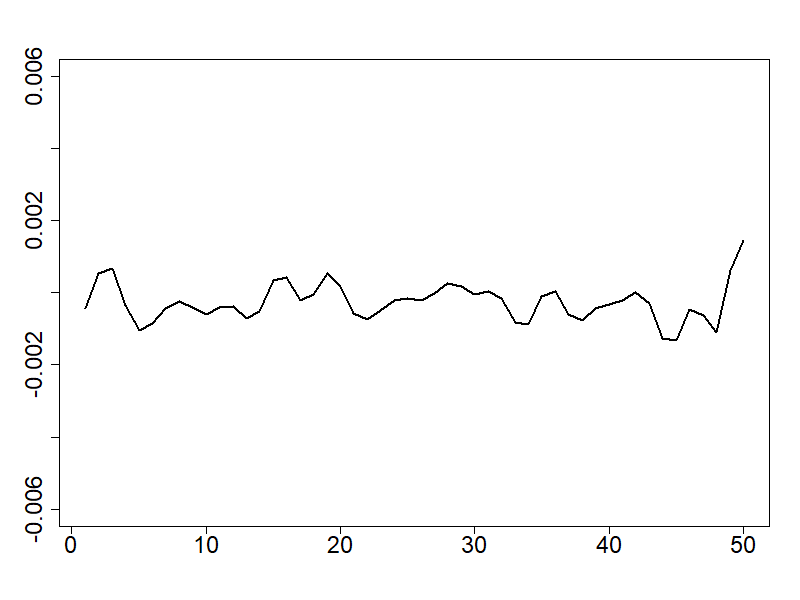}& \includegraphics[scale=.09, align=c]{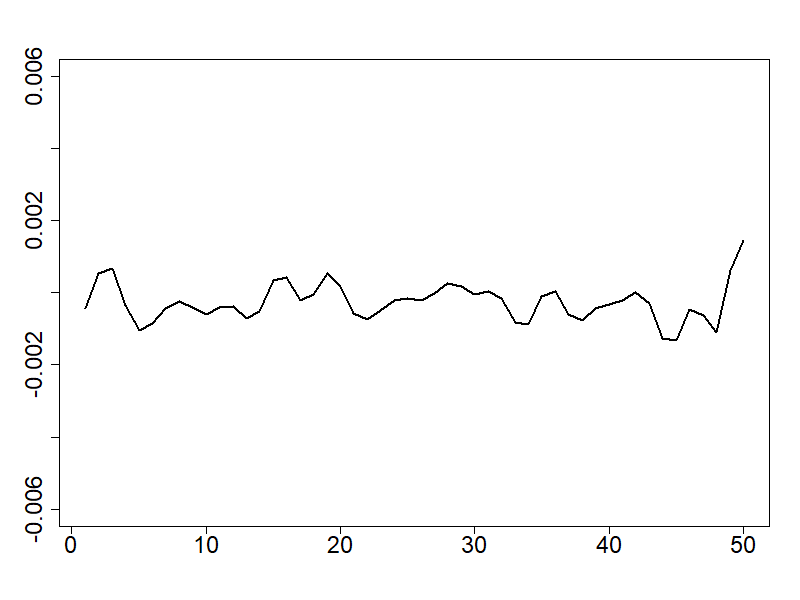} \\
     &  $\hat{\beta}^{(13)}=\hat{\beta}^{(14)}$ & $\hat{\beta}^{(14)}=\hat{\beta}^{(13)} $ & \begin{tabular}{c}
          $\hat{\beta}^{(15)}$  \\
          $H_0: \beta^{(15)}=\beta^{(14)}$\\ $\small (p=0.998)$ 
     \end{tabular} & \begin{tabular}{c}
          $\hat{\beta}^{(16)}$  \\
          $H_0: \beta^{(16)}=\beta^{(9)}$\\$\small (p=0.995)$  
     \end{tabular}  \\
    & \includegraphics[scale=.09, align=c]{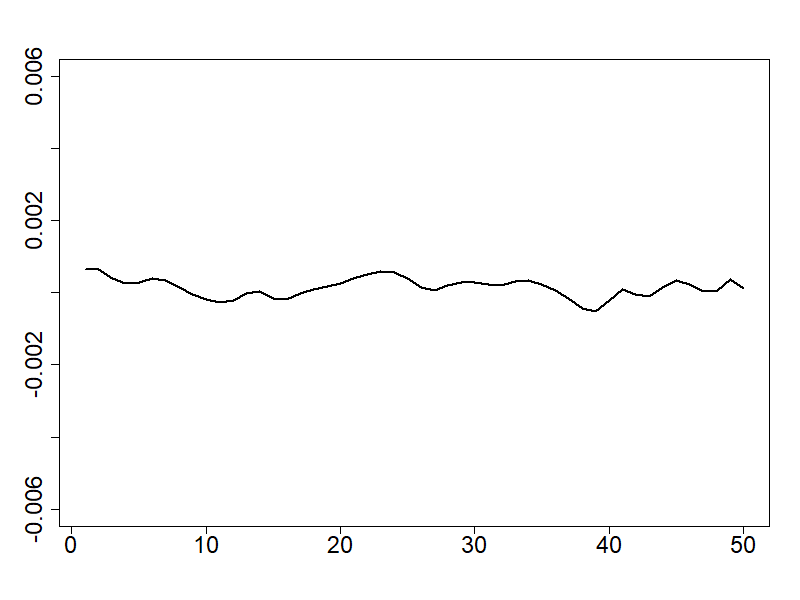} & \includegraphics[scale=.09, align=c]{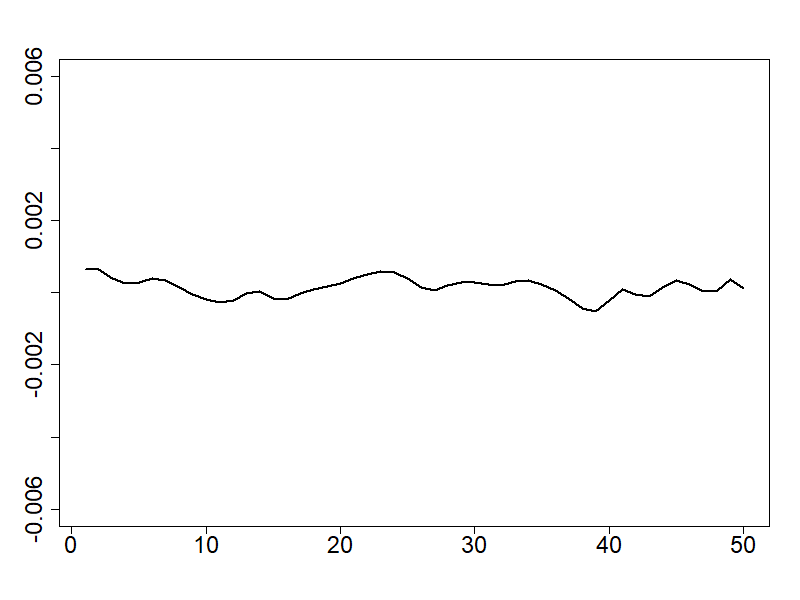} &    \includegraphics[scale=.09, align=c]{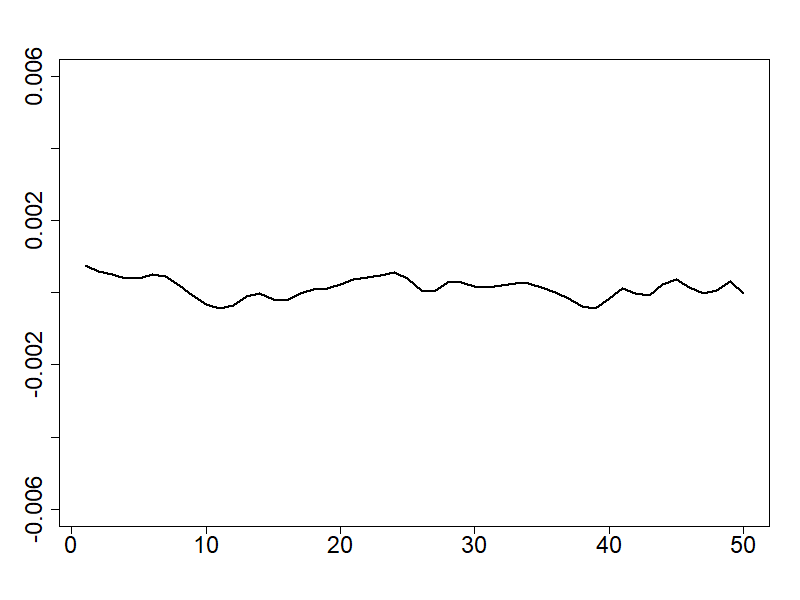} & \includegraphics[scale=.09, align=c]{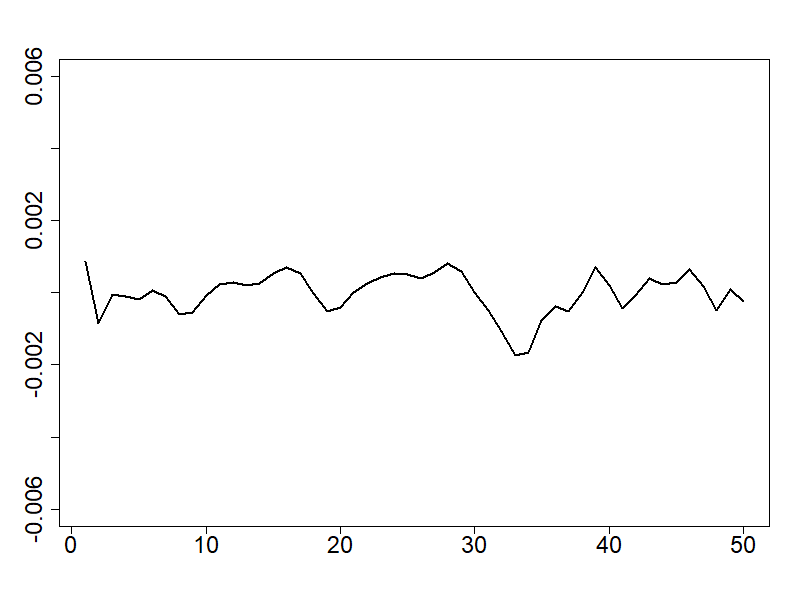} \\
     &  $\hat{\beta}^{(17)}$ & $\hat{\beta}^{(18)}=\hat{\beta}^{(19)} $ & $\hat{\beta}^{(19)}=\hat{\beta}^{(18)}$ & $\hat{\beta}^{(20)}=\hat{\beta}^{(21)}$  \\
    & \includegraphics[scale=.09, align=c]{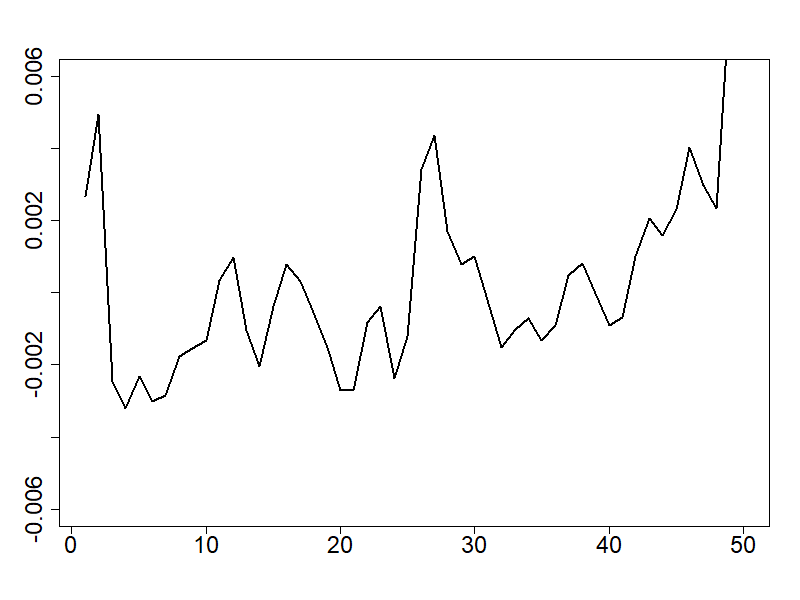} & \includegraphics[scale=.09, align=c]{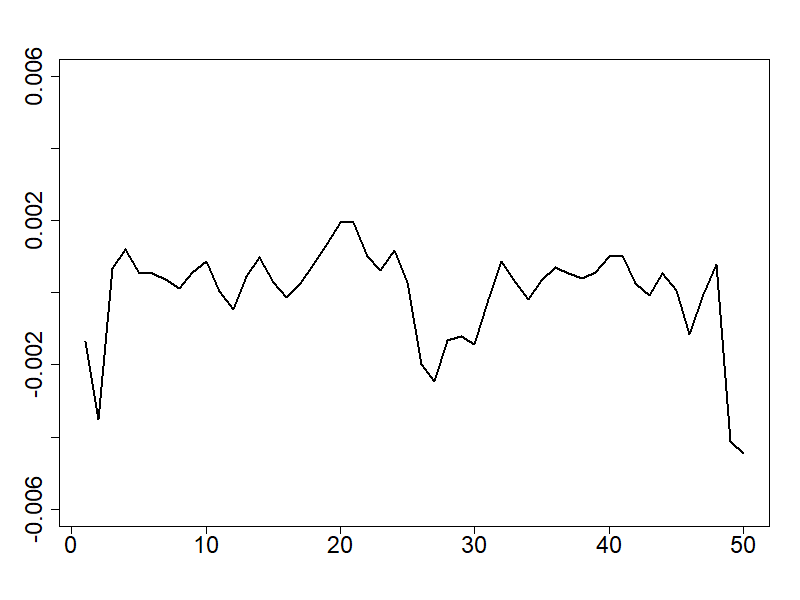} & \includegraphics[scale=.09, align=c]{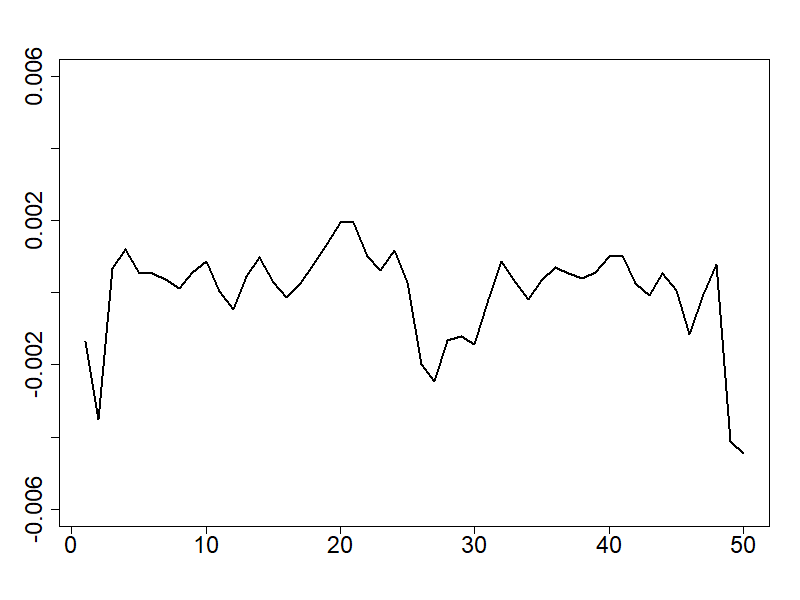} & \includegraphics[scale=.09, align=c]{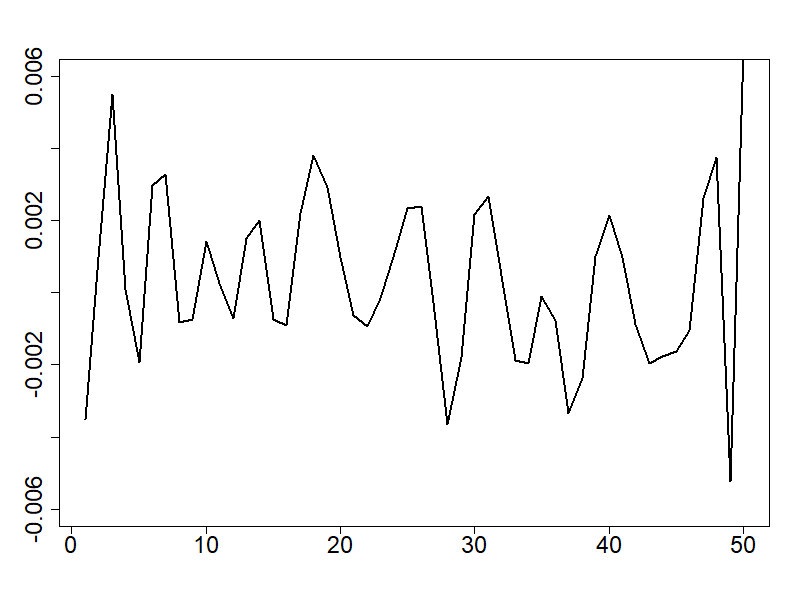} \\ 
 &$\hat{\beta}^{(21)}=\hat{\beta}^{(20)}$ &\begin{tabular}{c}
      $\hat{\beta}^{(22)} $ \\
      $H_0: \beta^{(22)}=\beta^{(21)}$ \\$\small (p=0.996)$  
 \end{tabular} & \begin{tabular}{c}
      $\hat{\beta}^{(23)}$ \\
      $H_0: \beta^{(23)}=\beta^{(16)}$ \\$\small (p=0.999)$  
 \end{tabular} & \begin{tabular}{c}
      $\hat{\beta}^{(24)}$  \\
      $H_0: \beta^{(24)}= \beta^{(25)}$ \\$\small (p=1)$
 \end{tabular}  \\
    
    & \includegraphics[scale=.09, align=c]{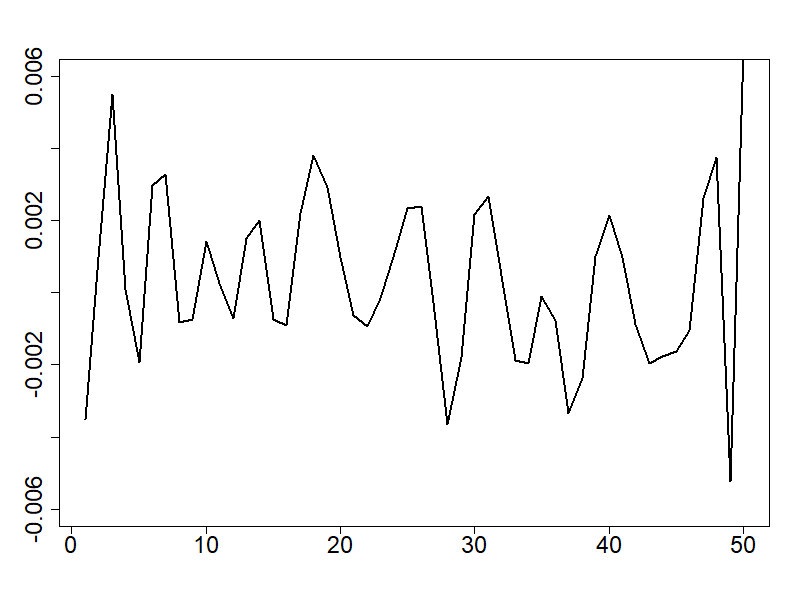} & \includegraphics[scale=.09, align=c]{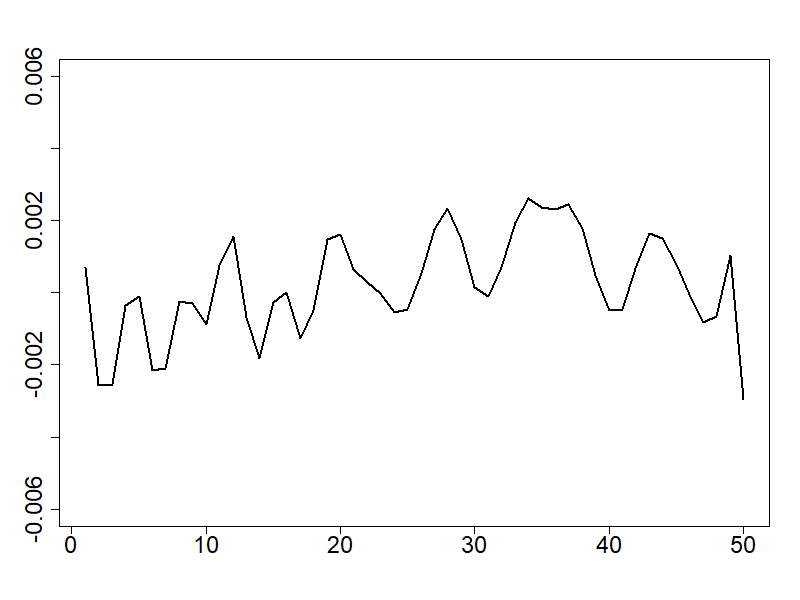} & \includegraphics[scale=.09, align=c]{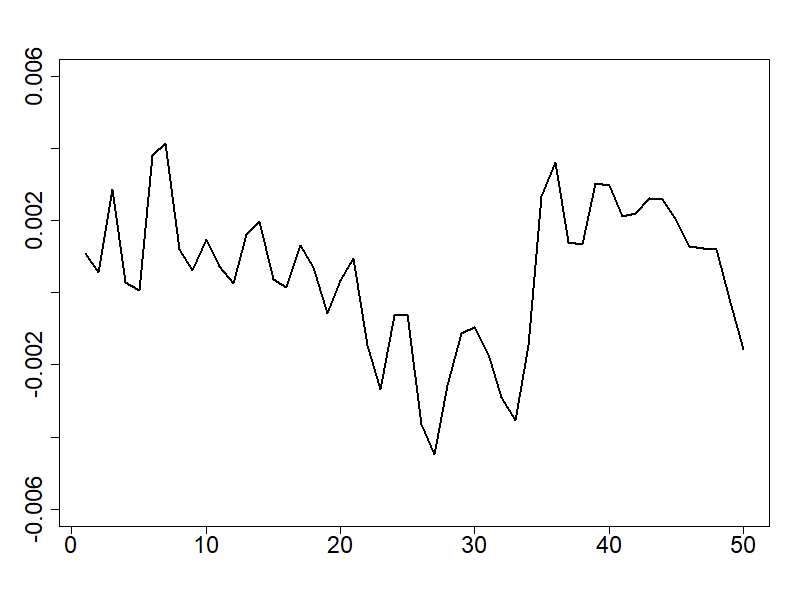} & \includegraphics[scale=.09, align=c]{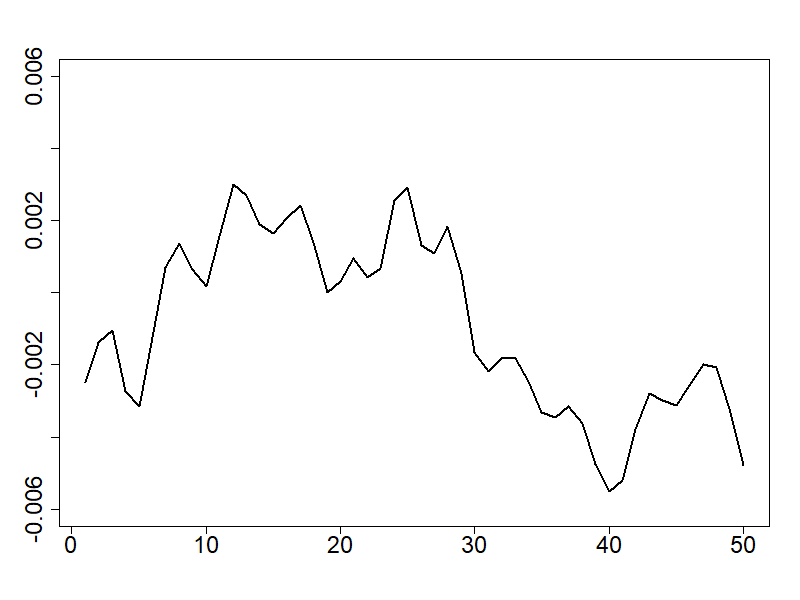} \\ 
&$\hat{\beta}^{(25)}=\hat{\beta}^{(26)} $ & $\hat{\beta}^{(26)}=\hat{\beta}^{(25)}$ & \begin{tabular}{c}
     $\hat{\beta}^{(27)}$   \\ 
     $H_0: \beta^{(27)}=\beta^{(22)}$ \\  $\small (p=0.952)$
\end{tabular}& \begin{tabular}{c}
     $\hat{\beta}^{(28)}$  \\
   $H_0: \beta^{(28)}=\beta^{(24)}$ \\ $\small (p=0.979)$ 
\end{tabular}  \\

    & \includegraphics[scale=.09, align=c]{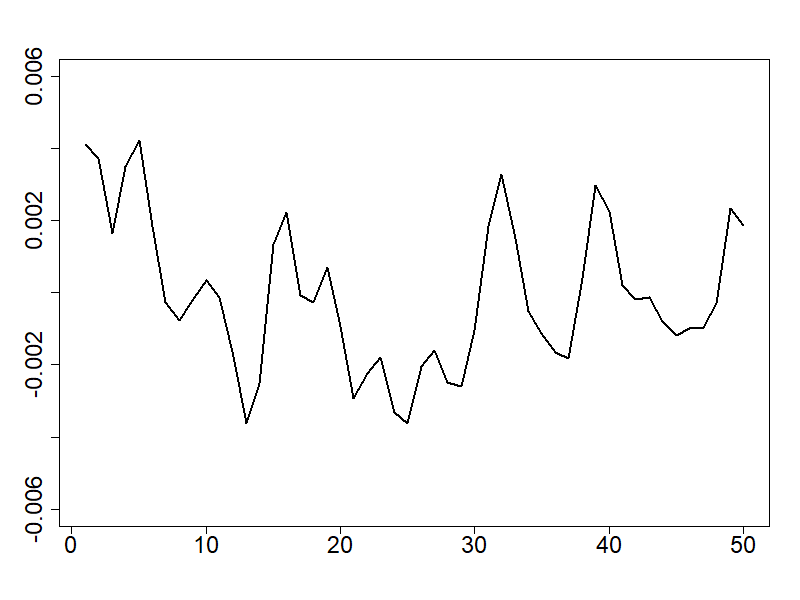} & \includegraphics[scale=.09, align=c]{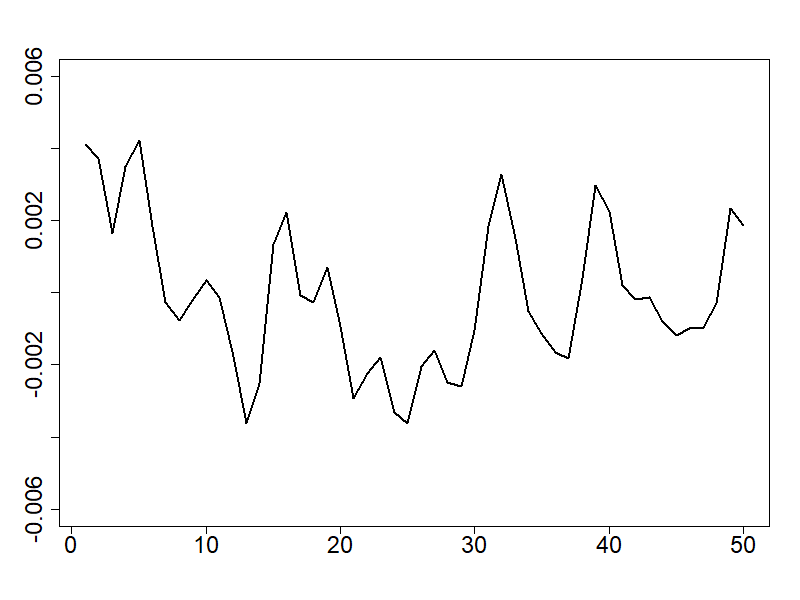} & \includegraphics[scale=.09, align=c]{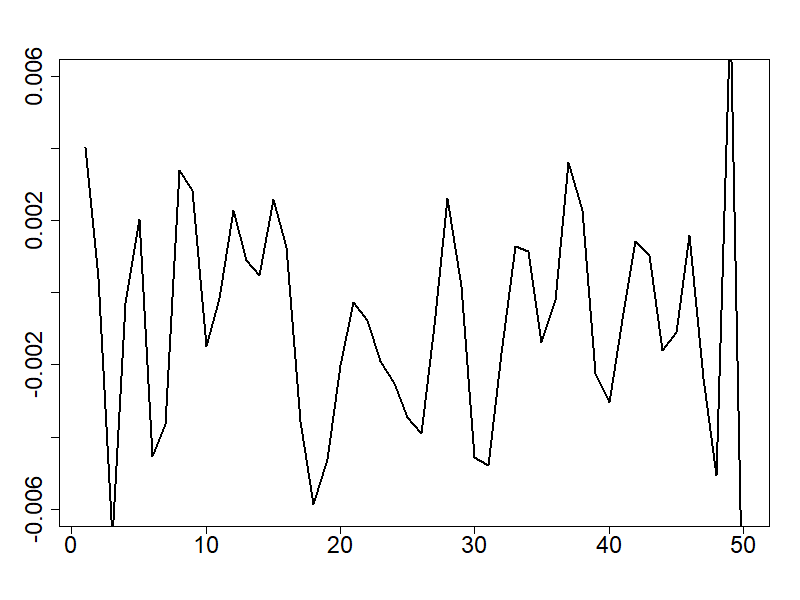} & \includegraphics[scale=.09, align=c]{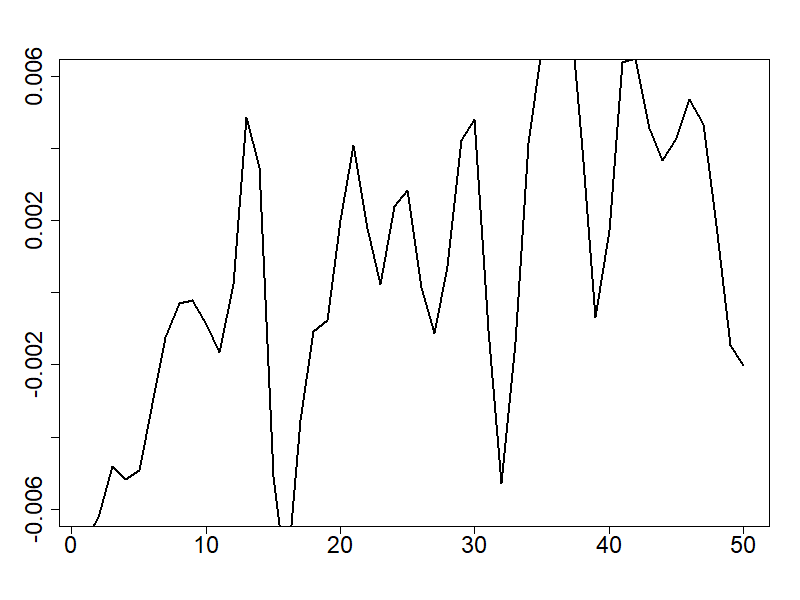}
    
    \end{tabular}
    }
    \caption{FU$_\text{Euclidean}$ estimated coefficients and associated p-values}
    \label{fig:my_label}
\end{figure}

\newpage

\begin{figure}[H]
\centering 
\resizebox*{!}{0.8\textheight}{%
\begin{tabular}{c c c c }
      \begin{tabular}{c c c c  c}
    & 
     \begin{tabular}{c}
          $\hat{\beta}^{(1)}$  \\
           $H_0: \beta^{(1)}=\beta^{(2)}$ \\ 
           (p=0.966)
     \end{tabular}&$\hat{\beta}^{(2)}=\hat{\beta}^{(9)}$ &$\hat{\beta}^{(3)}=\hat{\beta}^{(5)}$&  \begin{tabular}{c}
          $\hat{\beta}^{(4)}$   \\
           $H_0: \beta^{(4)}=\beta^{(9)}$ \\ 
           (p=0.919) 
     \end{tabular}\\ 
    & \includegraphics[scale=.09, align=c]{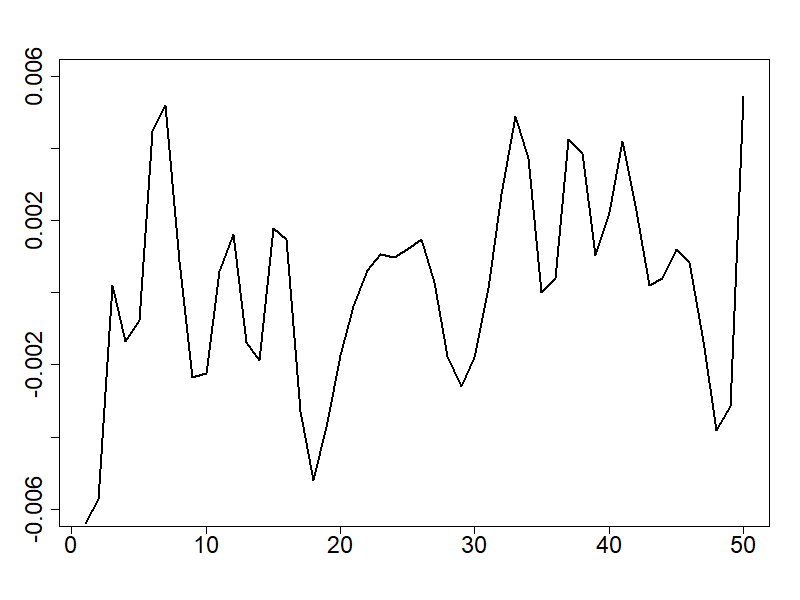}  &  \includegraphics[scale=.09, align=c]{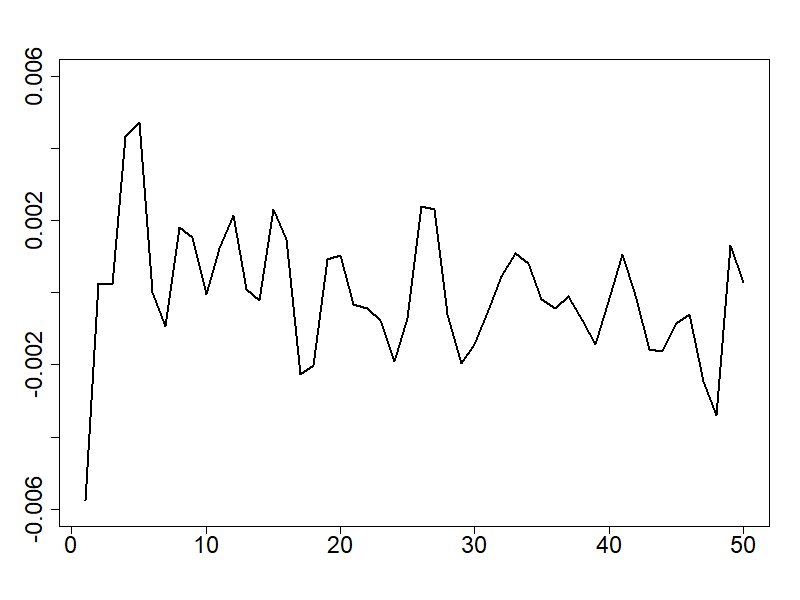} & \includegraphics[scale=.09, align=c]{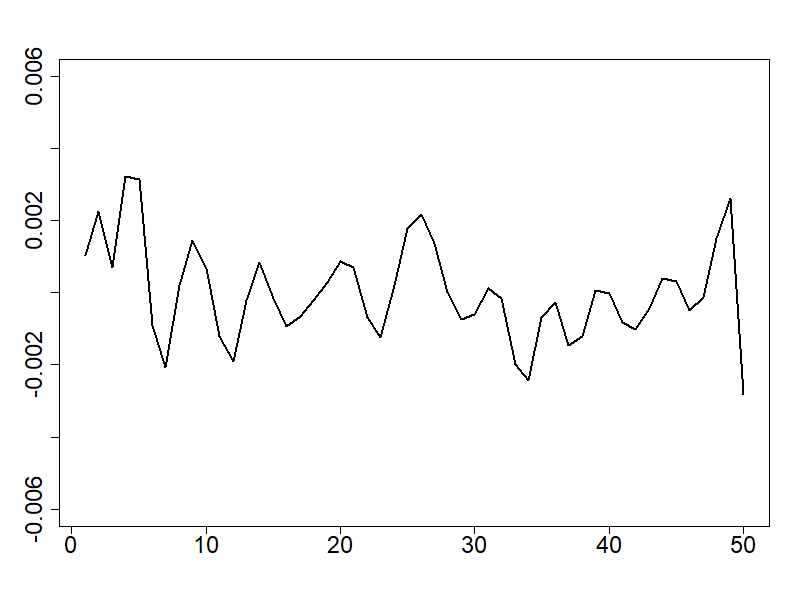} & \includegraphics[scale=.09, align=c]{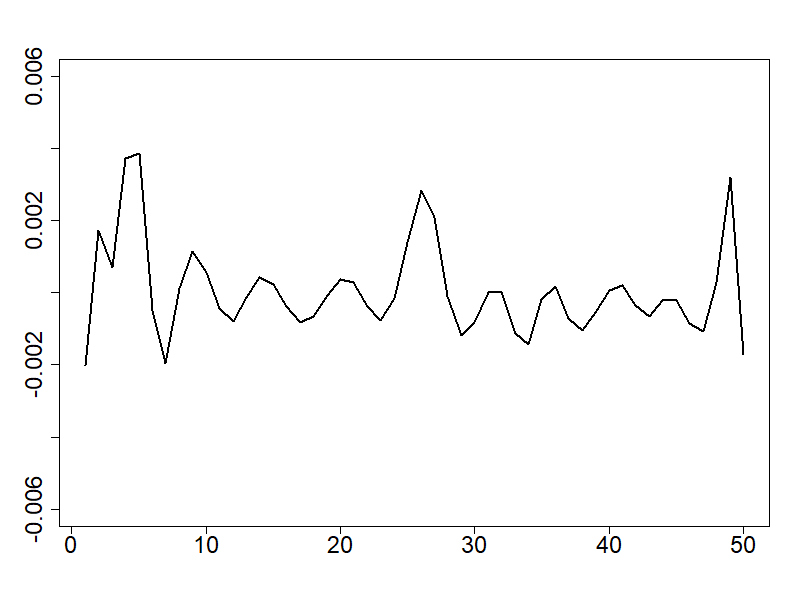} \\ 
     & $\hat{\beta}^{(5)}=\hat{\beta}^{(3)}$&\begin{tabular}{c}$\hat{\beta}^{(6)}$ \\  $H_0: \beta^{(6)}=\beta^{(1)}$ \\ 
          (p=0.672)
     \end{tabular} & \begin{tabular}{c}
     $\hat{\beta}^{(7)}$ \\ 
     $H_0: \beta^{(7)}=\beta^{(1)}$ \\
(p=0.857)
     \end{tabular}
     &  \begin{tabular}{c}
$\hat{\beta}^{(8)}$ \\
      $H_0: \beta^{(8)}=\beta^{(7)}$ \\ (p=0.938)
      \end{tabular}\\ 
    &\includegraphics[scale=.09, align=c]{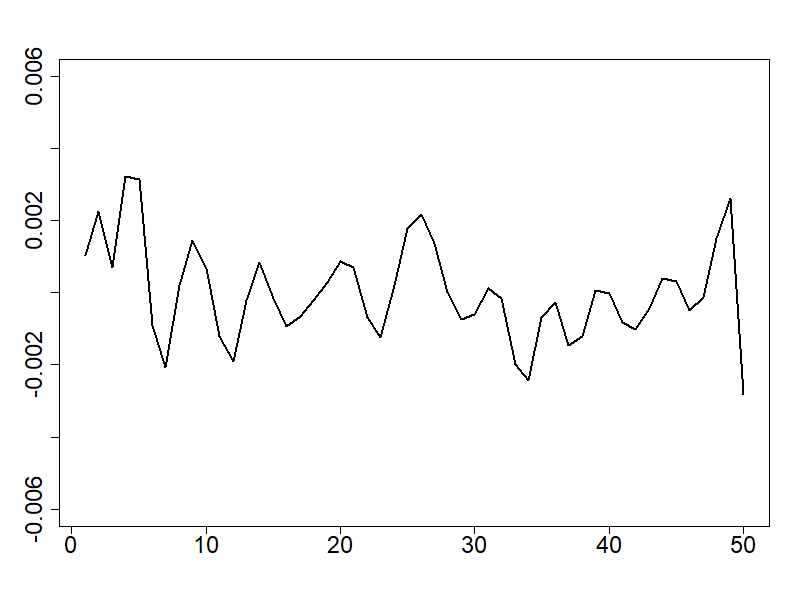} &  \includegraphics[scale=.09, align=c]{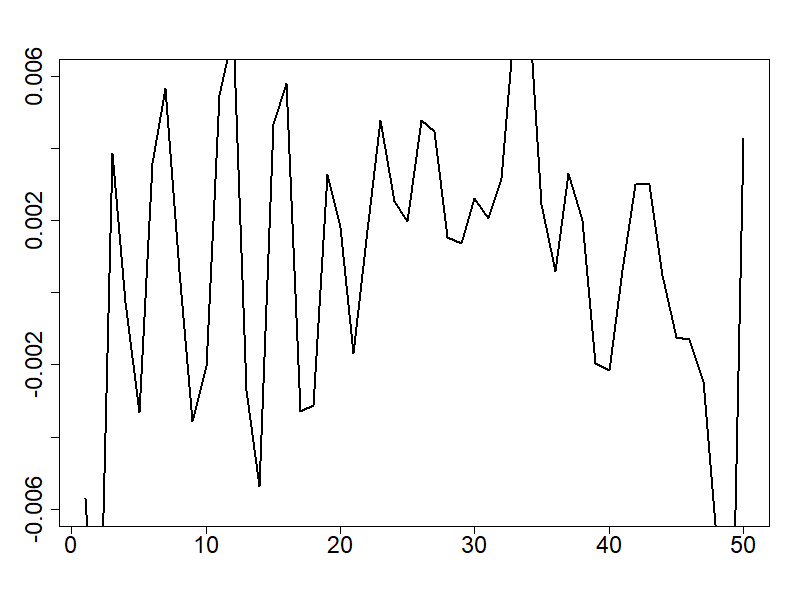} &  \includegraphics[scale=.09, align=c]{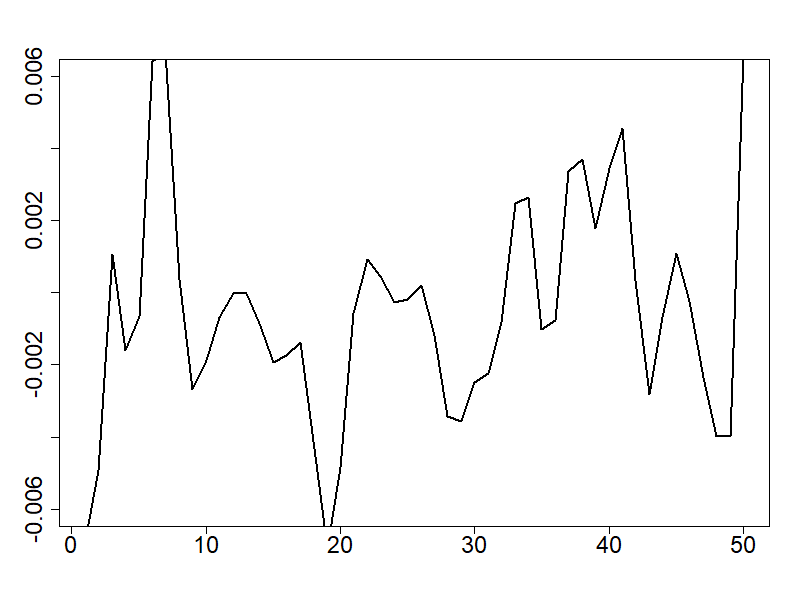} &  \includegraphics[scale=.09, align=c]{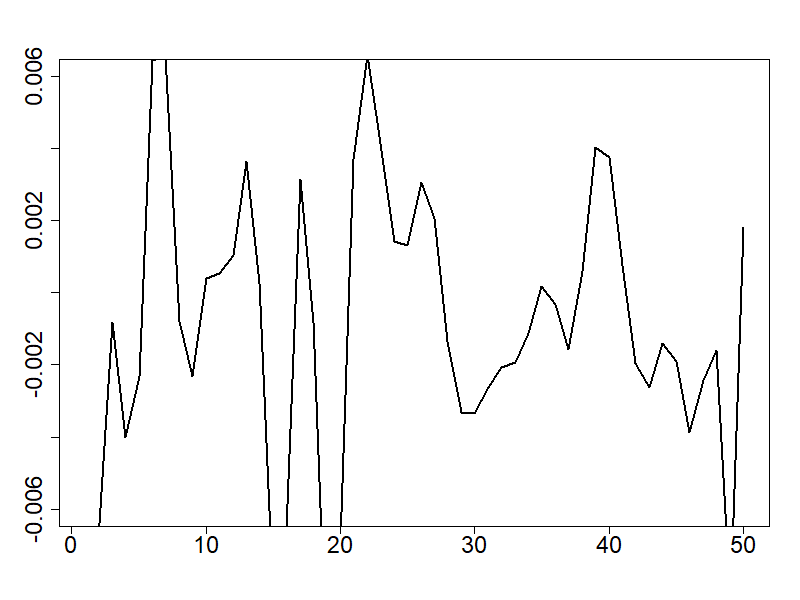} \\ 
      &  $\hat{\beta}^{(9)}=\hat{\beta}^{(2)}$&\begin{tabular}{c}
           $\hat{\beta}^{(10)}$  \\
           $H_0: \beta^{(10)}=\beta^{(11)}$ \\ 
           (p=0.983)
      \end{tabular} &\begin{tabular}{c}
           $\hat{\beta}^{(11)}$ \\
           $H_0: \beta^{(11)}=\beta^{(5)}$\\
           (p=0.980)
      \end{tabular}&  $\hat{\beta}^{(12)}=\hat{\beta}^{(19)}$\\ 
     & \includegraphics[scale=.09, align=c]{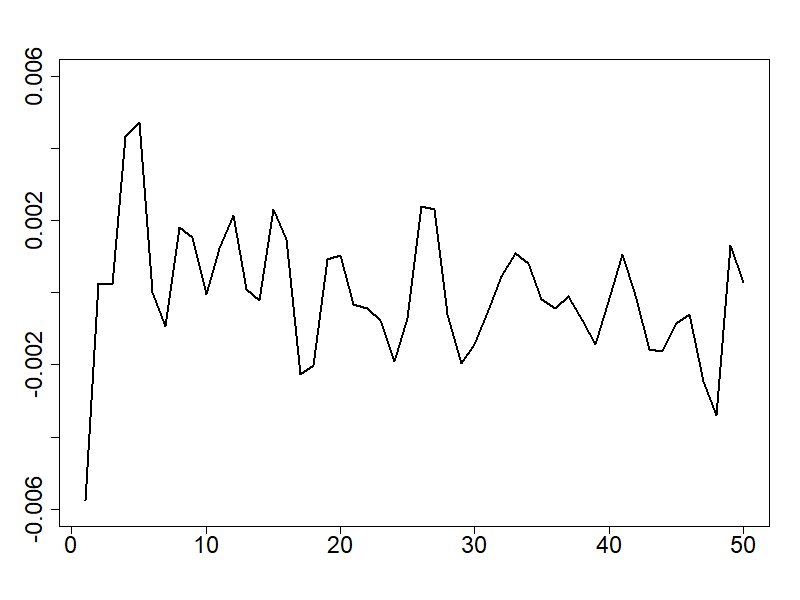} & \includegraphics[scale=.09, align=c]{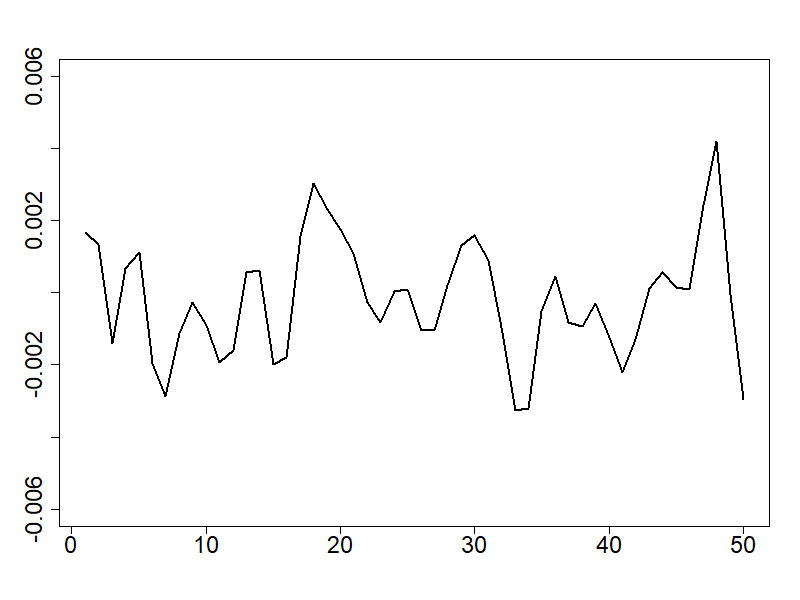} & 
    \includegraphics[scale=.09, align=c]{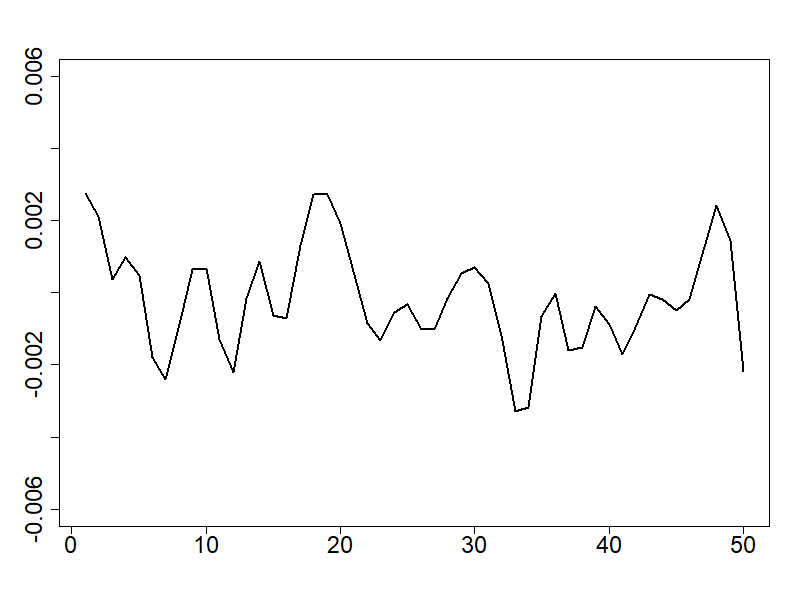}& \includegraphics[scale=.09, align=c]{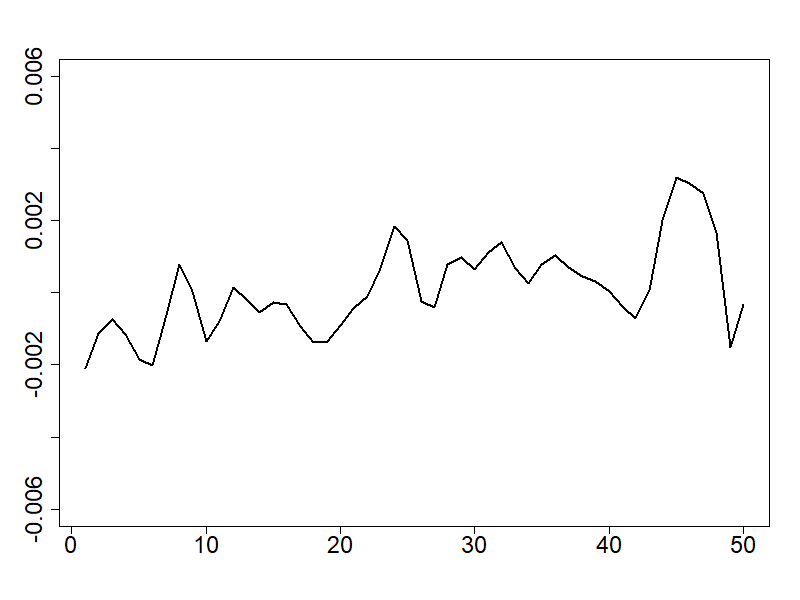} \\
     &  $\hat{\beta}^{(13)}=\hat{\beta}^{(20)}$ & $\hat{\beta}^{(14)}=\hat{\beta}^{(15)}=\hat{\beta}^{(21)} $ & $\hat{\beta}^{(15)}=\hat{\beta}^{(14)}=\hat{\beta}^{(21)}$ & \begin{tabular}{c}
          $\hat{\beta}^{(16)}$  \\
          $H_0 : \beta^{(16)}=\beta^{(22)} $\\ 
          (p=0.832)
     \end{tabular}  \\
    & \includegraphics[scale=.09, align=c]{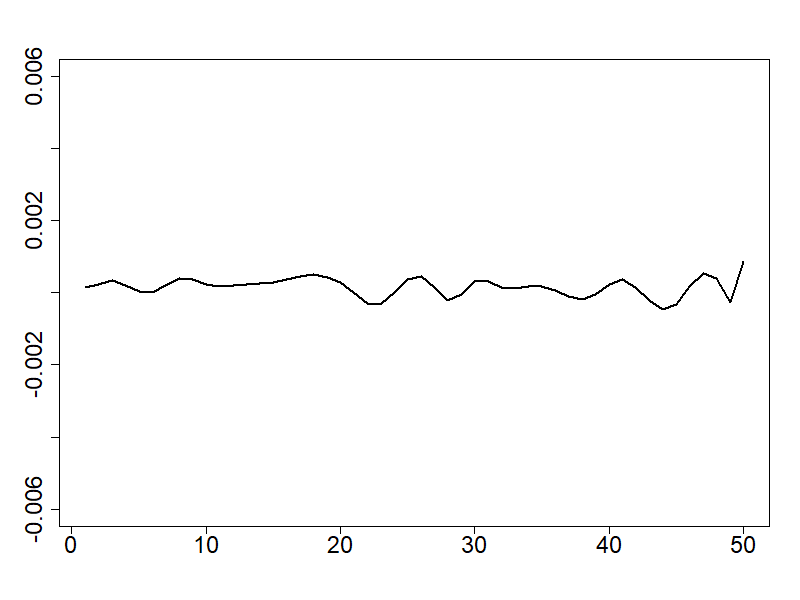} & \includegraphics[scale=.09, align=c]{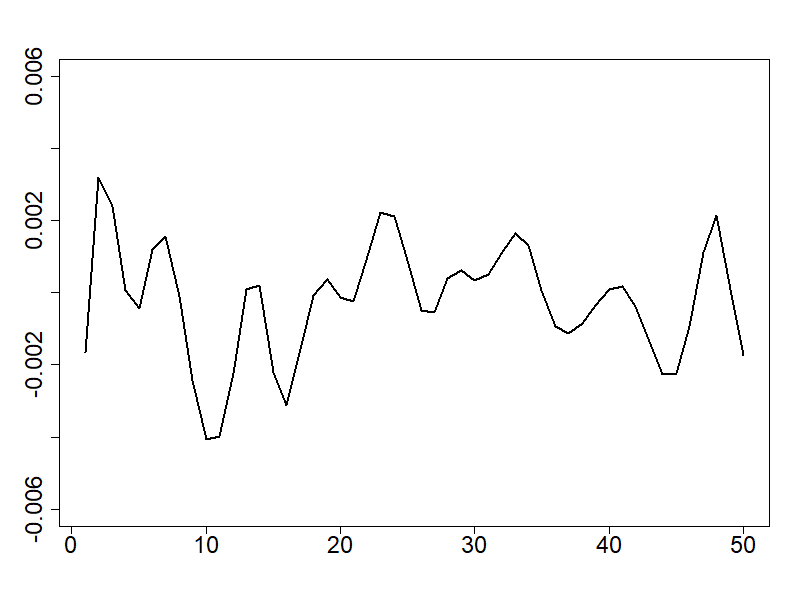} &    \includegraphics[scale=.09, align=c]{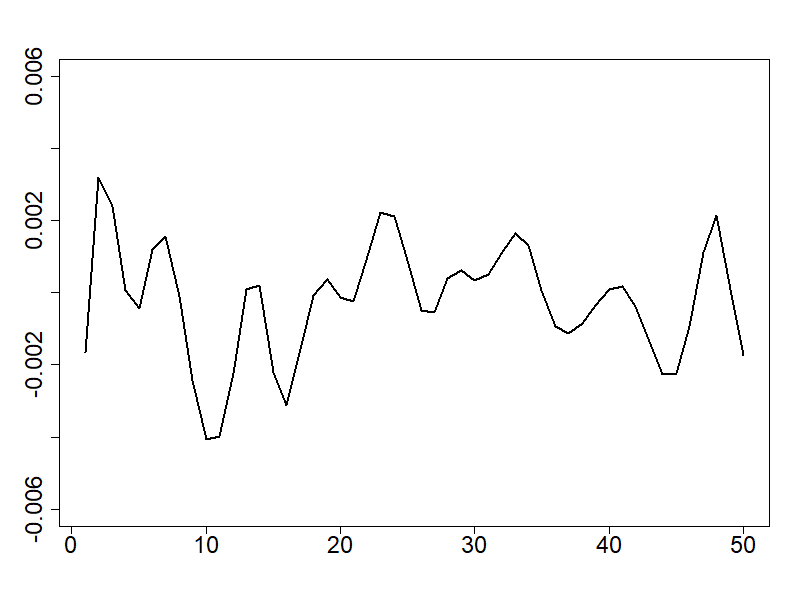} & \includegraphics[scale=.09, align=c]{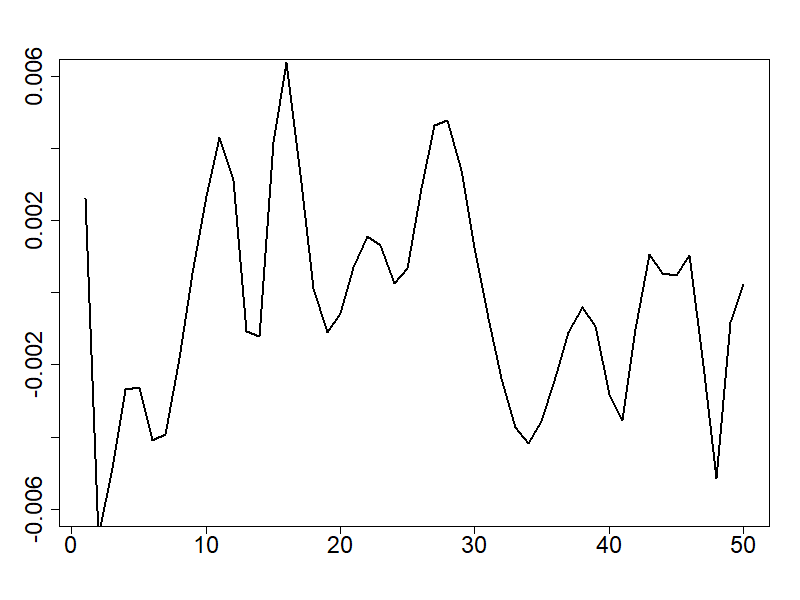} \\
     & \begin{tabular}{c}
          $\hat{\beta}^{(17)}$   \\
          $H_0: \beta^{(17)}=\beta^{(18)} $ \\ 
          (p=0.797)
     \end{tabular} & \begin{tabular}{c }
          $\hat{\beta}^{(18)} $ \\
          $H_0: \beta^{(18)}=\beta^{(11)}$ \\ 
          (p=0.530)
     \end{tabular} & $\hat{\beta}^{(19)}=\hat{\beta}^{(12)}$ & $\hat{\beta}^{(20)}=\hat{\beta}^{(13)}$  \\
    & \includegraphics[scale=.09, align=c]{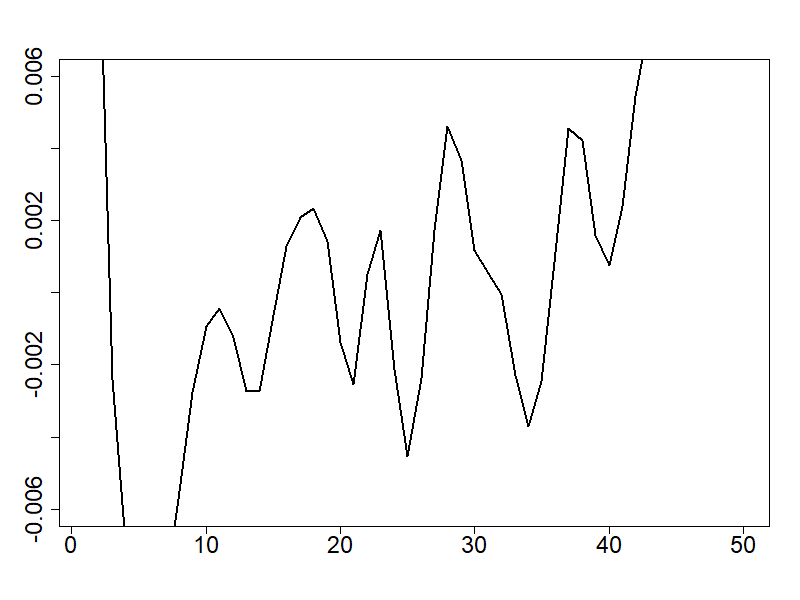} & \includegraphics[scale=.09, align=c]{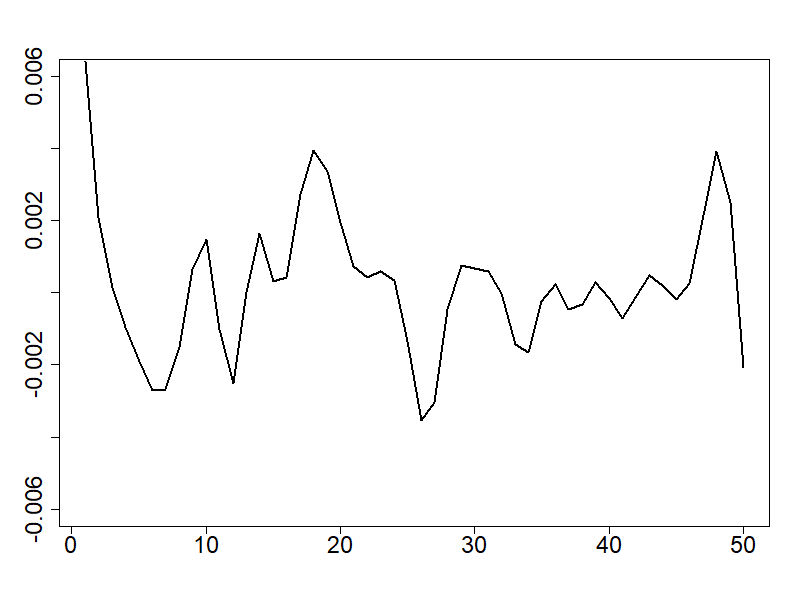} & \includegraphics[scale=.09, align=c]{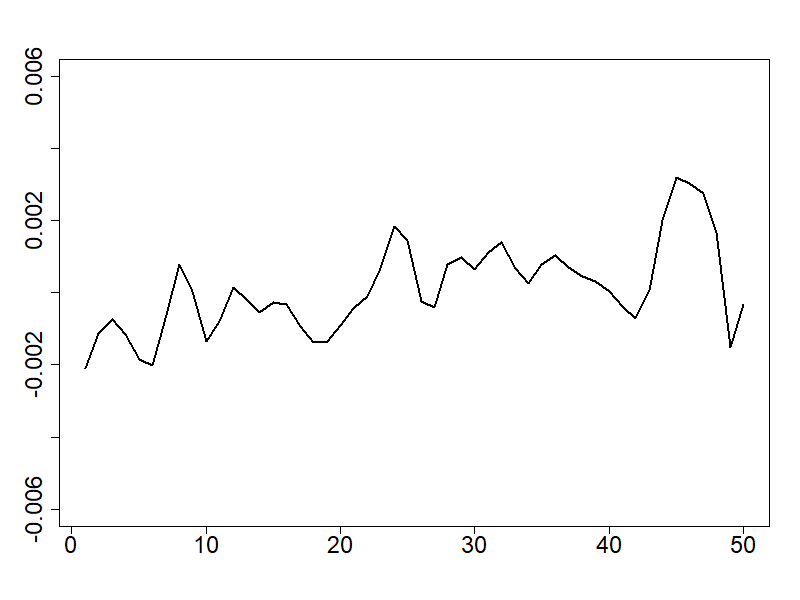} & \includegraphics[scale=.09, align=c]{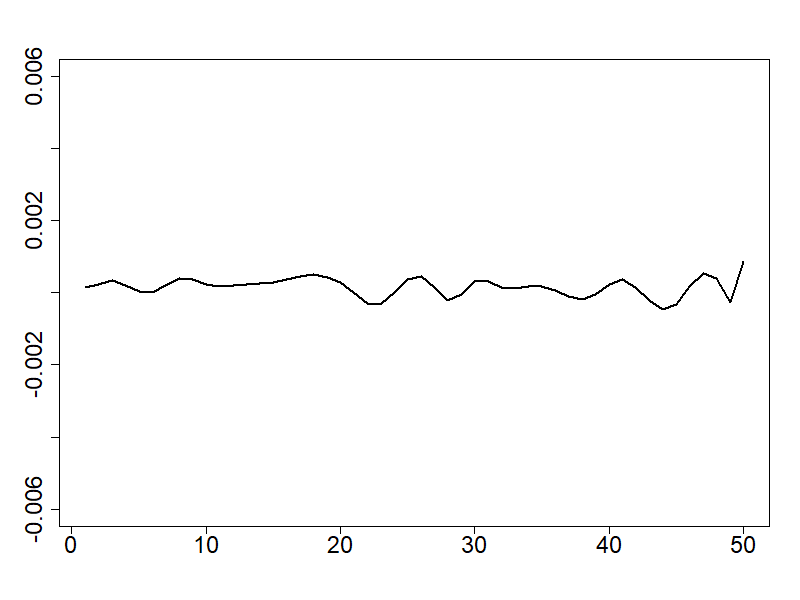} \\ 
 &$\hat{\beta}^{(21)}=\hat{\beta}^{(14)}=\hat{\beta}^{(15)}$ & \begin{tabular}{c}
      $\hat{\beta}^{(22)} $   \\
      $H_0:  \beta^{(22)}= \beta^{(21)}$ \\
      (p=0.644)
 \end{tabular}& \begin{tabular}{c }
     $\hat{\beta}^{(23)}$ \\ 
      $H_0:  \beta^{(23)}= \beta^{(22)}$ \\
      (p=0.848)
 \end{tabular}& \begin{tabular}{c}
     $\hat{\beta}^{(24)}$ \\
      $H_0: \beta^{(24)}=\beta^{(25)}$ \\ 0.838
 \end{tabular}  \\
    
    & \includegraphics[scale=.09, align=c]{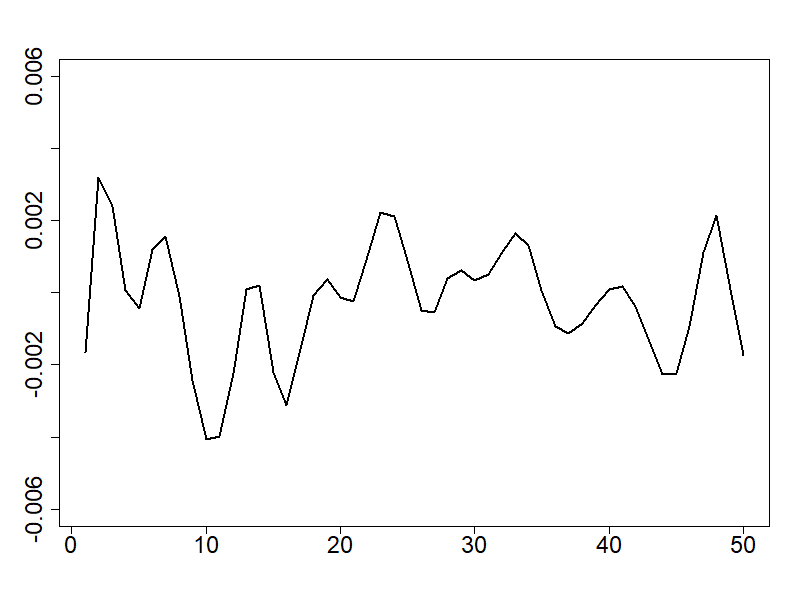} & \includegraphics[scale=.09, align=c]{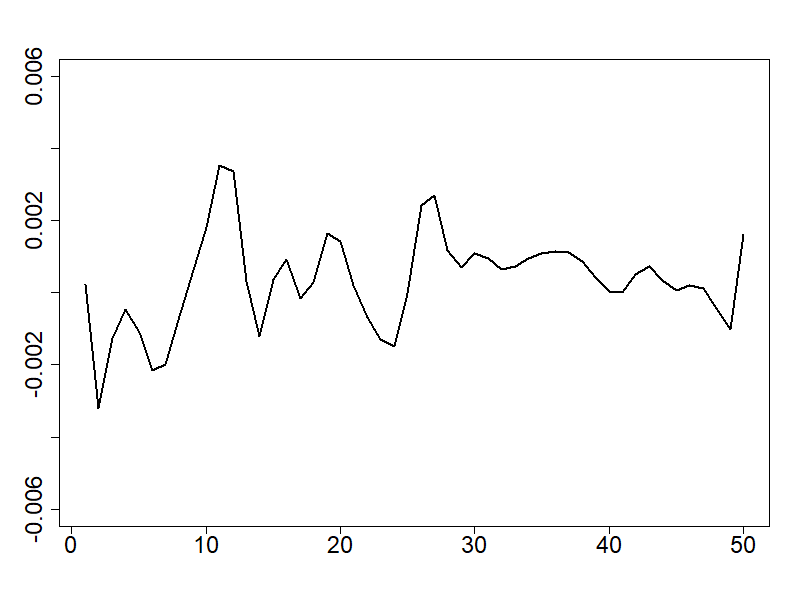} & \includegraphics[scale=.09, align=c]{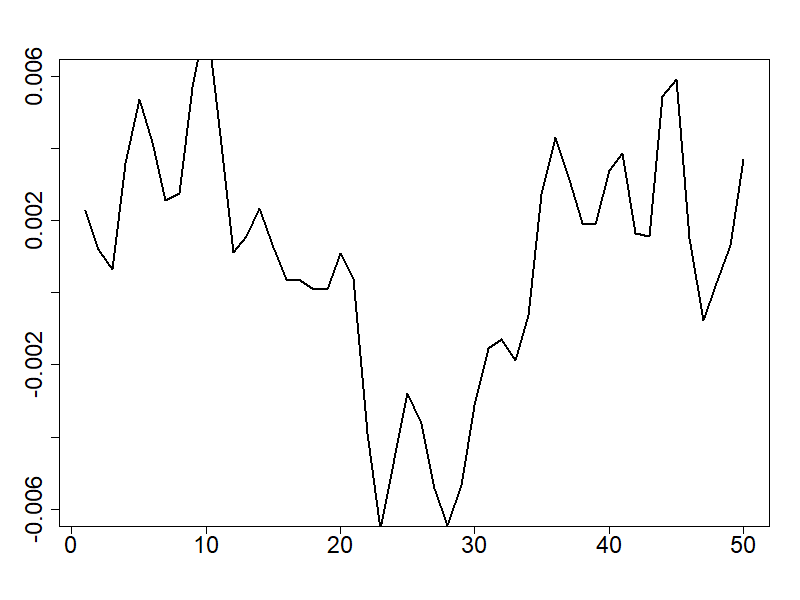} & \includegraphics[scale=.09, align=c]{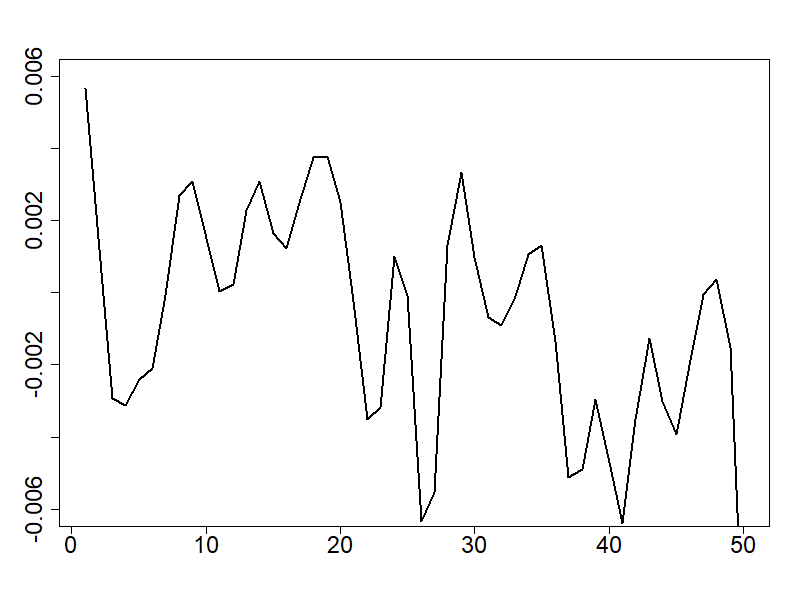} \\ 
&\begin{tabular}{c}
     $\hat{\beta}^{(25)} $ \\
     $H_0 :  \beta^{(25)}=\beta^{(18)}$\\ (p=0.986)
\end{tabular} & \begin{tabular}{c}
     $\hat{\beta}^{(26)}$\\
       $H_0 :  \beta^{(26)}=\beta^{(19)}$\\ (p=0.895)
\end{tabular} & $\hat{\beta}^{(27)}=\hat{\beta}^{(28)}$ & $\hat{\beta}^{(28)}=\hat{\beta}^{(27)}$   \\

    & \includegraphics[scale=.09, align=c]{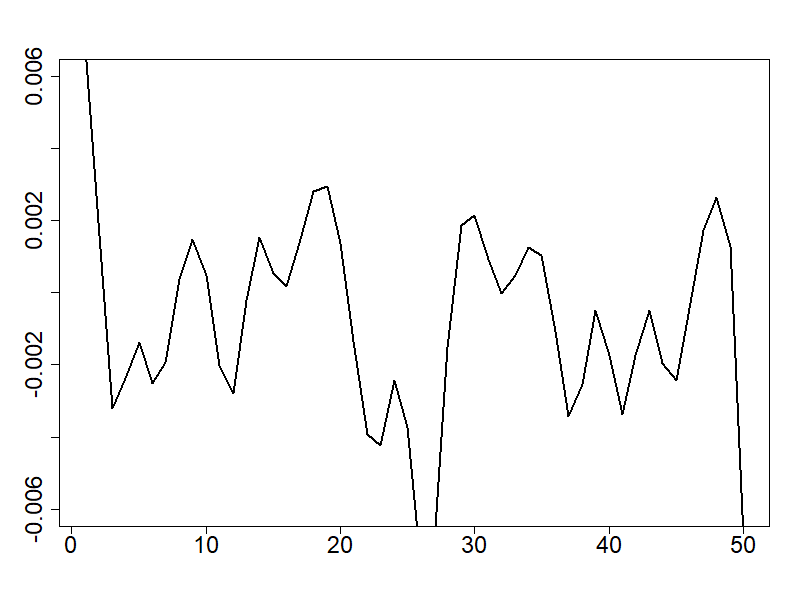} & \includegraphics[scale=.09, align=c]{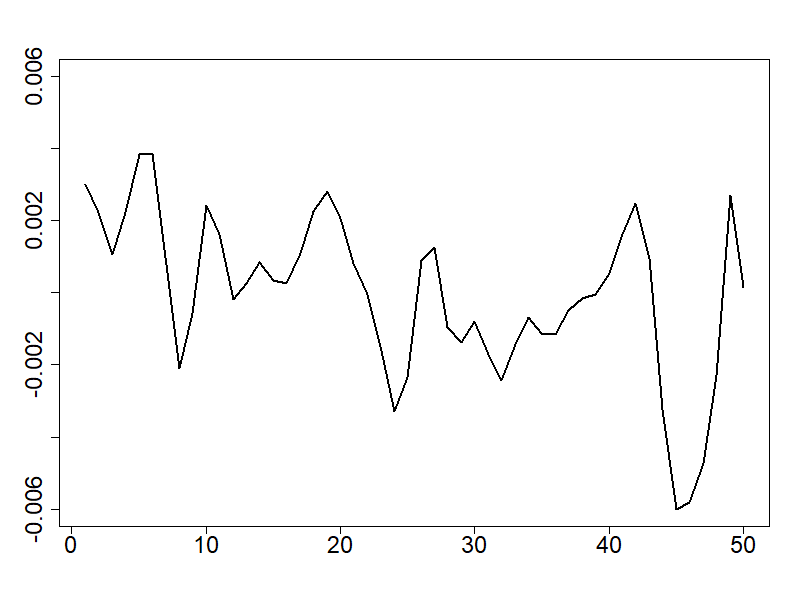} & \includegraphics[scale=.09, align=c]{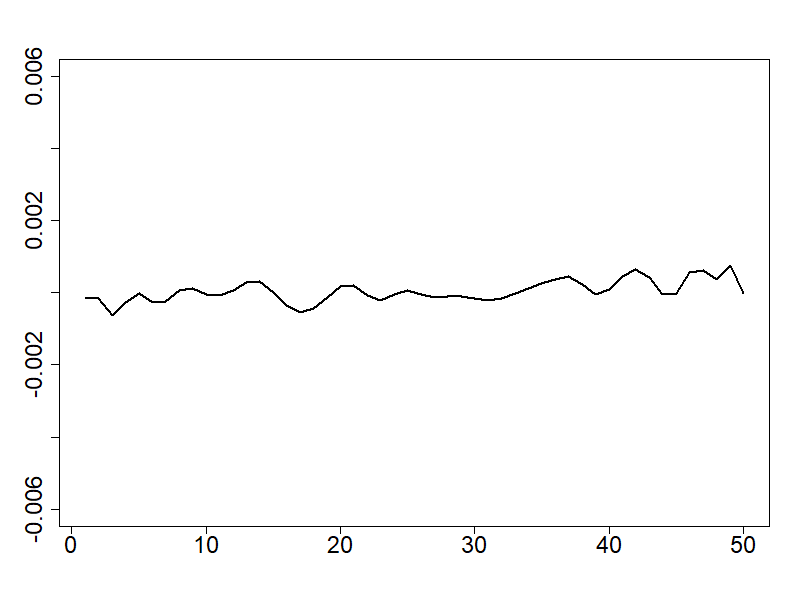} & \includegraphics[scale=.09, align=c]{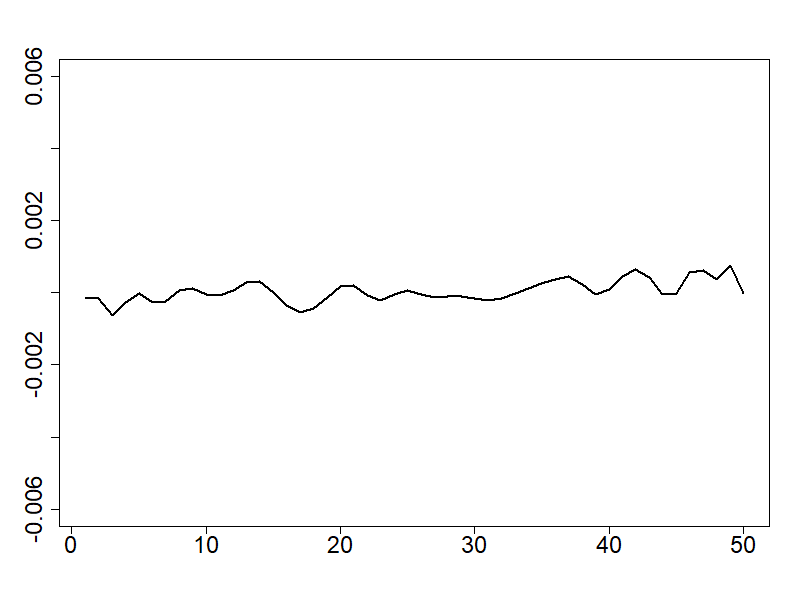}
    
    \end{tabular}
   
\end{tabular}
 }   \caption{ FU$_\text{Circle}$ estimated coefficients and associated p-values}
\end{figure}

\begin{figure}[ht]
\centering
    \begin{tabular}{c c c c c}
         Group 1& Group 2& Group 3 & Group 4 & Group 5  \\
         $\scriptstyle \hat{\beta}^{(1)} \neq \hat{\beta}^{(2)} $ & $\scriptstyle \hat{\beta}^{(4)}=\hat{\beta}^{(5)}$   & $ \scriptstyle \hat{\beta}^{(6)}\neq  \hat{\beta}^{(13)} \neq  \hat{\beta}^{(20)}$  & $\scriptstyle \hat{\beta}^{(7)} \neq \hat{\beta}^{(8)} \neq \hat{\beta}^{(14)} $ &   $ \scriptstyle \hat{\beta}^{(3)}\neq  \hat{\beta}^{(9)} \neq \hat{\beta}^{(16)} \neq  \hat{\beta}^{(23)}$   \\ 
\includegraphics[scale=.09]{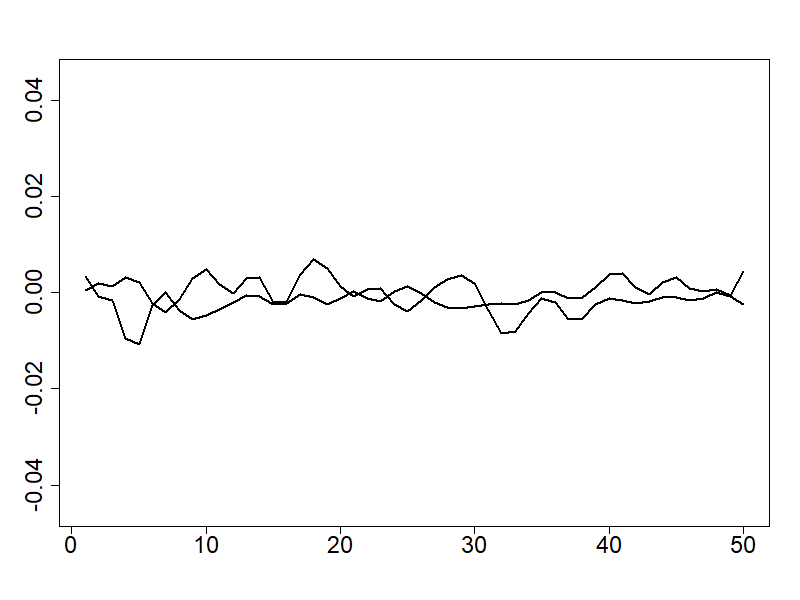} & \includegraphics[scale=.09]{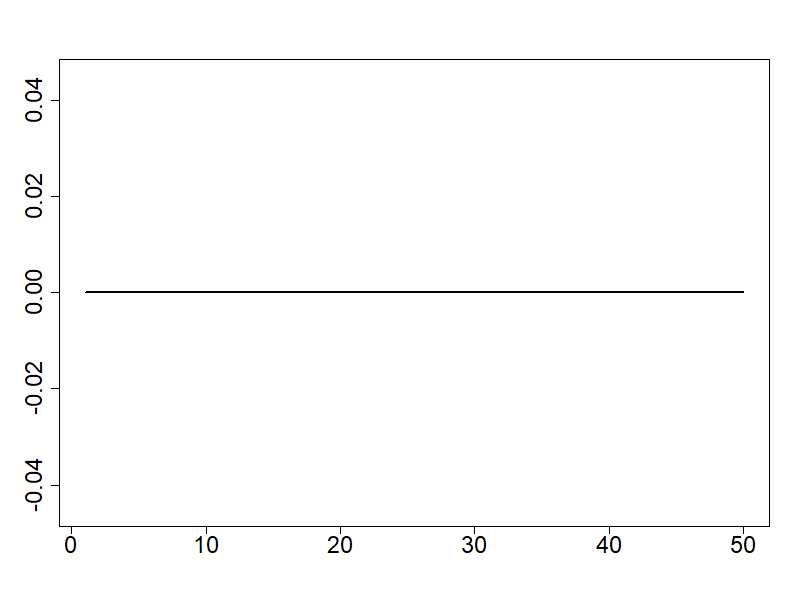} & \includegraphics[scale=.09]{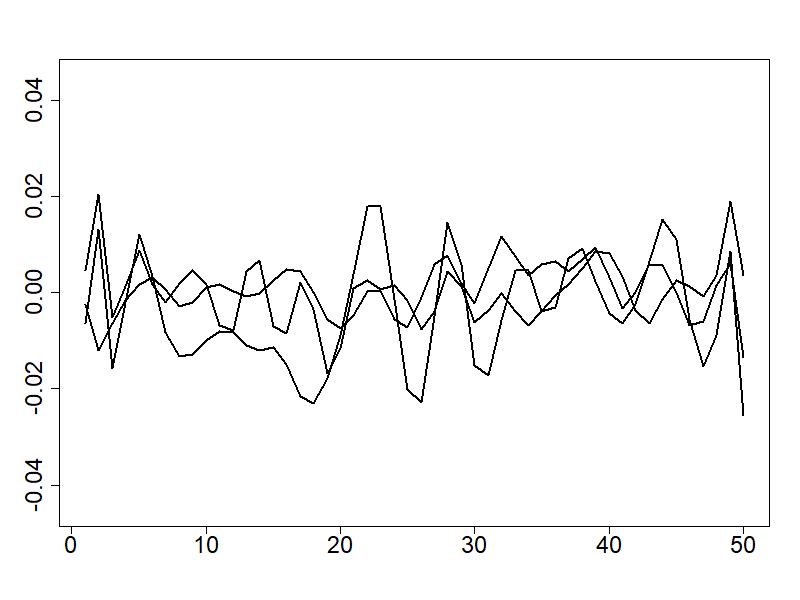} & 
\includegraphics[scale=.09]{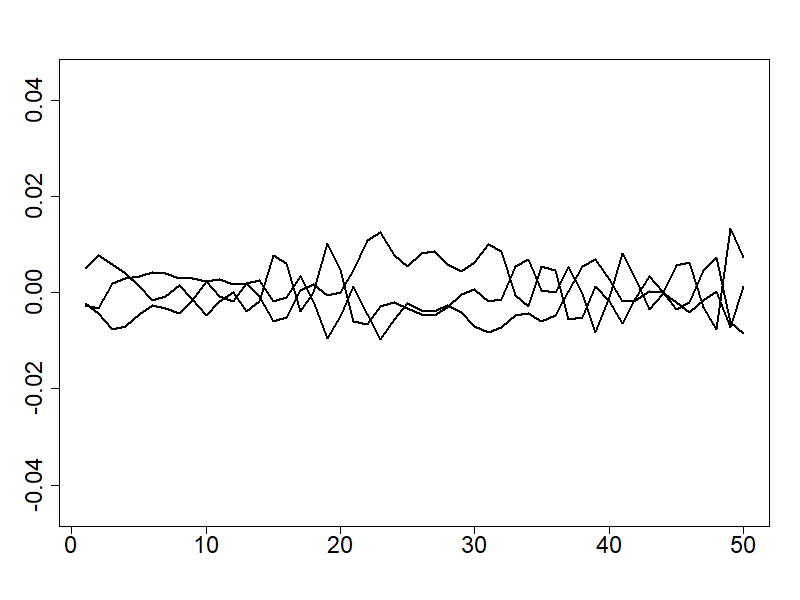} & 
\includegraphics[scale=.09]{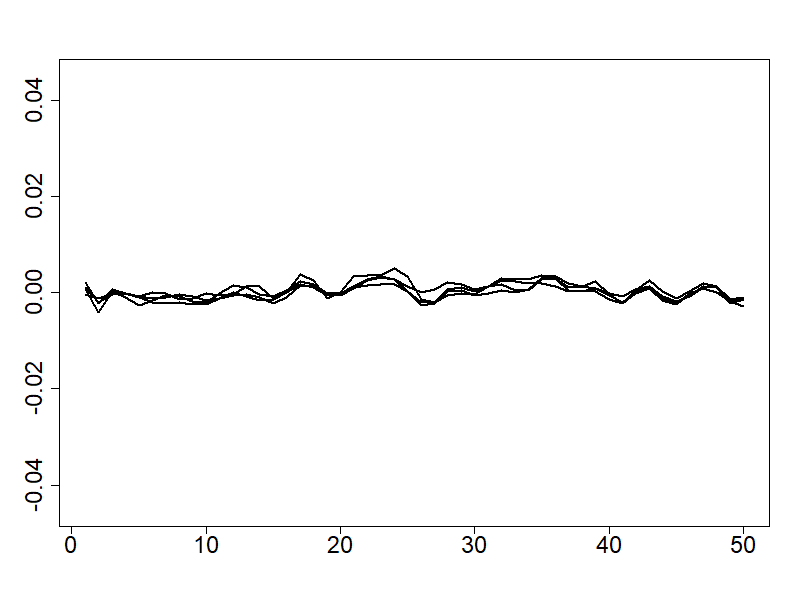} \\
Group 6& Group 7& Group 8 & Group 9 & Group 10  \\
$\scriptstyle \hat{\beta}^{(10)}\neq \hat{\beta}^{(11)} \neq  \hat{\beta}^{(17)} \neq \hat{\beta}^{(18)}$  &  $\scriptstyle \hat{\beta}^{(12)} \neq \hat{\beta}^{(19)} \neq \hat{\beta}^{(26)} $ &  $\scriptstyle \hat{\beta}^{(15)} \neq \hat{\beta}^{(21)} \neq \hat{\beta}^{(22)}$ & $ \scriptstyle \hat{\beta}^{(24)} \neq \hat{\beta}^{(25)}$ & $\scriptstyle \hat{\beta}^{(27)} \neq \hat{\beta}^{(28)}$  \\ 
\includegraphics[scale=.09]{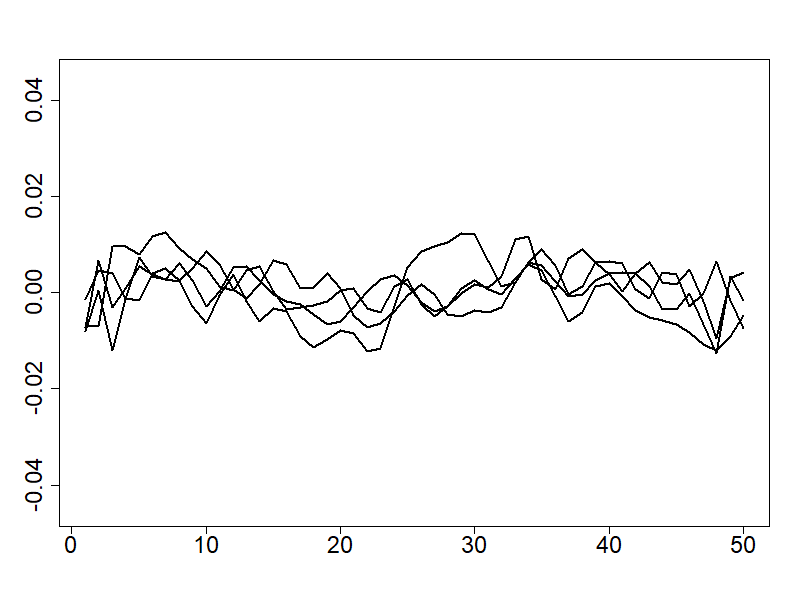} & \includegraphics[scale=.09]{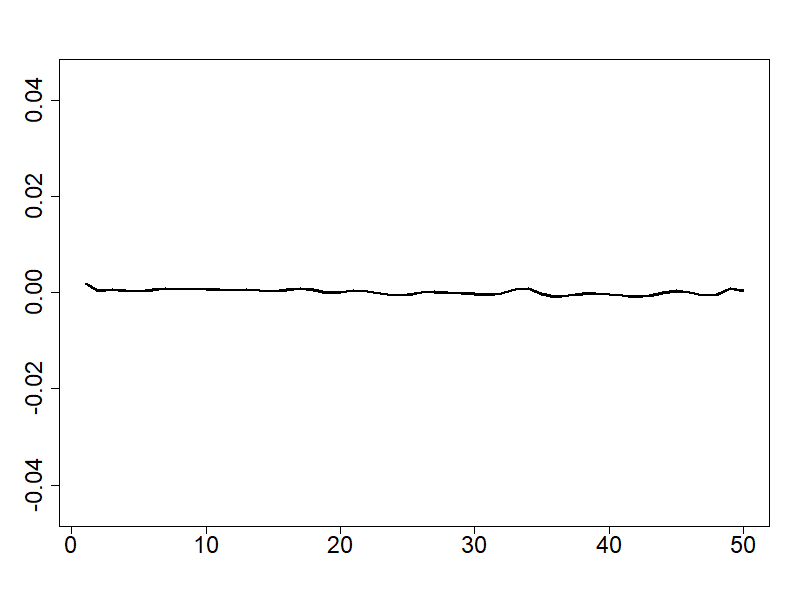} & \includegraphics[scale=.09]{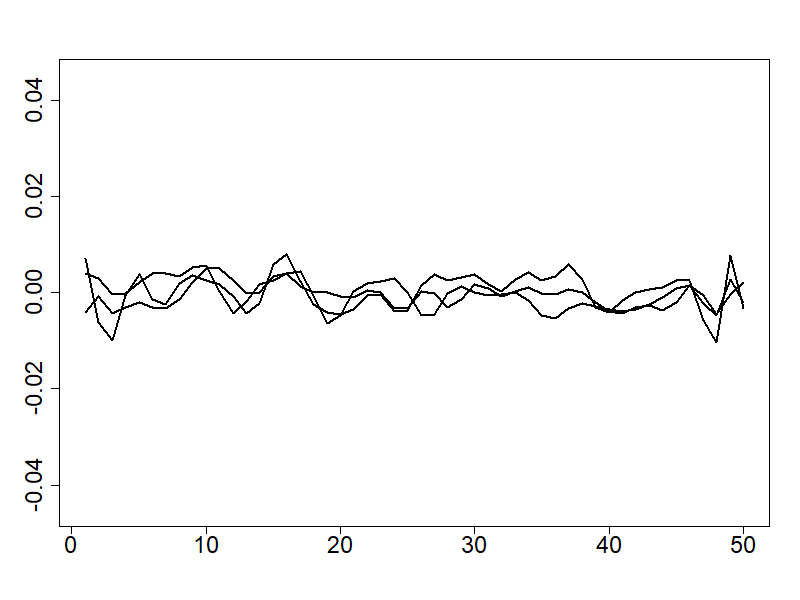} & 
\includegraphics[scale=.09]{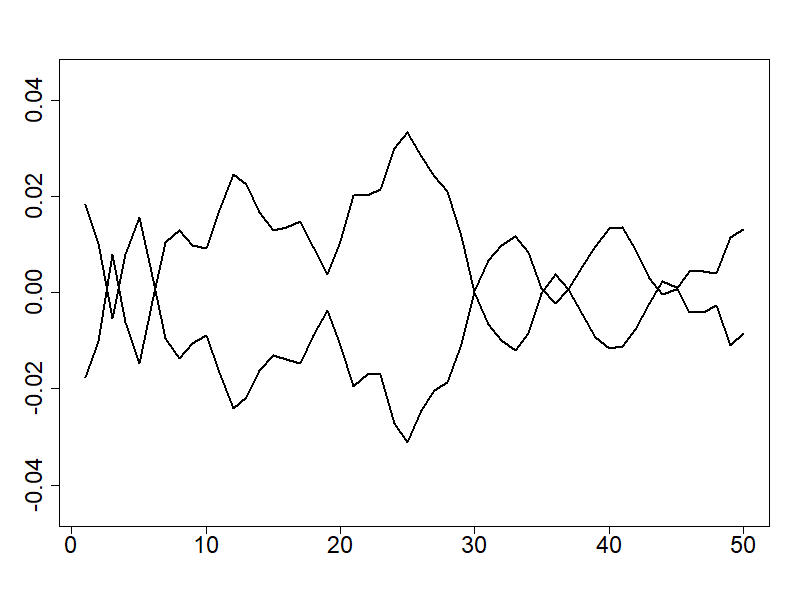} & 
\includegraphics[scale=.09]{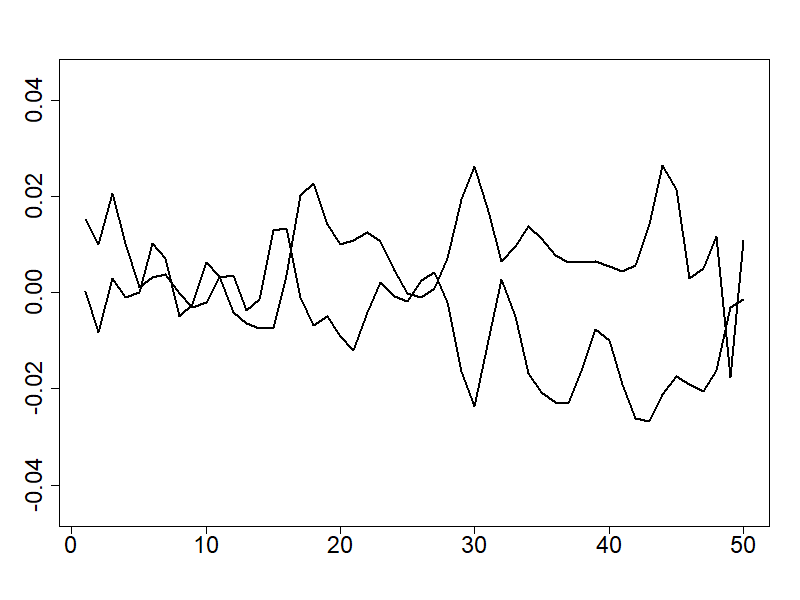} 
    \end{tabular}
  \caption{GFUL$_\text{Euclidean}$ estimated coefficients}
\end{figure}
\begin{figure}[H]
\centering
   \begin{tabular}{c c }
         Group 1& Group 2 \\ 
         $\scriptstyle \hat{\beta}^{(1)} \neq\hat{\beta}^{(2)} \neq\hat{\beta}^{(3)} \neq\hat{\beta}^{(4)} \neq\hat{\beta}^{(5)} \neq \hat{\beta}^{(6)} \neq\hat{\beta}^{(7)} \neq\hat{\beta}^{(8)} \neq\hat{\beta}^{(9)}$& $\scriptstyle \hat{\beta}^{(10)} \neq \hat{\beta}^{(11)} \neq\hat{\beta}^{(12)} \neq\hat{\beta}^{(17)}\neq \hat{\beta}^{(18)} \neq\hat{\beta}^{(19)} \neq\hat{\beta}^{(26)}$ \\ 
         
\includegraphics[scale=.09]{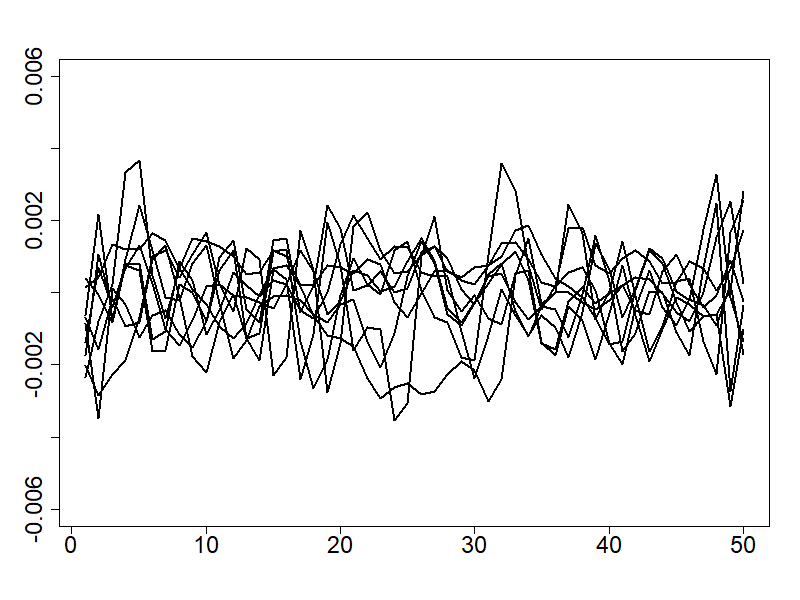} & 
\includegraphics[scale=.09]{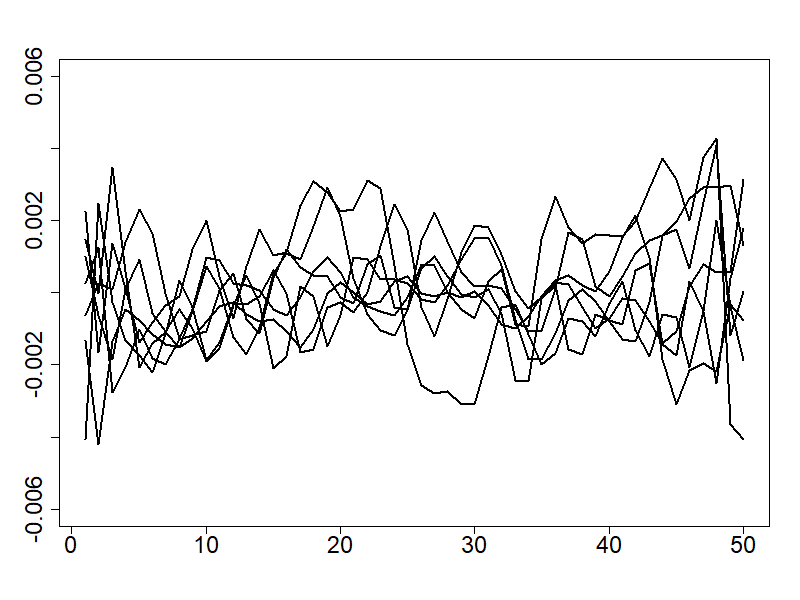} \\ 
         Group 3 & Group 4 \\ 
         $ \scriptstyle \hat{\beta}^{(13)}\neq \hat{\beta}^{(14)} \neq\hat{\beta}^{(15)} \neq\hat{\beta}^{(20)}\neq \hat{\beta}^{(21)} \neq\hat{\beta}^{(22)}$ &  $\scriptstyle \hat{\beta}^{(16)} \neq\hat{\beta}^{(23)} \neq\hat{\beta}^{(24)} \neq\hat{\beta}^{(25)}$ \\ 
         
\includegraphics[scale=.09]{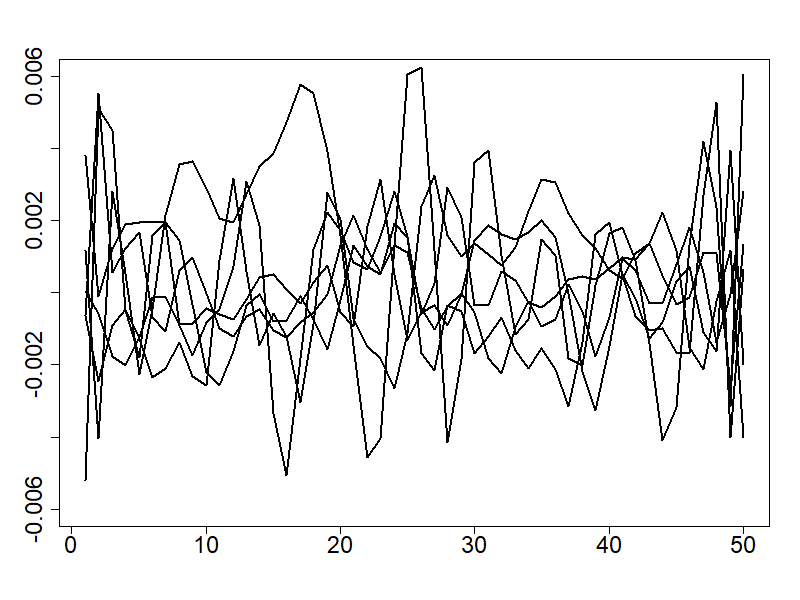} & 
\includegraphics[scale=.09]{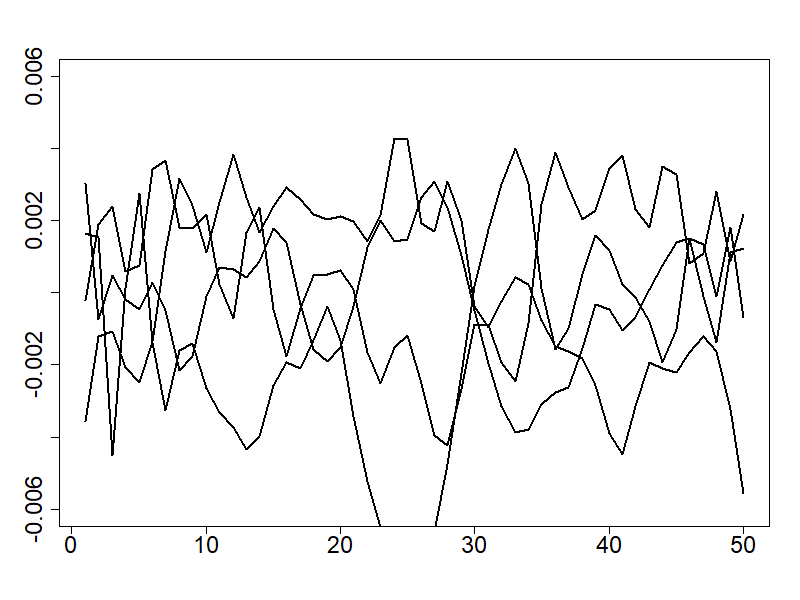} \\ 
         \multicolumn{2}{c}{Group 5}  \\
          \multicolumn{2}{c}{$\scriptstyle \hat{\beta}^{(27)}\neq \hat{\beta}^{(28)}$ } \\
\multicolumn{2}{c}{\includegraphics[scale=.09]{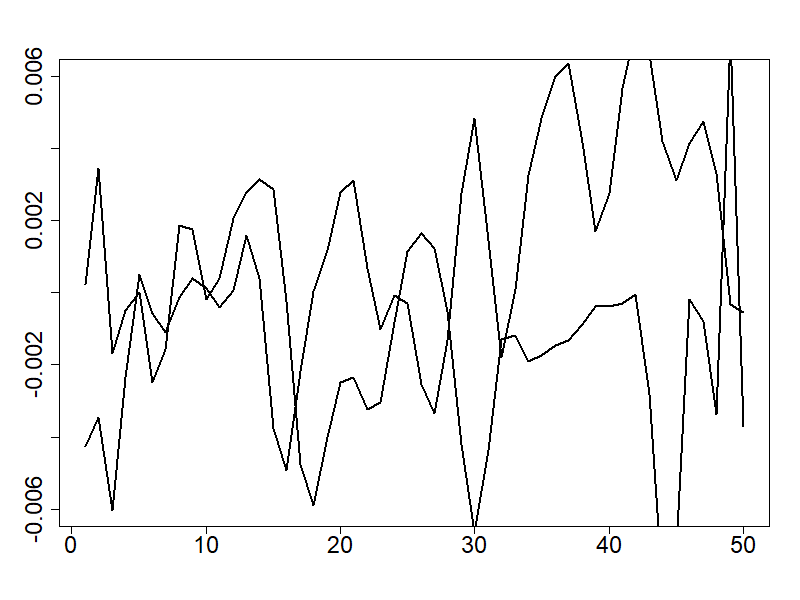}}   
   \end{tabular}
\caption{ GFUL$_\text{Circle}$ estimated coefficients}
\end{figure}

\newpage 
\begin{figure}[ht!]
    \centering
    \begin{tabular}{c c c c  c}
     & $\hat{\beta}^{(1)}$ & $\hat{\beta}^{(2)}$ &  $\hat{\beta}^{(3)}$& $\hat{\beta}^{(4)}$ \\
   &  (p=0.905) &(p=0.654) &(p=0.864) &(p=0.572) \\
    & \includegraphics[scale=.09, align=c]{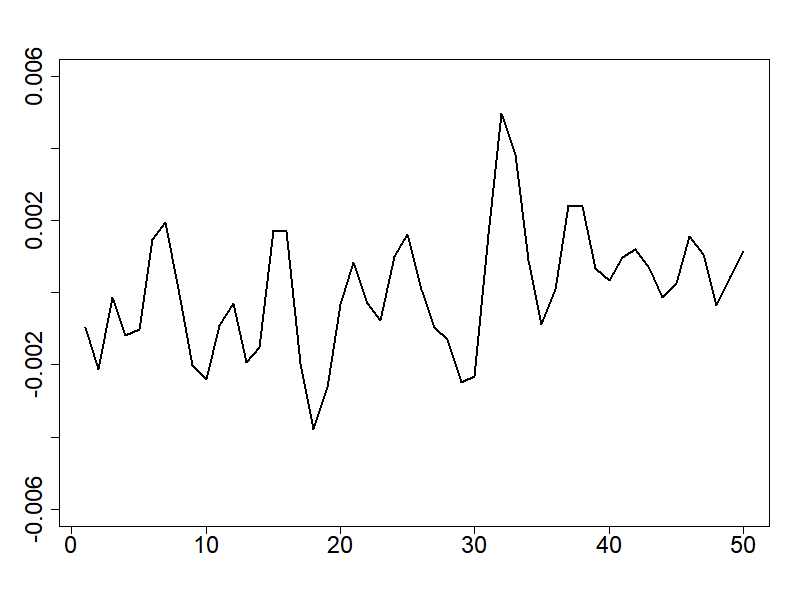}  &  \includegraphics[scale=.09, align=c]{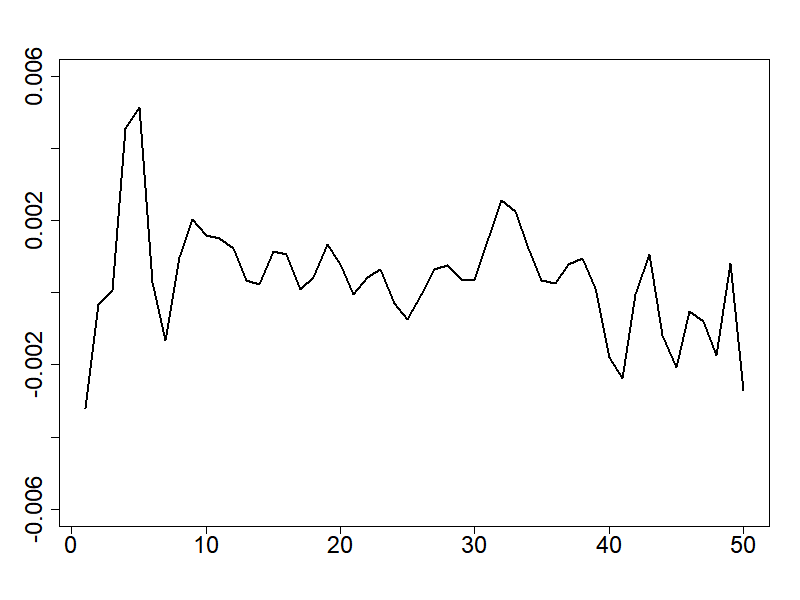} & \includegraphics[scale=.09, align=c]{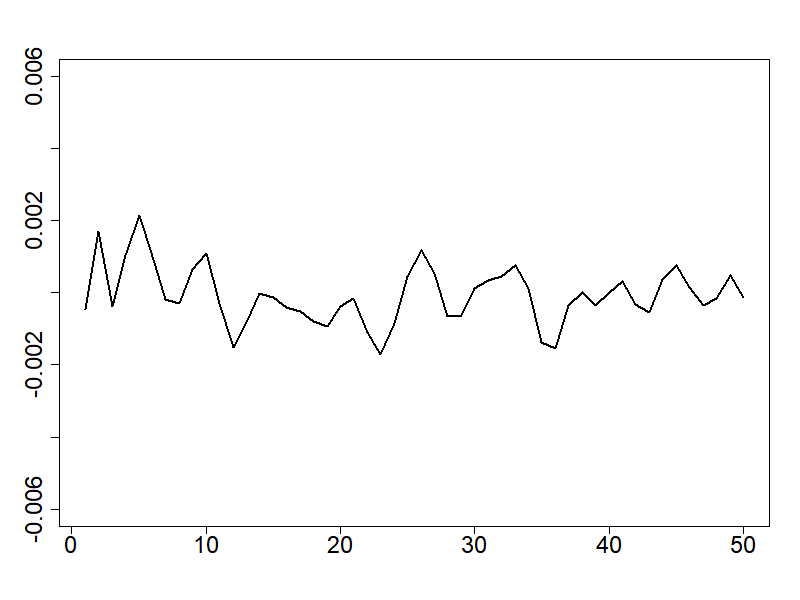} & \includegraphics[scale=.09, align=c]{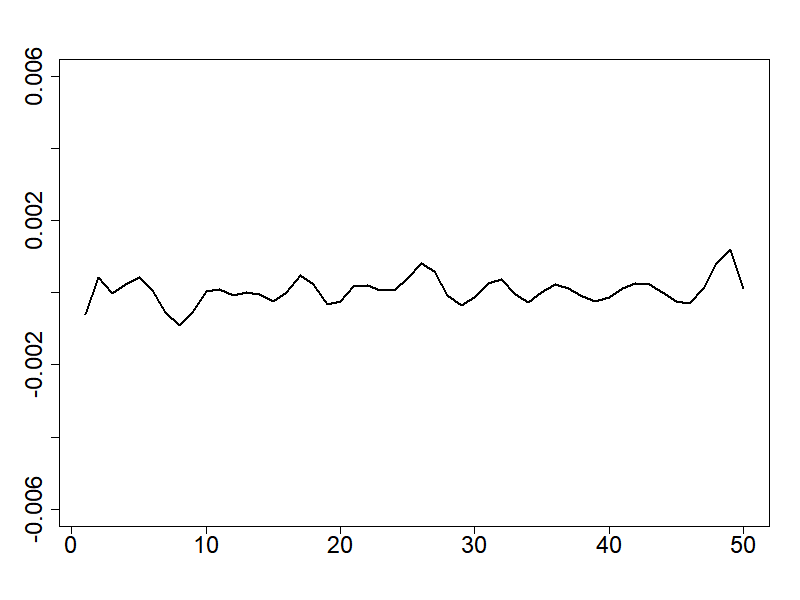} \\ 
     & $\hat{\beta}^{(5)}$ & $\hat{\beta}^{(6)}$ & $\hat{\beta}^{(7)}$& $\hat{\beta}^{(8)}$  \\
     & (p=0.996) &(p=0.245) &(p=0.858)& (p=0.680) \\ 
    &\includegraphics[scale=.09, align=c]{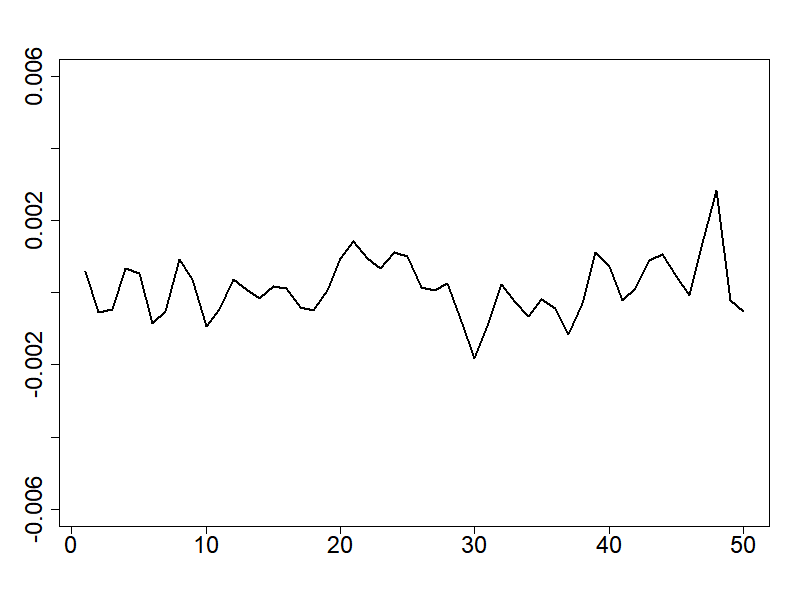} &  \includegraphics[scale=.09, align=c]{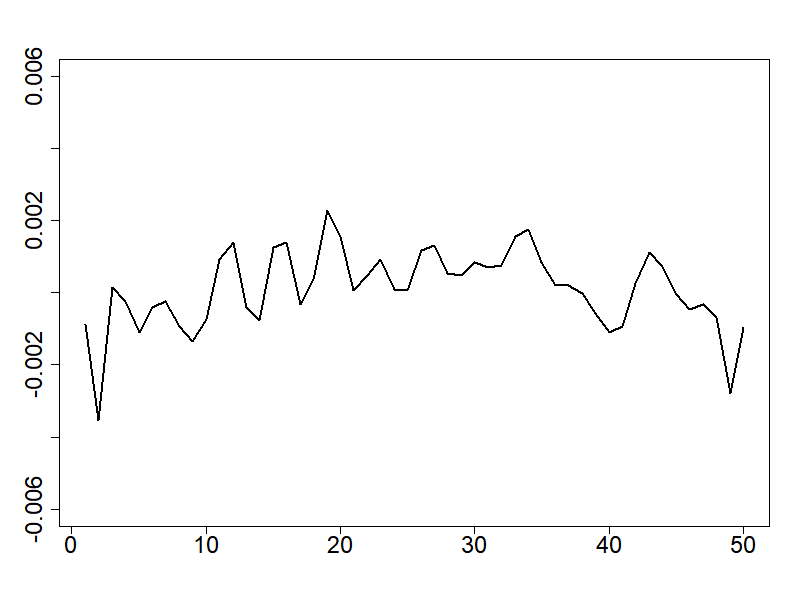} &  \includegraphics[scale=.09, align=c]{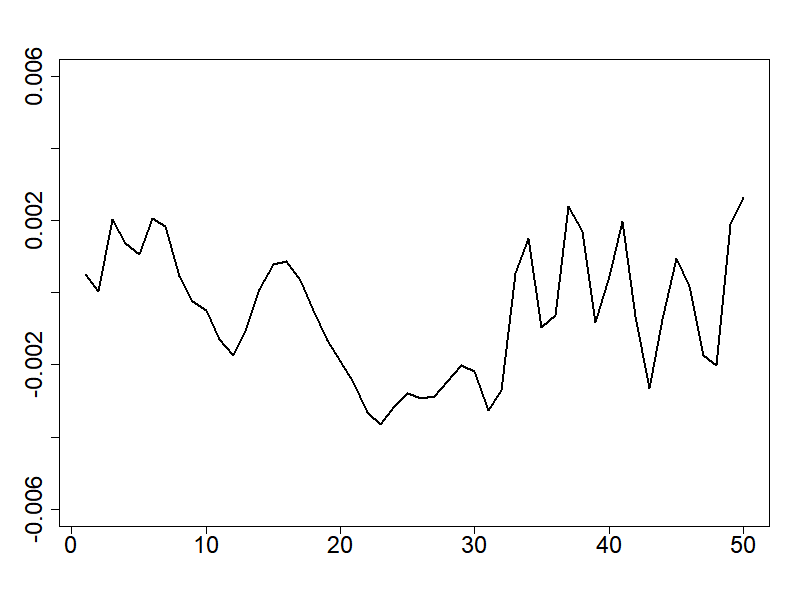} &  \includegraphics[scale=.09, align=c]{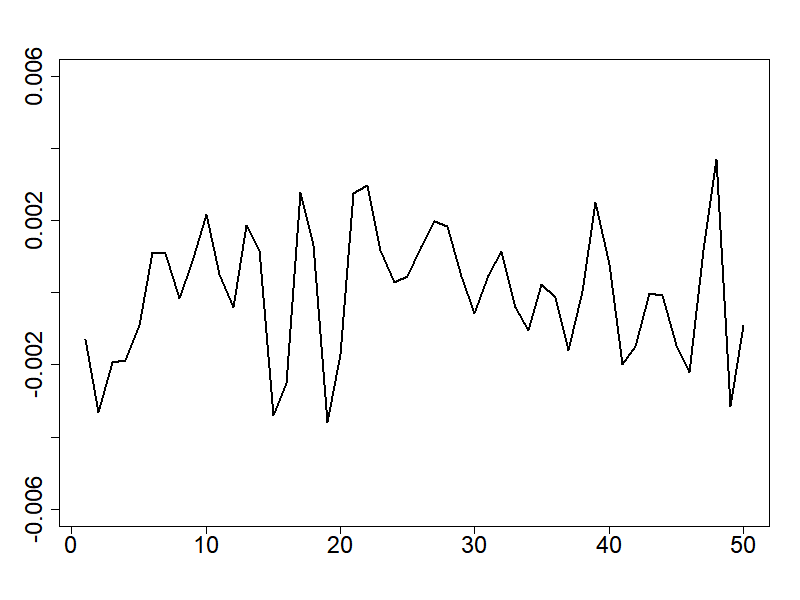} \\ 
      & $\hat{\beta}^{(9)}$& $\hat{\beta}^{(10)}$ &$\hat{\beta}^{(11)}$ & $\hat{\beta}^{(12)}$   \\ 
    &  (p=0.959)& (p=0.991) &(p=0.576) & (p=0.993) \\ 
      & \includegraphics[scale=.09, align=c]{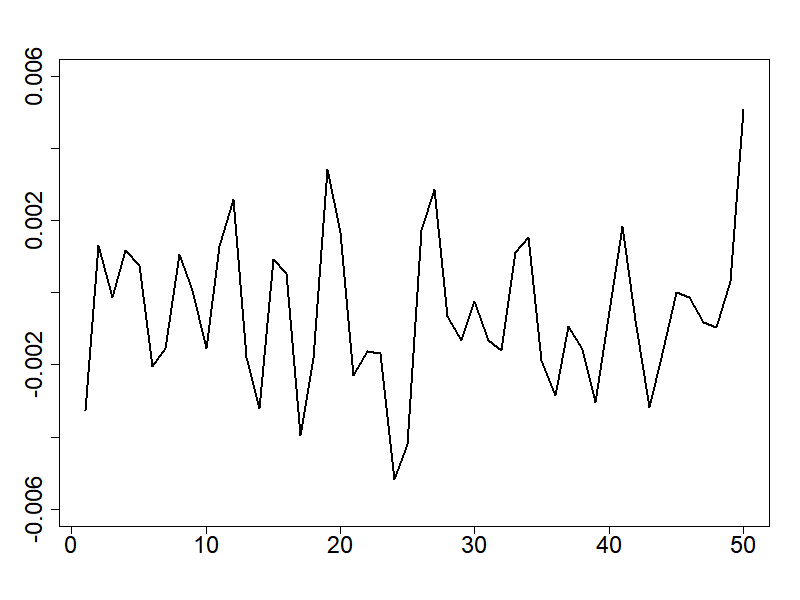} & \includegraphics[scale=.09, align=c]{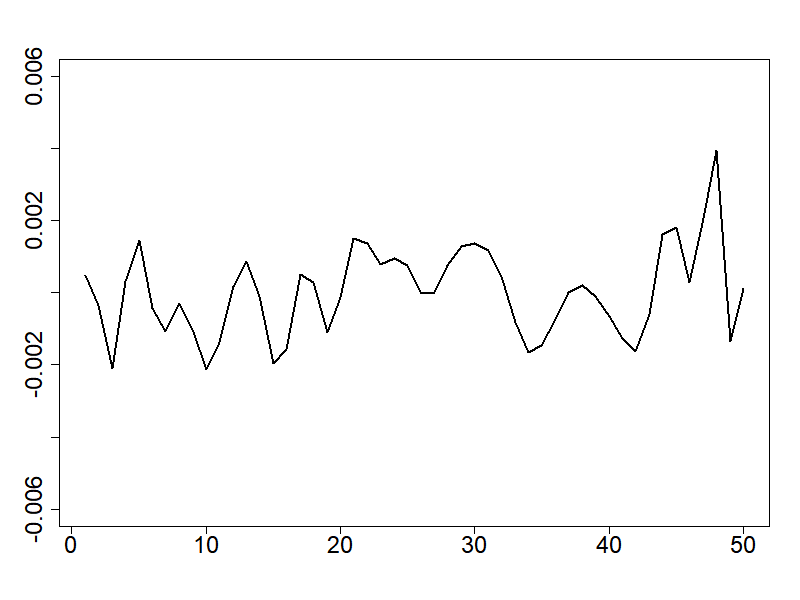} & 
    \includegraphics[scale=.09, align=c]{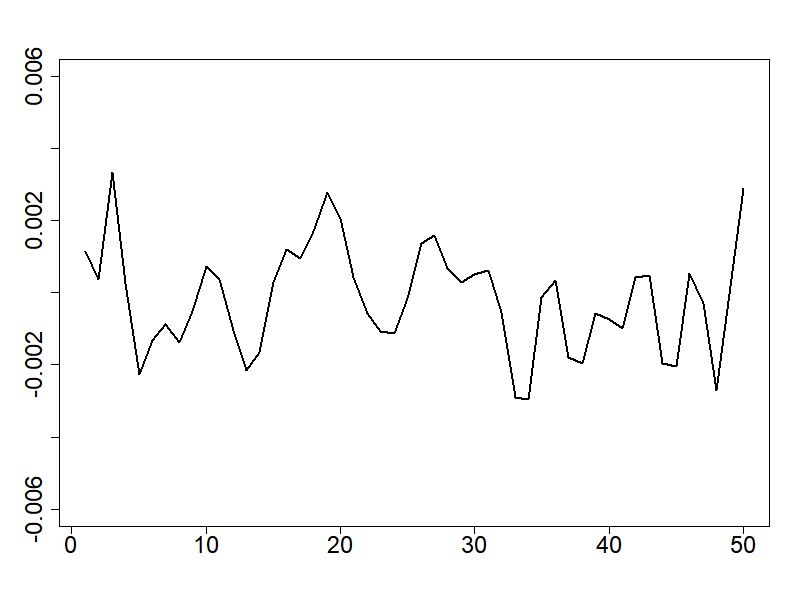}& \includegraphics[scale=.09, align=c]{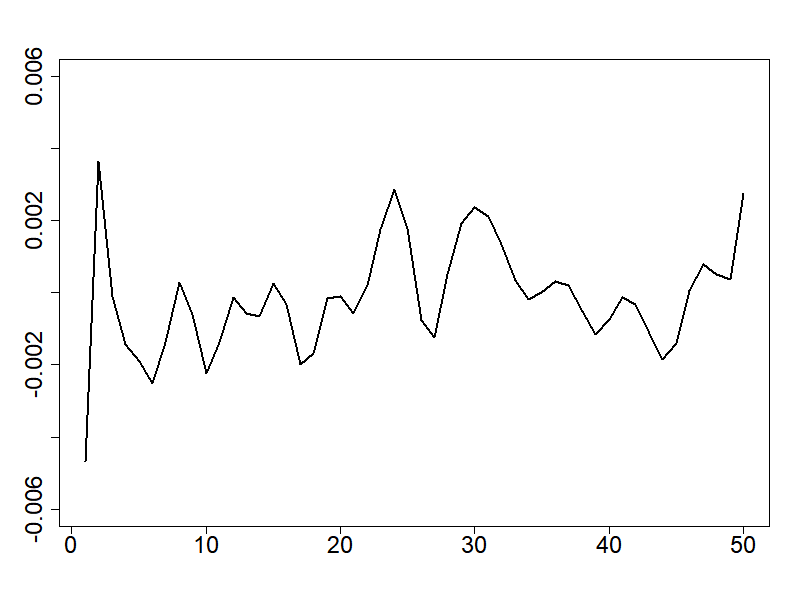} \\
     & $\hat{\beta}^{(13)}$ & $\hat{\beta}^{(14)}$ &  $\hat{\beta}^{(15)}$& $\hat{\beta}^{(16)}$\\
    & (p=0.789) &(p=0.259)& (p=0.645)& (p=0.896) \\ 
    & \includegraphics[scale=.09, align=c]{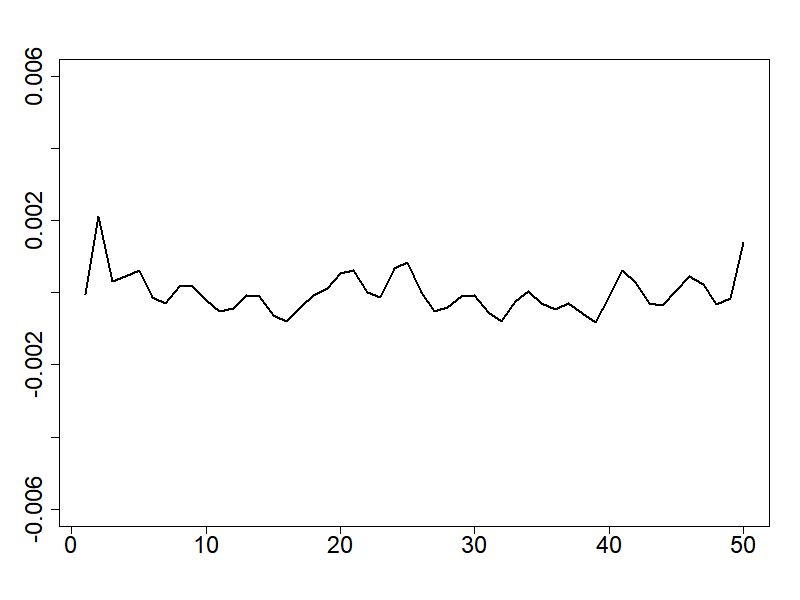} & \includegraphics[scale=.09, align=c]{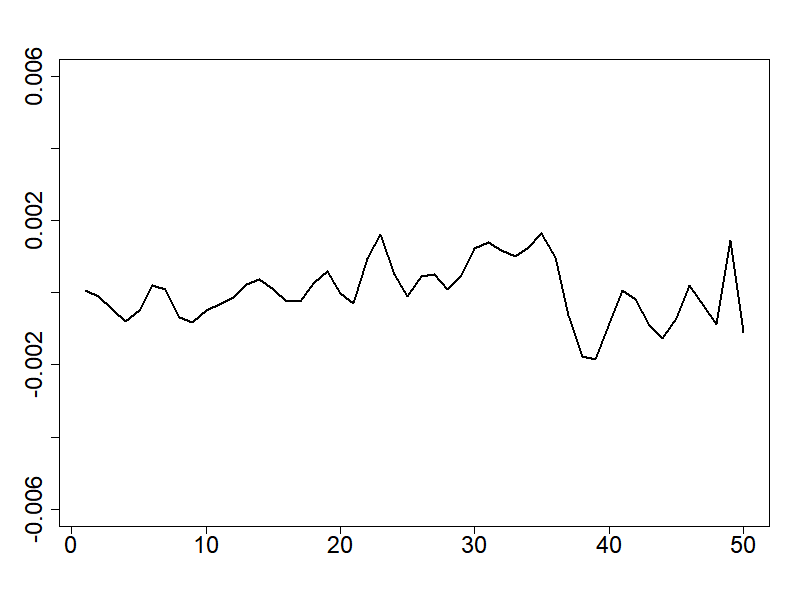} &    \includegraphics[scale=.09, align=c]{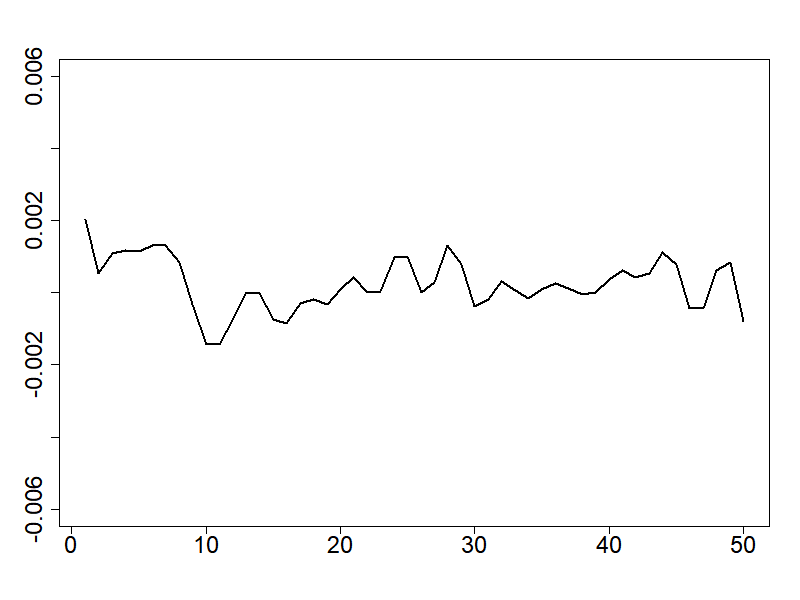} & \includegraphics[scale=.09, align=c]{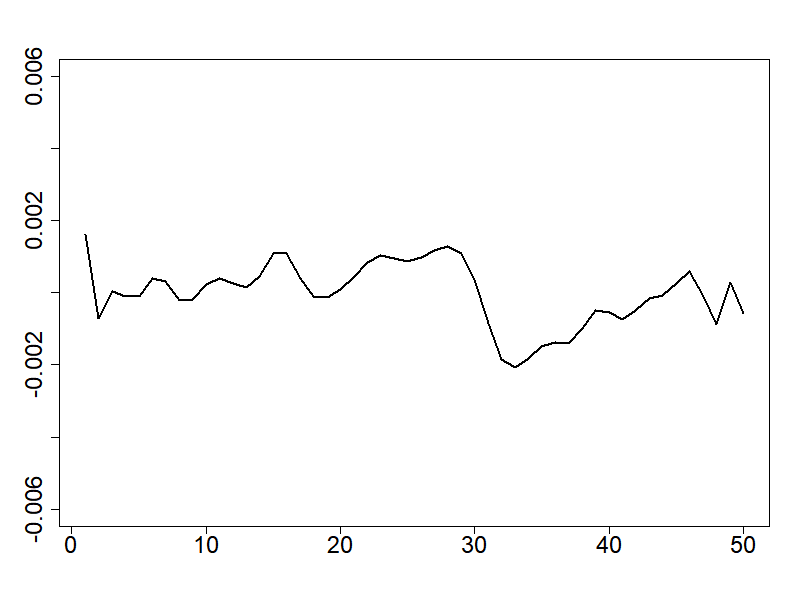} \\
     &  $\hat{\beta}^{(17)}$ & $\hat{\beta}^{(18)}$ & $\hat{\beta}^{(19)}$ & $\hat{\beta}^{(20)}$ \\
    &  (p=0.895)& (p=0.839) &(p=0.922) & (p=0.721) \\ 

    & \includegraphics[scale=.09, align=c]{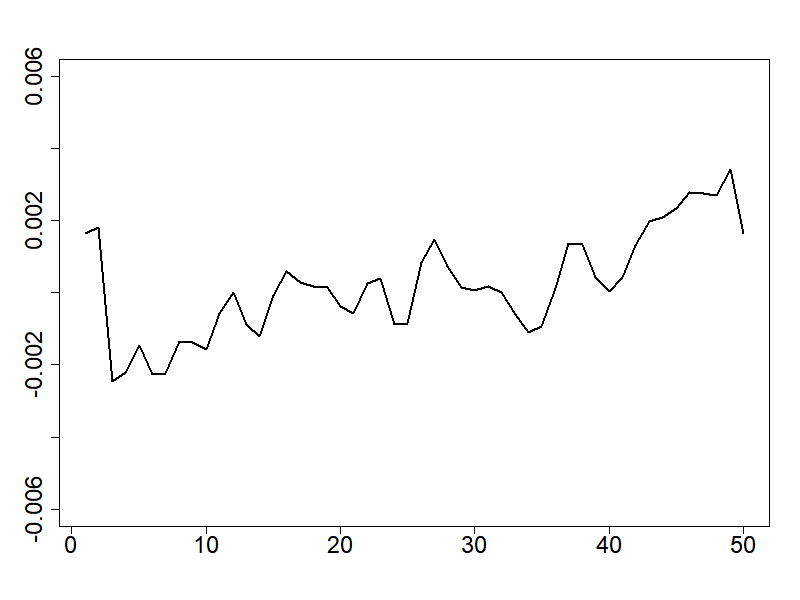} & \includegraphics[scale=.09, align=c]{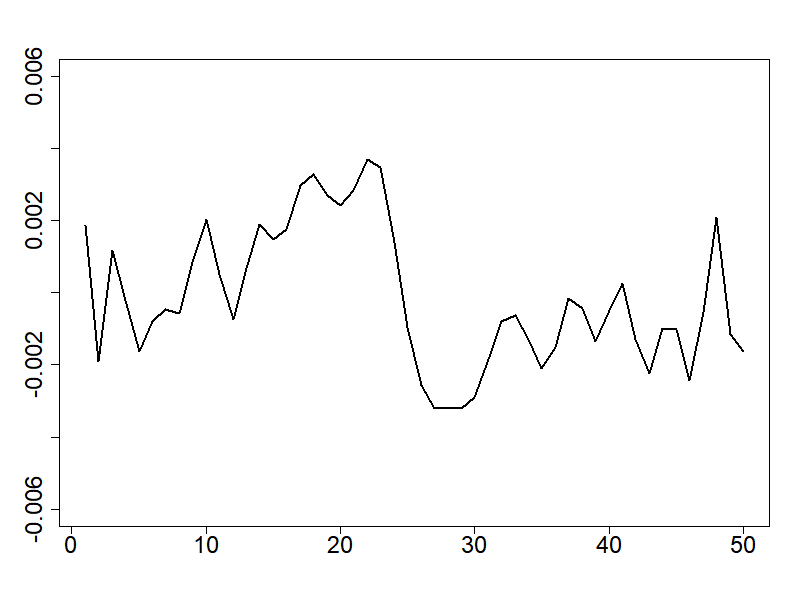} & 
    \includegraphics[scale=.09, align=c]{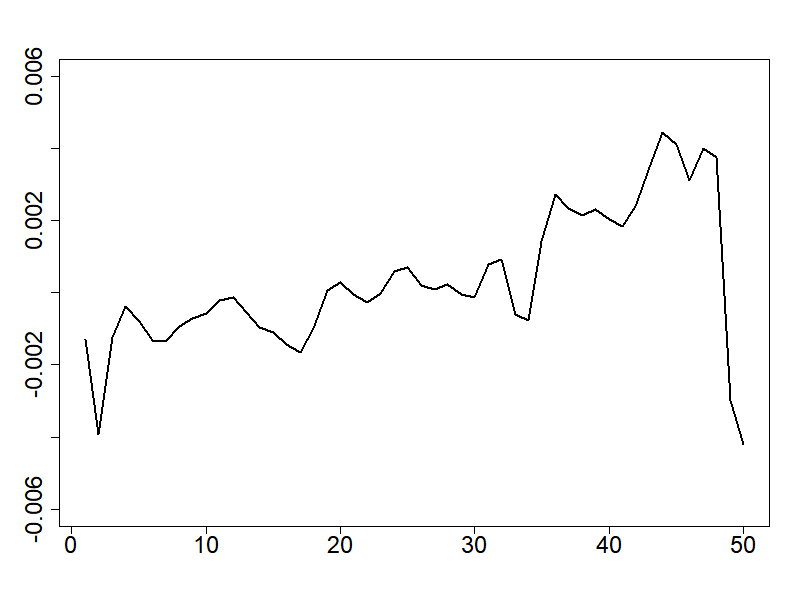} & \includegraphics[scale=.09, align=c]{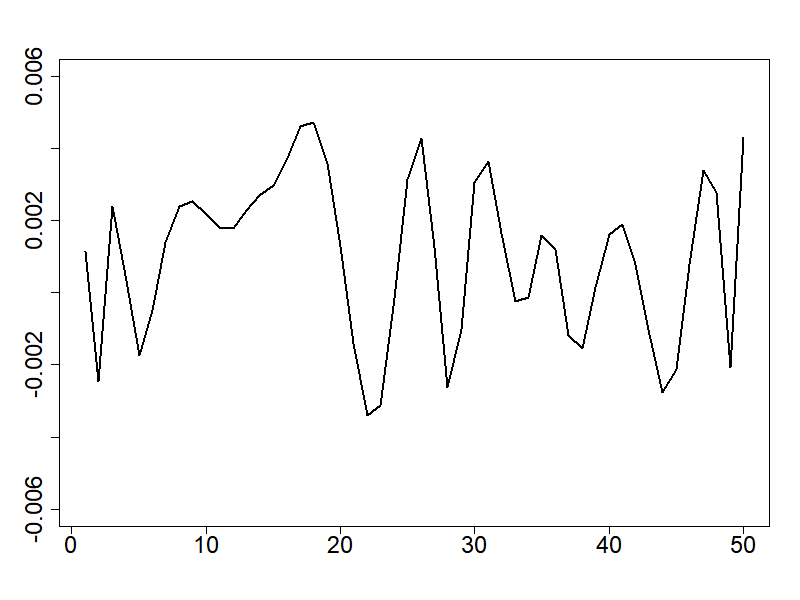} \\ 
    & $\hat{\beta}^{(21)}$ & $\hat{\beta}^{(22)}$ & $\hat{\beta}^{(23)}$  & $\hat{\beta}^{(24)}$  \\ 
    &(p=0.680) &(p=0.561)& (p=0.954)& (p=0.834) \\
    & \includegraphics[scale=.09, align=c]{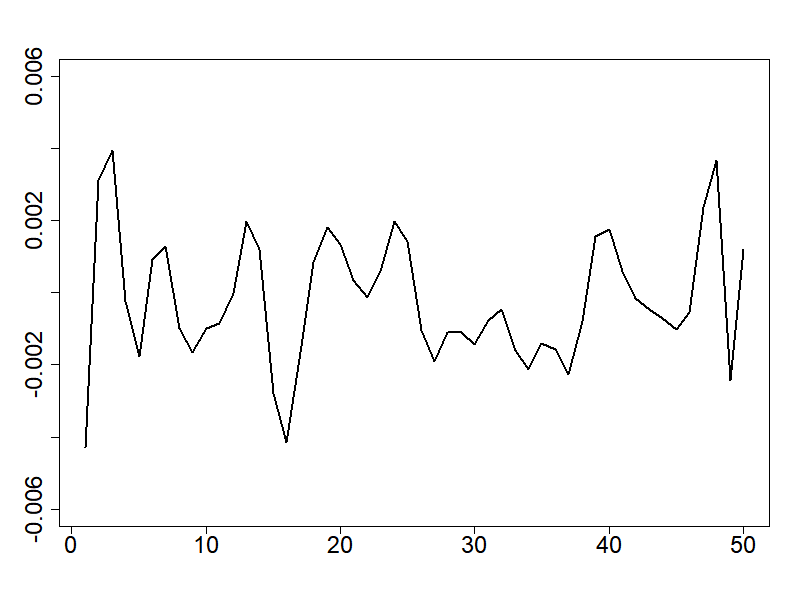} & \includegraphics[scale=.09, align=c]{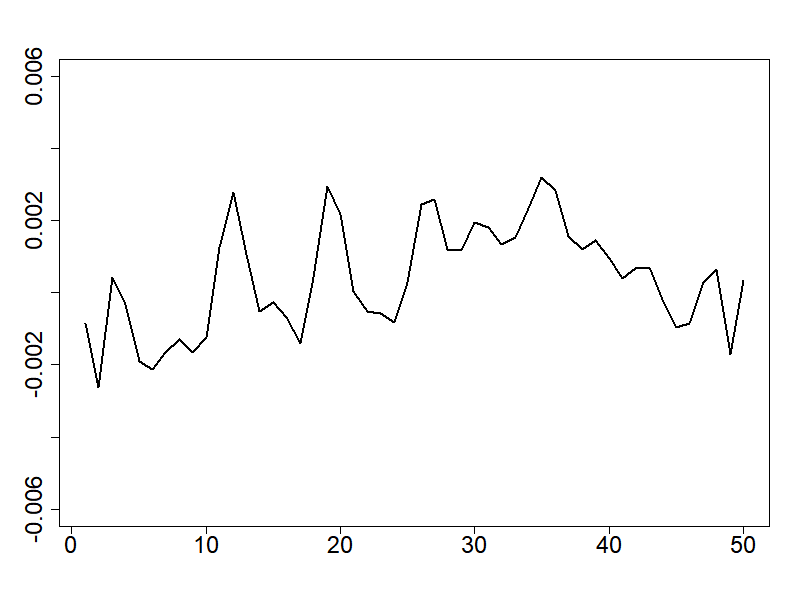} & \includegraphics[scale=.09, align=c]{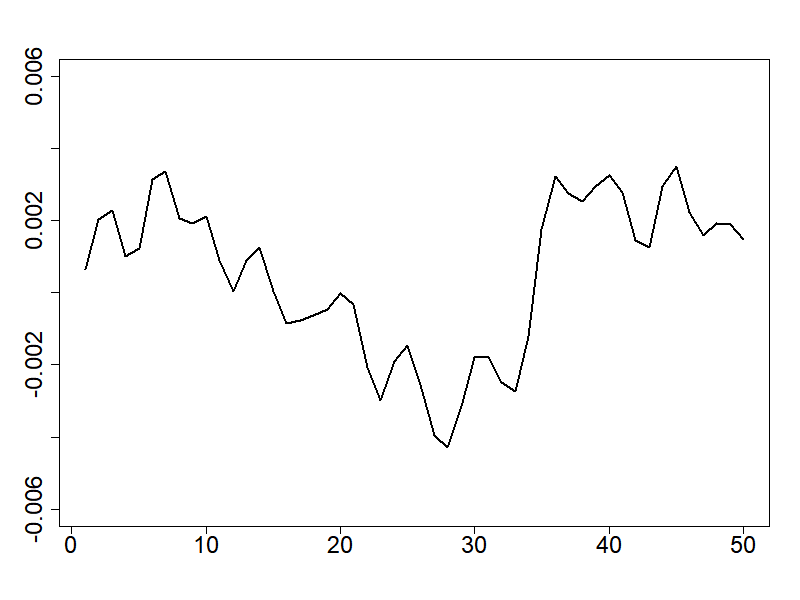} & \includegraphics[scale=.09, align=c]{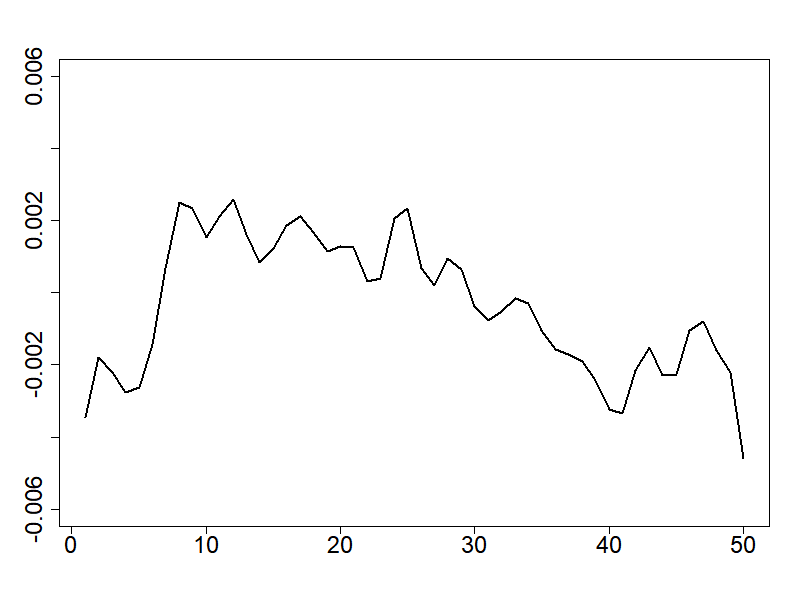}  \\    
    & $\hat{\beta}^{(25)}$ & $\hat{\beta}^{(26)}$ & $\hat{\beta}^{(27)}$ &  $\hat{\beta}^{(28)}$  \\ 
   & (p=0.984) &(p=0.991) &(p=0.845) &(p=0.577) \\ 
    & \includegraphics[scale=.09, align=c]{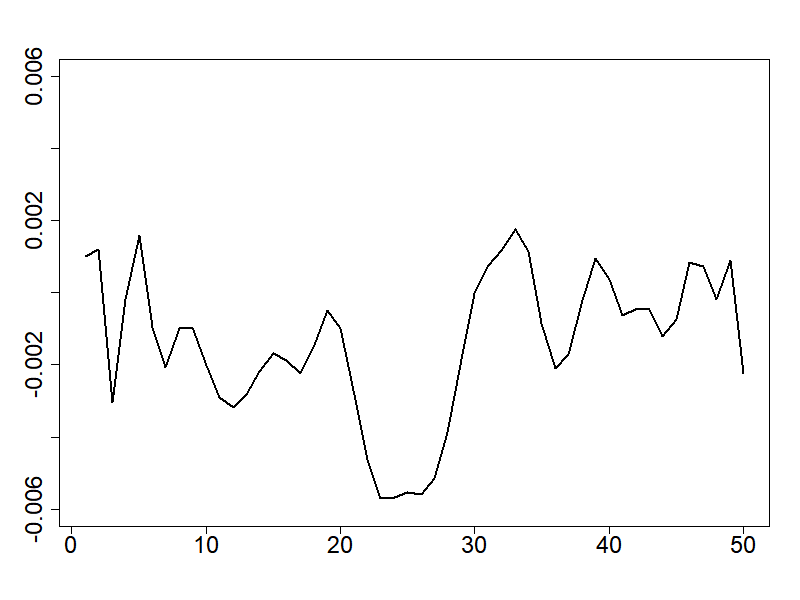} & \includegraphics[scale=.09, align=c]{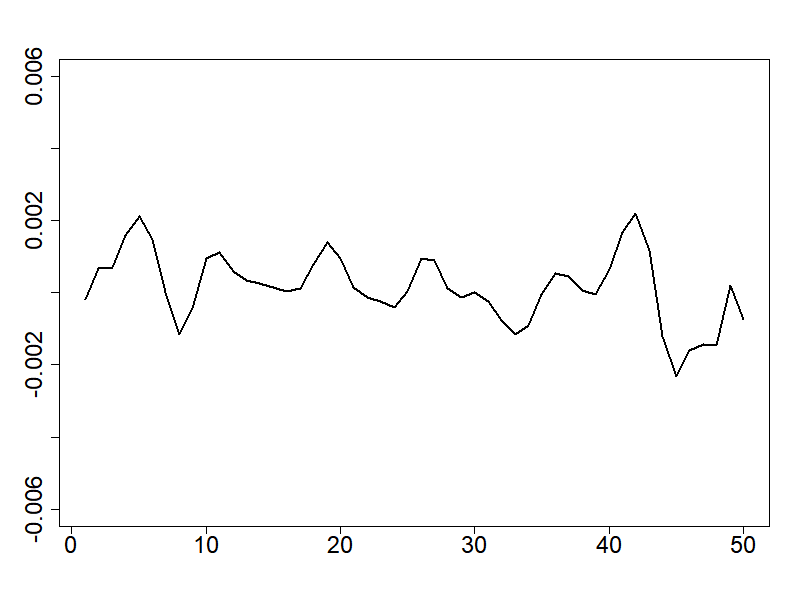} & \includegraphics[scale=.09, align=c]{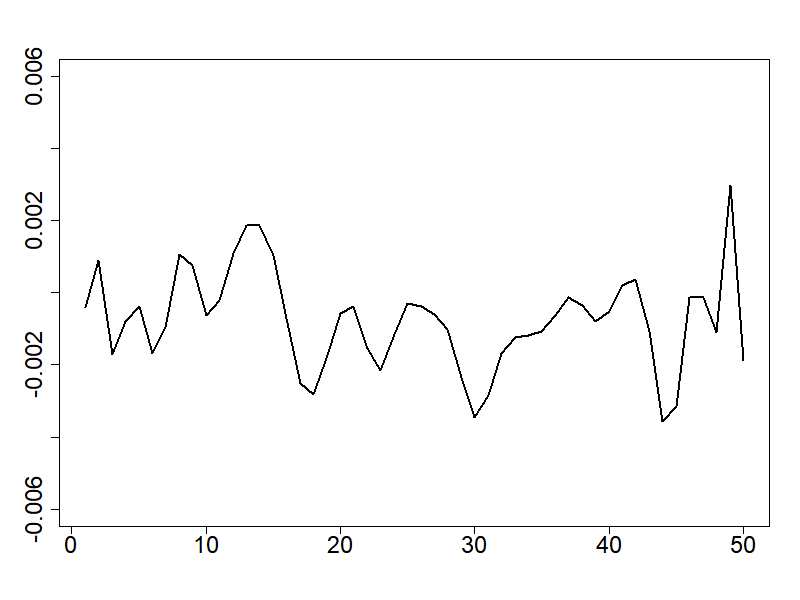} & \includegraphics[scale=.09, align=c]{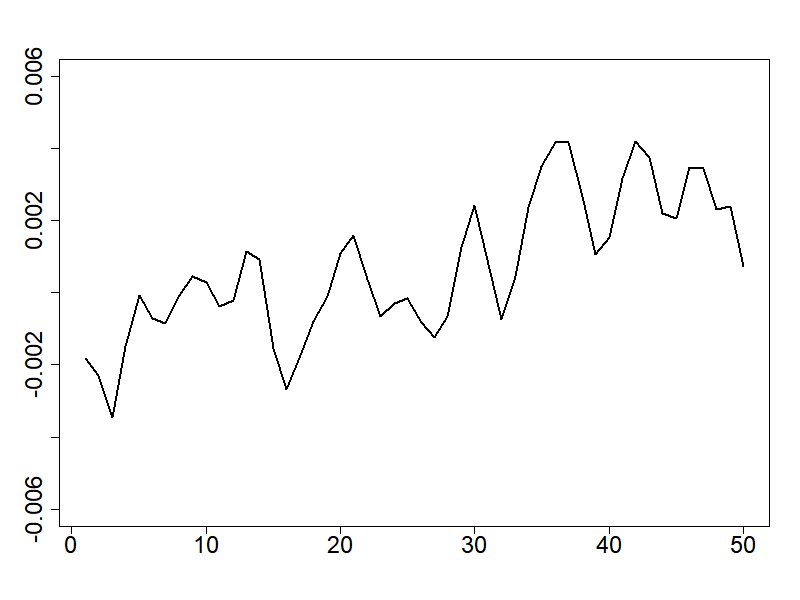}  
    \end{tabular}
    \caption{GL1 estimated coefficients and p-values ($H_0$: the coefficient is zero)}
\end{figure}
\newpage 
\begin{figure}[ht!]
    \centering
\resizebox{1\textwidth}{!}{
    \begin{tabular}{c c c c c}
\textbf{Group 1}& \textbf{Group 2} & \textbf{Group 3} & \textbf{Group 4} & \textbf{Group 5} \\
    $\scriptstyle \beta^{(1)} \neq 0, \ \beta^{(2)} \neq 0$ & $\scriptstyle \beta^{(4)} \neq 0, \ \beta^{(5)} \neq 0$ &  $ \scriptstyle \beta^{(6)}\neq 0, \ \beta^{(13)} \neq 0, \  \beta^{(20)} \neq 0$ & $\scriptstyle \beta^{(7)} \neq 0, \  \beta^{( 8)} \neq 0, \ \beta^{(14)} \neq 0$ & $\scriptstyle \beta^{(3)} \neq 0, \ \beta^{(9)} \neq 0, \  \beta^{(16)} \neq 0, \ \beta^{(23)} \neq 0$\\  
    (p=1) &(p=0.998) &(p=0.841)& (p=0.257) &(p=0.078) \\ 
     
    \includegraphics[scale=.09, align=c]{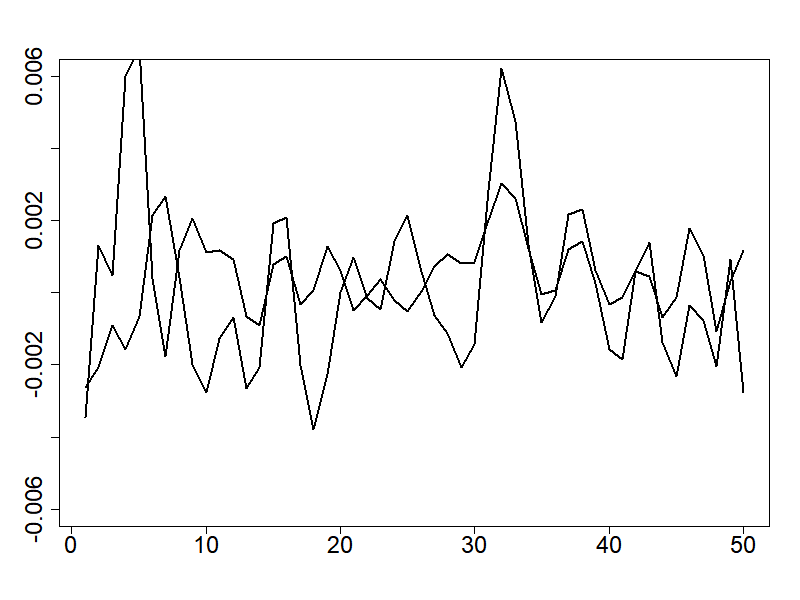} & 
    \includegraphics[scale=.09, align=c]{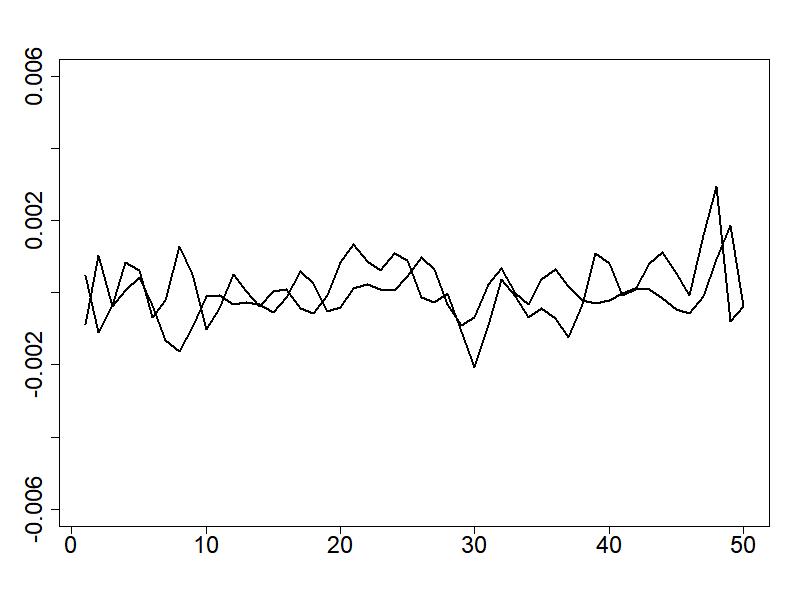} & 
    \includegraphics[scale=.09, align=c]{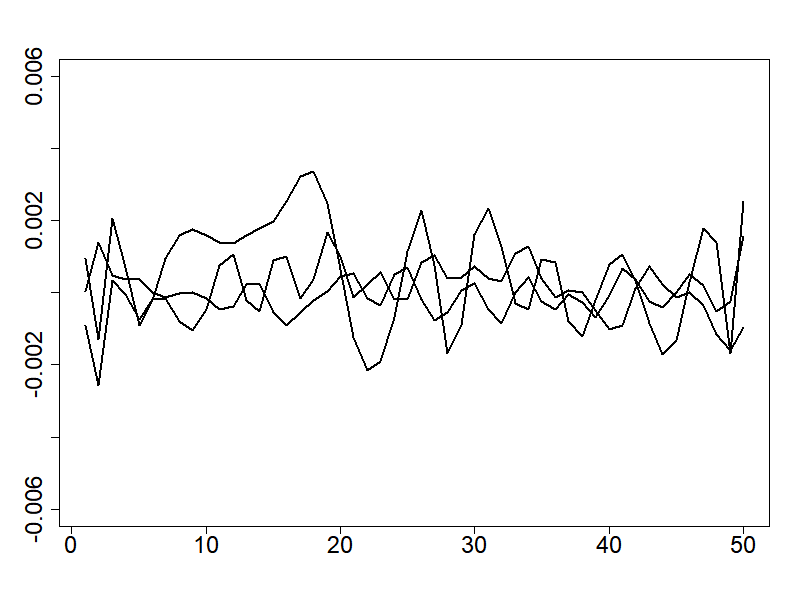} & 
    \includegraphics[scale=.09, align=c]{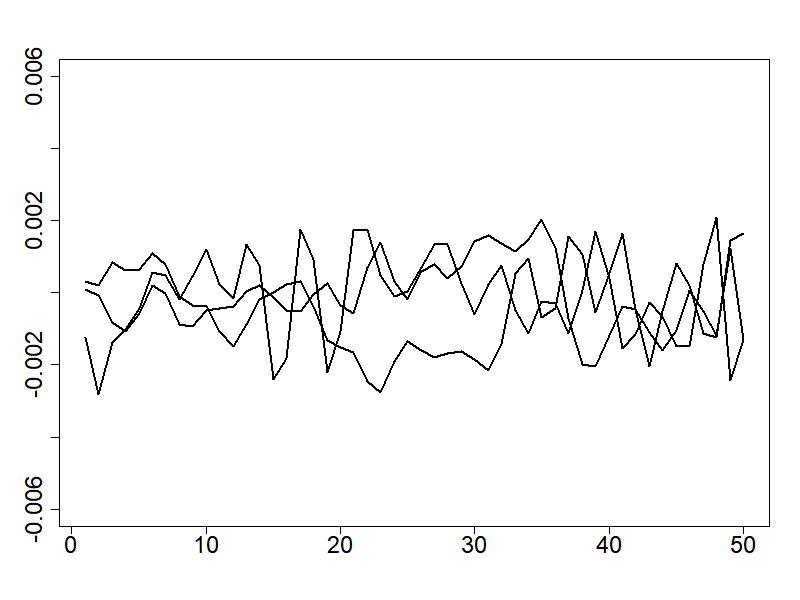} & 
    \includegraphics[scale=.09, align=c]{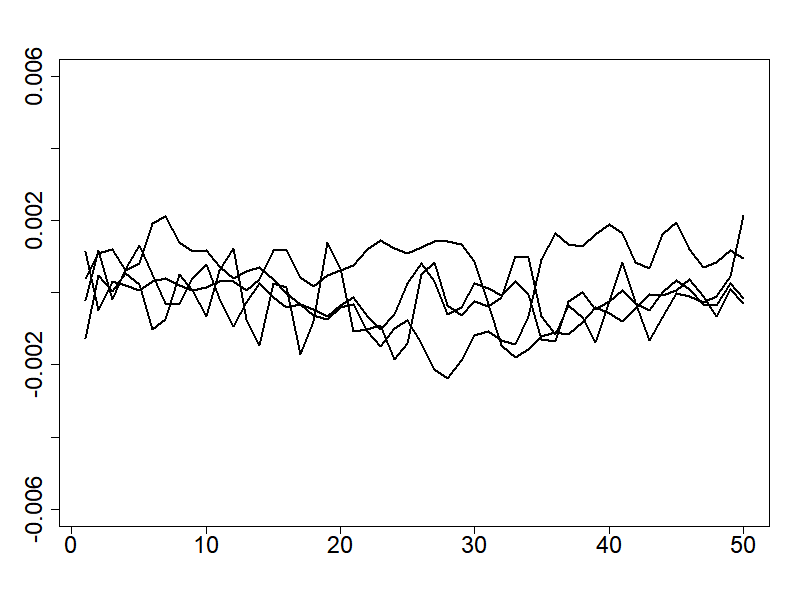} \\
    \textbf{Group 6} & \textbf{Group 7} & \textbf{Group 8} & \textbf{Group 9} & \textbf{Group 10} \\
    $\scriptstyle \beta^{(10)}\neq 0, \  \beta^{(11)} \neq 0, \ \beta^{(17)} \neq 0, \ \beta^{(18)}\neq 0$ & $\scriptstyle \beta^{(12)} \neq 0, \  \beta^{(19)} \neq 0, \ \beta^{(26)} \neq 0$ & $\scriptstyle \beta^{(15)} \neq 0, \ \beta^{(21)} \neq 0, \  \beta^{(22)} \neq 0$ & $\scriptstyle \beta^{(24)} \neq 0, \ \beta^{(25)} \neq 0$ &  $\scriptstyle \beta^{(27)} \neq 0, \ \beta^{(28)} \neq 0$ \\
    (p=0.066) &(p=0.797)& (p=0.440)& (p=0.999) &(p=0.999)\\
        \includegraphics[scale=.09, align=c]{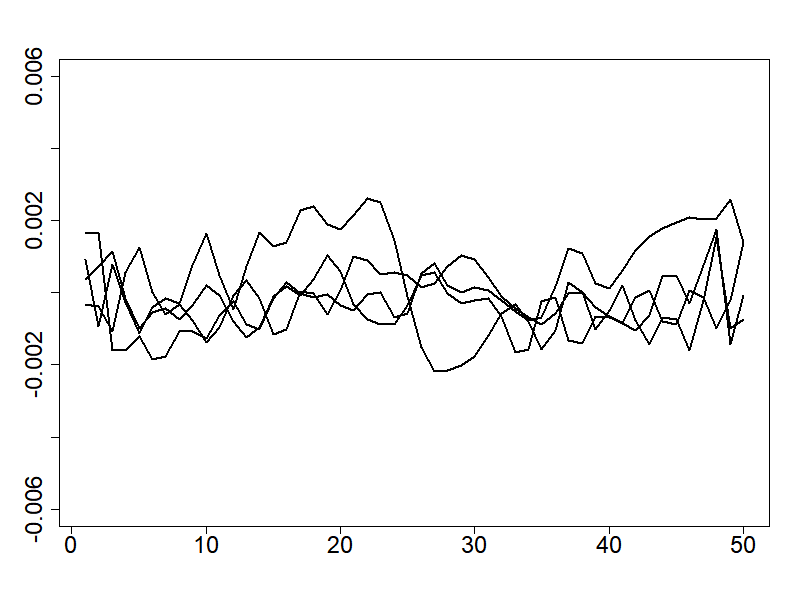} & 
    \includegraphics[scale=.09, align=c]{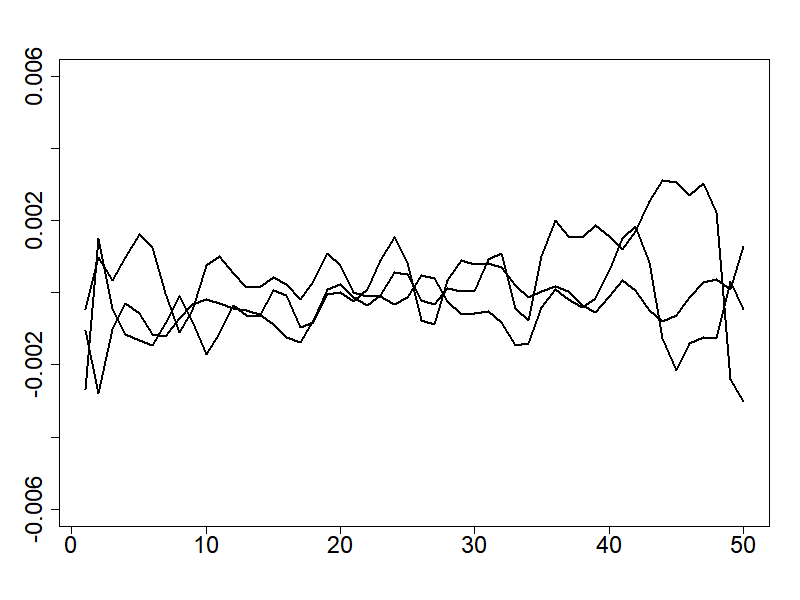} & 
    \includegraphics[scale=.09, align=c]{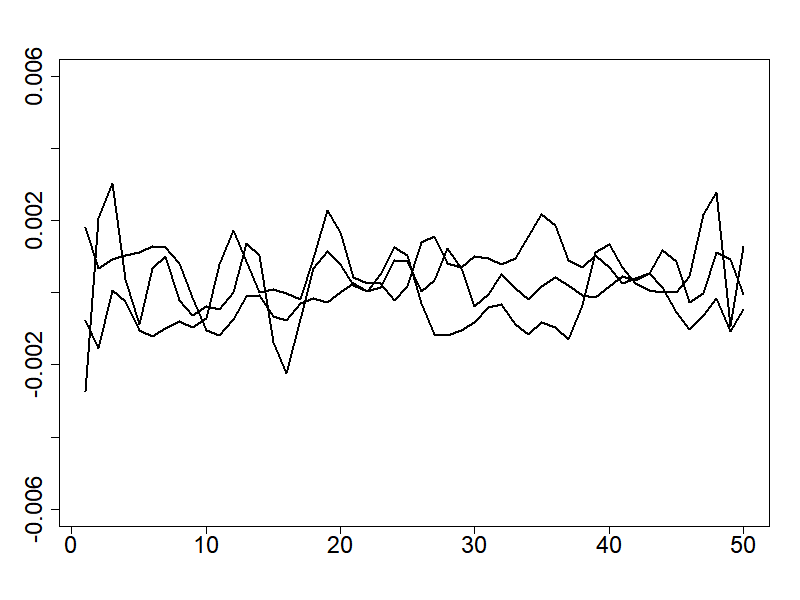} & 
    
    \includegraphics[scale=.09, align=c]{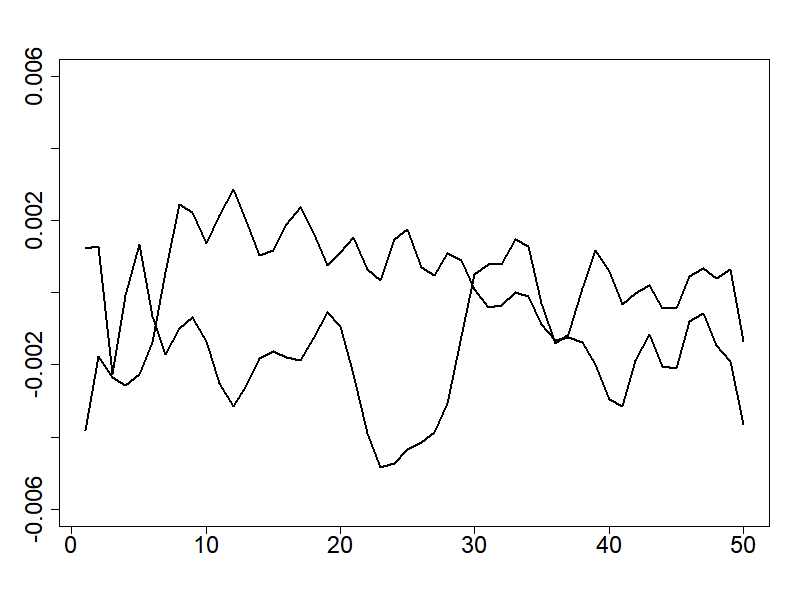} & 
    \includegraphics[scale=.09, align=c]{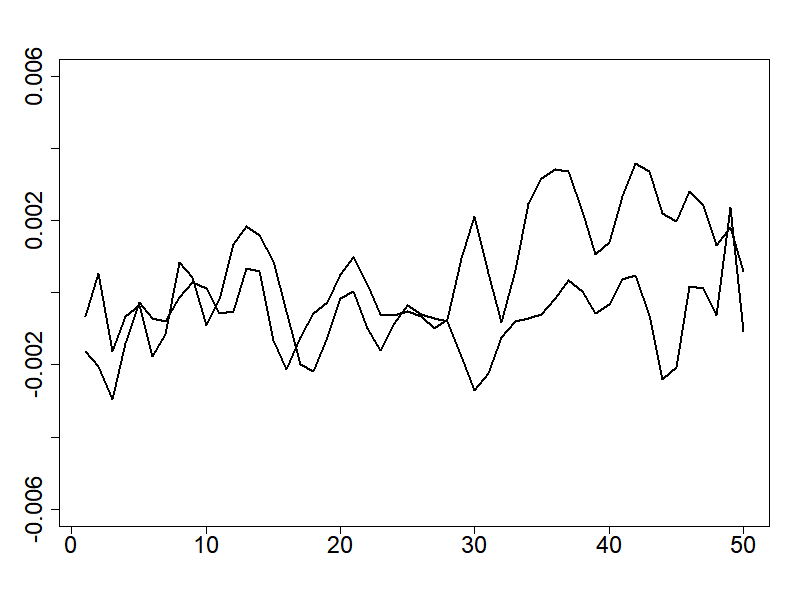}
    \end{tabular}
    }
    \caption{GL2$_\text{Euclidean}$ estimated coefficient model by groups and associated p-values ($H_0:$ all coefficients in the group are zero)}
    \label{fig_la}
\end{figure}
\newpage 

\begin{figure}[ht!]
    \centering
    \begin{tabular}{c c c c c}
    \textbf{Group 1} &\textbf{ Group 2} & \textbf{Group 3} & \textbf{Group 4} & \textbf{Group 5} \\ 
     1 2 3 4 5 6 7 8 9 &  10 11 12 17 18 19 26 &  13 14 15 20 21 22 &  16 23 24 25& 27 28 \\
     (p $<$ 0.001)& (p=0.124)& (p=0.840)& (p=1) &(p=1) \\ 
     \includegraphics[scale=.09, align=c]{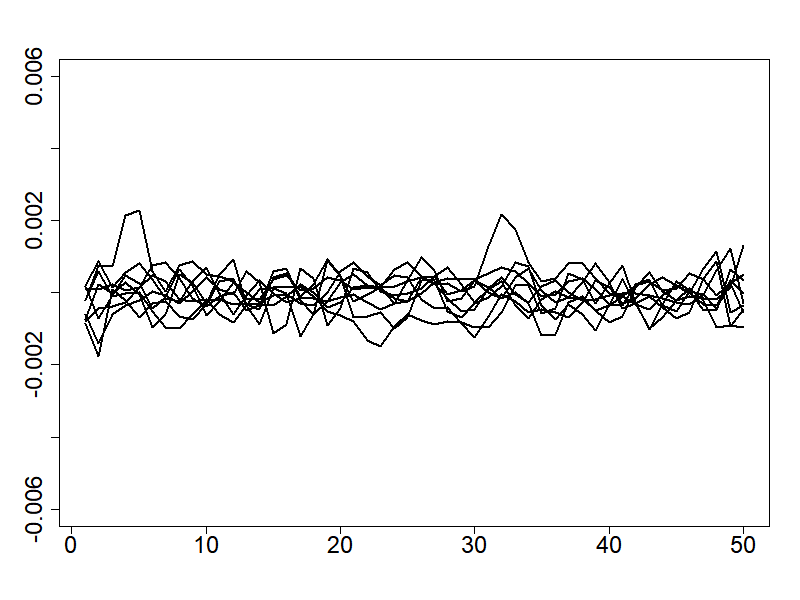} & \includegraphics[scale=.09, align=c]{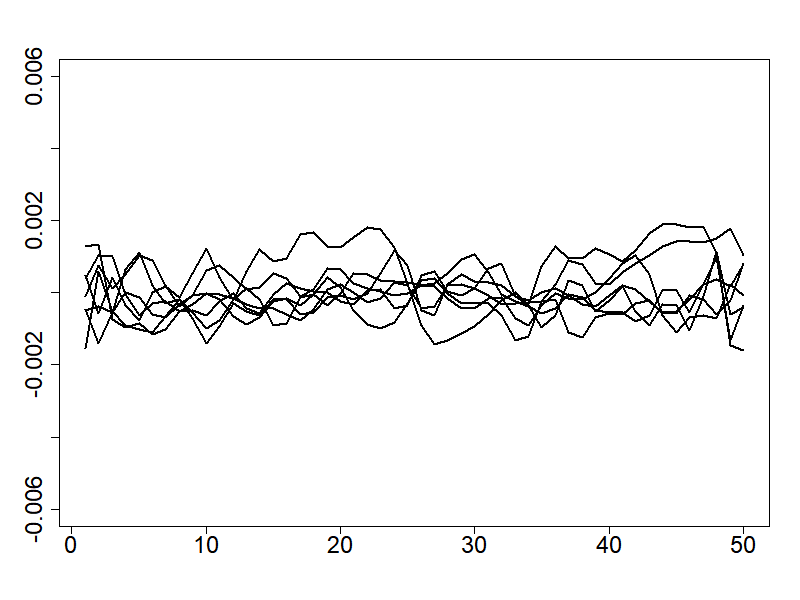} & \includegraphics[scale=.09, align=c]{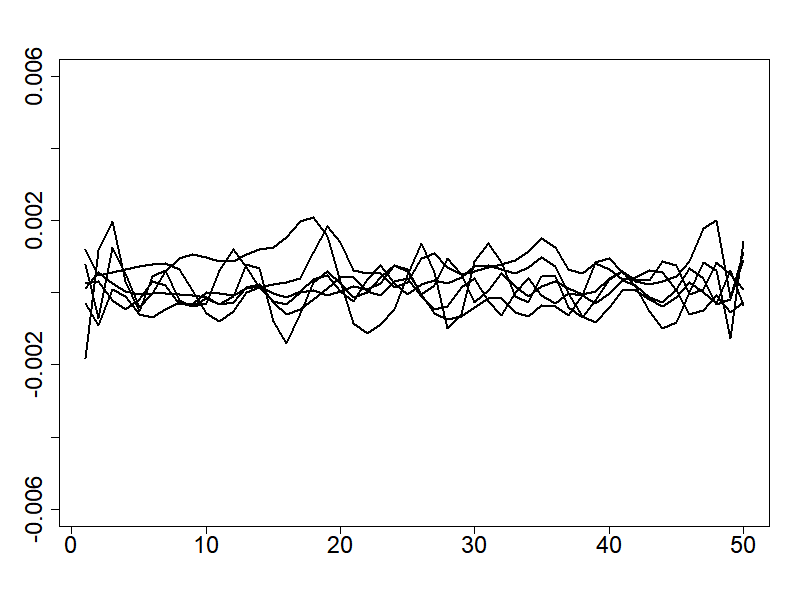} & \includegraphics[scale=.09, align=c]{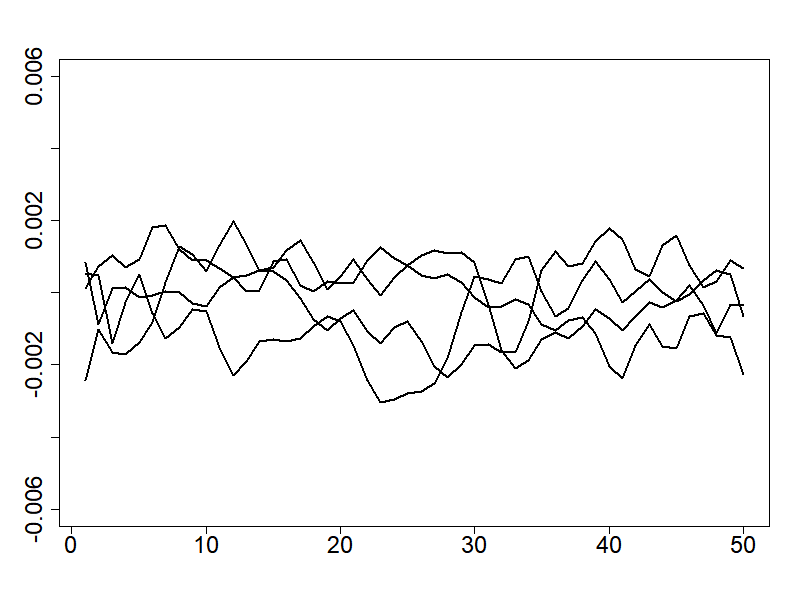} & \includegraphics[scale=.09, align=c]{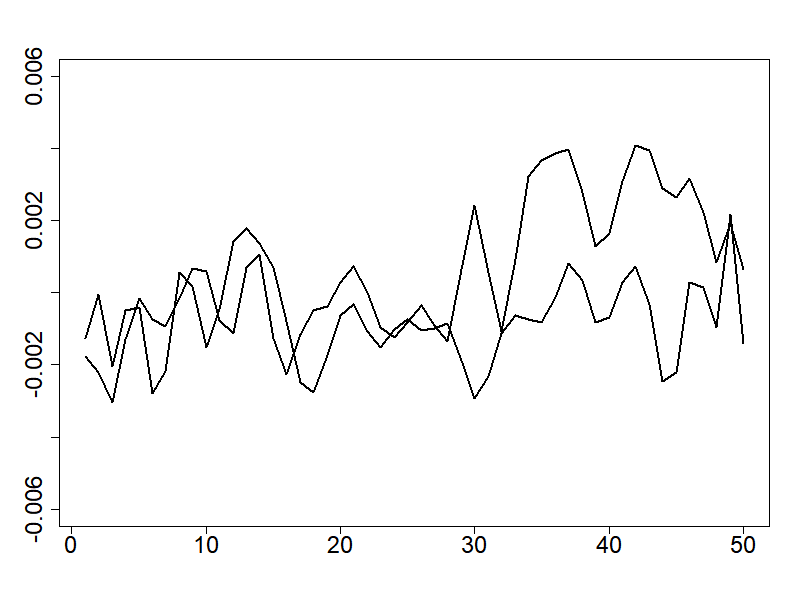}
    \end{tabular}

    \caption{ GL2$_\text{Circle}$ estimated coefficient model}
    \label{fig_la}
\end{figure}

\section{Proofs}

\begin{proof}[\textbf{Proof of Lemma} \ref{lem1}] Let the neighbor function $v$: $\{\mathcal{C}_1, \mathcal{C}_2, \ldots, \mathcal{C}_p\} \to \{1, 2, \ldots, p\}$,  as defined in Section \ref{FL} and  $\mathcal{V}_0$, $\mathcal{V}_1$ defined as:  $$\mathcal{V}_0= \{i \in  \{1, 2, \ldots, p\}, v^2(\mathcal{C}_i)=i  \} $$ and $$ \mathcal{V}_1= \{i \in \{1, 2, \ldots, p\}, i>v(\mathcal{C}_i),  v^2(\mathcal{C}_i) =i  \},$$ with $v^2(\mathcal{C}_i)$= $v(\mathcal{C}_{v(\mathcal{C}_i)} )$, for $i=1, \ldots, p$. \\ 
Observe that $i \in \mathcal{V}_0$ is equivalent to $v(\mathcal{C}_i) \in \mathcal{V}_0$, and $i \in \mathcal{V}_1$ implies that $v(\mathcal{C}_i) \not \in \mathcal{V}_1$. In other words,  $\mathcal{V}_0$ is the set of indexes corresponding to conditions for which a $2$-cycle structure is present in the $1$-NN graph. $\mathcal{V}_1$ is a subset of $\mathcal{V}_0$ with  $\text{card}\mathcal{V}_1= \frac{1}{2}\text{card}(\mathcal{V}_0)$. \\
Then, for all $\boldsymbol{f} \in {\mathcal{H}^p}$, we have
\begin{align*}
    \sum_{j=1}^ {p}||(\mathbf{L}\boldsymbol{f})^{(j)}||_2 &= \sum_{j\in \mathcal{V}_0}||f^{(j)}-f^{(v(\mathcal{C}_j))}||_2+\sum_{j \not \in \mathcal{V}_0}||f^{(j)}-f^{(v(\mathcal{C}_j))}||_2  \\ 
    &=\sum_{j\in \mathcal{V}_1}||2(f^{(j)}-f^{(v(\mathcal{C}_j))})||_2+  \sum_{j \not \in \mathcal{V}_0}||f^{(j)}-f^{(v(\mathcal{C}_j))}||_2.
\end{align*}
Then, there exists a matrix $\mathbf{L}_0 \in \mathbb{R}^{r \times p}$, with $r=p-\frac{1}{2}\text{card}(\mathcal{V}_0)$, such as 
\begin{align*}
    \sum_{j=1}^ {p}||(\mathbf{L}\boldsymbol{f})^{(j)}||_2&=\sum_{j=1}^{r} ||(\mathbf{L}_0\boldsymbol{f})^{(j)}||_{2}.
\end{align*}
Since $v$ is constructed by the one-nearest neighbor graph--only 2-cycle structures can occur \citep{eppstein1997nearest}-- then $\mathbf{L}_0$ is a full rank matrix.
\end{proof}
\begin{proof}[\textbf{Proof of Proposition \ref{prop_1}}] {\color{black} We borrow some reasoning from \cite{tibshirani2011} (the full-rank matrix case). In their paper, these authors were interested in the case where the penalty is defined using the $l1$ norm and a linear transformation of the coefficient in the multivariate case. We extend their reasoning to the $||\cdot||_{L_2, 1}$ norm and the setting of multivariate functional coefficients.} \par 
Notice that $$||\mathbf{D}\boldsymbol{f}||_{L_2, 1}= \sum_{j=1}^{r} ||( \mathbf{L}_0 \boldsymbol{f})^{(j)}||_{L_2}+  \sum_{j=r+1}^{p} ||( \mathbf{T} \boldsymbol{f})^{(j)}||_{L_2}, \  \boldsymbol{f} \in {\mathcal{H}^p}.$$ Using the definition of $\mathbf{L}_0$, $$||\mathbf{D}\boldsymbol{f}||_{L_2, 1}=|| \mathbf{L} \boldsymbol{f}||_{L_2, 1}+  \sum_{j=r+1}^{p} ||( \mathbf{T} \boldsymbol{f})^{(j)}||_{L_2, 1}. $$ Hence, the problem \eqref{f_lasso} can be written as 
\begin{equation}
\hat{\boldsymbol \beta}_{\lambda}= \argmin_{\boldsymbol{f} \in {\mathcal{H}^p}}  \frac{1}{2} \sum_{i=1}^{n}\left( y_i-\langle \boldsymbol{x}_i,\boldsymbol{f} \rangle_{\mathcal{H}^p}   \right)^2+ \sum_{j=1}^{p}\lambda \mathbb{I}(j\leq r) || (\mathbf{D}\boldsymbol{f})^{(j)}||_{ L_2}
\end{equation}
with $\mathbb{I}(\cdot)$ is the indicator function. The non-singularity of $\mathbf{D}$ implies   
\begin{equation}
\hat{\boldsymbol \psi}_{\lambda}= \argmin_{\boldsymbol{f} \in {\mathcal{H}^p}}  \frac{1}{2} \sum_{i=1}^{n}\left( y_i-\langle \boldsymbol{x}_i,\mathbf{D}^{-1} \boldsymbol{f}\rangle_{\mathcal{H}^p}   \right)^2+ \sum_{j=1}^{p}\lambda \mathbb{I}(j\leq r) || f^{(j)}||_{ L_2},
\end{equation}
with $\hat{\boldsymbol \psi}_{\lambda}=\mathbf{D} \hat{\boldsymbol \beta}_{\lambda}$. The equality
$\langle (\mathbf{D}^{-1})^{\T} \boldsymbol{x}_i, \boldsymbol{f}\rangle_{\mathcal{H}^p} = \langle \boldsymbol{x}_i,\mathbf{D}^{-1} \boldsymbol{f}\rangle_{\mathcal{H}^p}$ concludes the proof. 
\end{proof}

\begin{proof}[\textbf{Proof of Lemma} \ref{lemma2}] To simplify the notation, let denote by $\boldsymbol{f}_{\mathcal{I}_k}$ the function composed only of the set of dimensions of $\boldsymbol{f}\in {\mathcal{H}^p}$ which belong to $\mathcal{I}_k$. \par 
The proof of the lemma relies on the following statements:
\begin{enumerate}
    \item[(a)]  For $\boldsymbol{f} \in {\mathcal{H}^p}$,  $$||\boldsymbol{f}_{\mathcal{I}_{p_k}}-\bar{f}_{\mathcal{I}_k} \Vec{1}_{p_k} ||_{L_2, 2}= || \mathrm{R}_k \boldsymbol{f}_{\mathcal{I}_k }||_{L_2, 2}.$$
    for all $k\in \{1, \ldots, K\}$. 
    \item[(b)] The matrices $\bar{\mathbf{M}}$ and $\mathbf{R}$ are such that    $\mathbf{R}\bar{\mathbf{M}}^\top=\mathbf{0}_{(p-K)\times K} $
\end{enumerate}
For the first point (a), direct calculation shows that  
\begin{align}
\boldsymbol{f}_{\mathcal{I}_k}-\bar {f}_{\mathcal{I}_k} \Vec{1}_{p_k}& =\underbrace{\left[ \mathbb{I}_{p_k \times p_k}-\displaystyle\frac{1}{p_k} \mathbf{1}_{p_k\times p_k} \right]}_{\mathbf{P}_k}\boldsymbol{f}_{\mathcal{I}_k}.
    \end{align}
The rank of  $\mathbf{P}_k$ is $p_k-1$.  Let $\mathbf{R}_k $ be the R reduced rank matrix of size $(p_k-1)\times |p_k|$  obtained by the QR decomposition of  $\mathbf{P}_k$. Since  $||\cdot ||_{L_2, 2}$ is the Frobenius function norm, we have (a), i.e. $||\mathbf{P}_k\boldsymbol{f}_{\mathcal{I}_k}||_{L_2, 2}=||\mathbf{R}_k\boldsymbol{f}_{\mathcal{I}_k}||_{L_2, 2}$.\\ 
For point (b), without loss of generality, we assume that 
\begin{equation*}
    \mathbf{M} = \begin{pmatrix}  \Vec{1}_{p_1}^{\T} & \Vec{0}_{p_2}^{\T} & ... & 0 \\ 
    \Vec{0}_{p_1}^{\T}  & \Vec{1}_{p_2}^{\T} & ... & 0 \\
    ...  &  &  &  ...\\ 
    \Vec{0}_{p_1}^{\T}  & \Vec{0}_{p_2}^{\T}  & ...& \Vec{1}_{p_K}^{\T} 
    \end{pmatrix}.  
\end{equation*}
Note that $\Vec{1}_{p_k}$ belongs to the kernel of $\mathbf{P}_k$, i.e $\mathrm{P}_k\Vec{1}_{p_k} = \Vec{0}_{p_k}$, for all $k \in \{1, \ldots, K\}$. From the definition of $\mathbf{R}_k$,it follows that $\mathbf{P}_k= \mathbf{Q}_k \begin{pmatrix} \mathbf{R}_k\\ 
     \mathbf{0}_{p_k}^{\T} \end{pmatrix}$, where $\mathbf{Q}_k$ is an orthogonal matrix. Then,  $\mathbf{P}_k\Vec{1}_k= \mathbf{0}_{p_k}$ implies that $\mathbf{R}_k\Vec{1}_k= \mathbf{0}_{p_k-1}$. As $\mathbf{R}_k\Vec{1}_k= \mathbf{0}_{p_k-1}$ for all $k\in \{1, \ldots, K\}$, we have    
$$
\mathbf{R}\bar{\mathbf M}^\top= \mathbf{0}_{(p-K)\times K}.  
$$ 
Finally, as a direct consequence of (a), the matrix $\mathbf{G}_\alpha$ satisfies the relation \eqref{eq_prop_2}. Observe that $\mathbf{G}_\alpha$ is non-singular as a consequence of (b). This concludes the proof.  
\end{proof}


\begin{proof}[\textbf{Proof of Proposition} \ref{prop3}] 
\
\begin{itemize}
    \item[\textbf{1.}] The equation \eqref{b_ex} implies that for each dimension $j$, $j =1, \ldots , p$, we have  
\begin{align*}
\beta^{(j)}(t)&= (b^{(j)})^{\T} \phi(t),   
\end{align*}
where $ b^{(j)}\in \mathbb{R}^{M}$ $\;  t \in [0,T]$. \\ 
 Define   $\mathrm{F}$= $\{ \langle \phi_i, \phi_j \rangle \}_{i, j}$ and $\mathrm{F}=(\mathrm{F}^{1/2})^{\T} \mathrm{F}^{1/2}$. Thus, we have $$|| \beta^{(j)}||_{L_2}=||(\mathrm{F}^{1/2})^{\T}b^{(j)}||_{2}= ||(b^{(j)})^{\T}\mathrm{F}^{1/2}||_{2}.$$ Moreover, 
$$\mathbf{B}\text{F}^{1/2}= \begin{pmatrix} (b^{(1)})^{\T} \\ 
    (b^{(2)})^{\T} \\ 
    ... \\ 
    (b^{(p)})^{\T}
    \end{pmatrix} \text{F}^{1/2}=\begin{pmatrix} (b^{(1)})^{\T}\text{F}^{1/2} \\ 
    (b^{(2)})^{\T}\text{F}^{1/2} \\ 
    ... \\ 
    (b^{(p)})^{\T}\text{F}^{1/2}
    \end{pmatrix},$$ and 
$||\beta||_{L_2, 1} = ||\mathbf{B}\mathrm{F}^{1/2}||_{2,1}$. 
\item[\textbf{2.}] Notice that $a_i=\text{vec}(\mathbf{A}_i^{\T})$, $b=\text{vec}(\mathbf{B}^{\T})$ with vec($\cdot$) denotes the vectorization operator. It follows that  
\begin{align*}
        b_0&=\text{vec}((\mathbf{Z}\mathbf{B})^{\T} ) = \text{vec}(\mathbf{B}^{\T}\mathbf{Z}^{\T}) \\ 
        &= (\mathbf{Z} \otimes  \mathrm{I}_{M\times M}) \text{vec}(\mathbf{B}^{\T} ) = (\mathbf{Z} \otimes  \mathrm{I}_{M\times M})b,  
    \end{align*}
with $\otimes$ denotes the Kronecker Product.
\end{itemize}
\end{proof}

\end{appendices}

\bibliography{bib.bib}

\begin{thebibliography}{}

\bibitem[Aguilera et~al., 2006]{aguilera2006using}
Aguilera, A.~M., Escabias, M., and Valderrama, M.~J. (2006).
\newblock Using principal components for estimating logistic regression with
  high-dimensional multicollinear data.
\newblock {\em Computational Statistics \& Data Analysis}, 50(8):1905--1924.

\bibitem[Beyaztas and Lin~Shang, 2022]{beyaztas2022}
Beyaztas, U. and Lin~Shang, H. (2022).
\newblock A robust functional partial least squares for
  scalar-on-multiple-function regression.
\newblock {\em Journal of Chemometrics}, 36(4):e3394.

\bibitem[Bleakley and Vert, 2011]{bleakley2011}
Bleakley, K. and Vert, J.-P. (2011).
\newblock The group fused lasso for multiple change-point detection.
\newblock {\em arXiv preprint arXiv:1106.4199}.

\bibitem[Cardot et~al., 1999]{cardot1999}
Cardot, H., Ferraty, F., and Sarda, P. (1999).
\newblock Functional linear model.
\newblock {\em Statistics \& Probability Letters}, 45(1):11--22.

\bibitem[Chen and M{\"u}ller, 2012]{chen2012}
Chen, K. and M{\"u}ller, H.-G. (2012).
\newblock Modeling repeated functional observations.
\newblock {\em Journal of the American Statistical Association},
  107(500):1599--1609.

\bibitem[Eppstein et~al., 1997]{eppstein1997nearest}
Eppstein, D., Paterson, M.~S., and Yao, F.~F. (1997).
\newblock On nearest-neighbor graphs.
\newblock {\em Discrete \& Computational Geometry}, 17(3):263--282.

\bibitem[Escabias et~al., 2005]{escabias2005}
Escabias, M., Aguilera, A., and Valderrama, M. (2005).
\newblock Modeling environmental data by functional principal component
  logistic regression.
\newblock {\em Environmetrics: The official journal of the International
  Environmetrics Society}, 16(1):95--107.

\bibitem[Godwin, 2013]{groupfda}
Godwin, J. (2013).
\newblock Group lasso for functional logistic regression.
\newblock Master's thesis.

\bibitem[G{\'o}recki et~al., 2015]{gorecki2015}
G{\'o}recki, T., Krzy{\'s}ko, M., and Wo{\l}y{\'n}ski, W. (2015).
\newblock Classification problems based on regression models for
  multi-dimensional functional data.
\newblock {\em Statistics in Transition new series}, 16(1).

\bibitem[Ismail~Fawaz et~al., 2020]{inception}
Ismail~Fawaz, H., Lucas, B., Forestier, G., Pelletier, C., Schmidt, D.~F.,
  Weber, J., Webb, G.~I., Idoumghar, L., Muller, P.-A., and Petitjean, F.
  (2020).
\newblock Inceptiontime: Finding alexnet for time series classification.
\newblock {\em Data Mining and Knowledge Discovery}, 34(6):1936--1962.

\bibitem[Jacob et~al., 2009]{jacob2009}
Jacob, L., Obozinski, G., and Vert, J.-P. (2009).
\newblock Group lasso with overlap and graph lasso.
\newblock In {\em Proceedings of the 26th annual international conference on
  machine learning}, pages 433--440.

\bibitem[Jacques and Preda, 2014]{jacques2014}
Jacques, J. and Preda, C. (2014).
\newblock Model-based clustering for multivariate functional data.
\newblock {\em Computational Statistics \& Data Analysis}, 71:92--106.

\bibitem[Land and Friedman, 1997]{fusion_variable}
Land, S.~R. and Friedman, J.~H. (1997).
\newblock Variable fusion: A new adaptive signal regression method.
\newblock {\em Dept. Statistics, Carnegie Mellon Univ. Pittsburgh, Pittsburgh,
  PA, USA, Rep}, 656.

\bibitem[Meier, 2009]{grplasso}
Meier, L. (2009).
\newblock grplasso: Fitting user specified models with group lasso penalty.
\newblock {\em R package version 0.4-2}.

\bibitem[Meier et~al., 2008]{group_rl}
Meier, L., Van De~Geer, S., and B{\"u}hlmann, P. (2008).
\newblock The group lasso for logistic regression.
\newblock {\em Journal of the Royal Statistical Society: Series B (Statistical
  Methodology)}, 70(1):53--71.

\bibitem[Moindji{\'e} et~al., 2024]{moindjie2024classification}
Moindji{\'e}, I.-A., Dabo-Niang, S., and Preda, C. (2024).
\newblock Classification of multivariate functional data on different domains
  with partial least squares approaches.
\newblock {\em Statistics and Computing}, 34(1):5.

\bibitem[Preda and Saporta, 2002]{PLS}
Preda, C. and Saporta, G. (2002).
\newblock R{\'e}gression pls sur un processus stochastique.
\newblock {\em Revue de statistique appliqu{\'e}e}, 50(2):27--45.

\bibitem[Ramsey and Silverman, 2005]{ramsay2008}
Ramsey, J.~O. and Silverman, B.~W. (2005).
\newblock {\em Functional Data Analysis}.
\newblock Springer-Verlag, 2 edition.

\bibitem[Ruiz et~al., 2021]{great}
Ruiz, A.~P., Flynn, M., Large, J., Middlehurst, M., and Bagnall, A. (2021).
\newblock The great multivariate time series classification bake off: a review
  and experimental evaluation of recent algorithmic advances.
\newblock {\em Data Mining and Knowledge Discovery}, 35(2):401--449.

\bibitem[Tibshirani et~al., 2005]{fused_lasso}
Tibshirani, R., Saunders, M., Rosset, S., Zhu, J., and Knight, K. (2005).
\newblock Sparsity and smoothness via the fused lasso.
\newblock {\em Journal of the Royal Statistical Society: Series B (Statistical
  Methodology)}, 67(1):91--108.

\bibitem[Tibshirani and Taylor, 2011]{tibshirani2011}
Tibshirani, R.~J. and Taylor, J. (2011).
\newblock The solution path of the generalized lasso.
\newblock {\em The annals of statistics}, 39(3):1335--1371.

\bibitem[Yang et~al., 2016]{post_inference}
Yang, F., Foygel~Barber, R., Jain, P., and Lafferty, J. (2016).
\newblock Selective inference for group-sparse linear models.
\newblock {\em Advances in neural information processing systems}, 29.

\bibitem[Yi et~al., 2022]{yi2022}
Yi, Y., Billor, N., Liang, M., Cao, X., Ekstrom, A., and Zheng, J. (2022).
\newblock Classification of eeg signals: an interpretable approach using
  functional data analysis.
\newblock {\em Journal of Neuroscience Methods}, 376:109609.

\bibitem[Yuan et~al., 2011]{yuan2011_2}
Yuan, L., Liu, J., and Ye, J. (2011).
\newblock Efficient methods for overlapping group lasso.
\newblock {\em Advances in neural information processing systems}, 24.

\bibitem[Yuan and Lin, 2006]{yuan2006model}
Yuan, M. and Lin, Y. (2006).
\newblock Model selection and estimation in regression with grouped variables.
\newblock {\em Journal of the Royal Statistical Society: Series B (Statistical
  Methodology)}, 68(1):49--67.

\bibitem[Zou and Hastie, 2005]{zou2005}
Zou, H. and Hastie, T. (2005).
\newblock Regularization and variable selection via the elastic net.
\newblock {\em Journal of the Royal Statistical Society Series B: Statistical
  Methodology}, 67(2):301--320.

\end{thebibliography}
\bibliographystyle{apalike}

\end{document}